\documentclass{statsoc}

\usepackage[a4paper, margin=1.2 in]{geometry}
\usepackage{natbib}
\RequirePackage{amsmath,float,amssymb,bm,graphics,graphicx,latexsym,rotate,lscape,color,epsfig,epsf,psfrag,multirow,rotate,
float,mathtools,enumitem,subcaption,caption}
\RequirePackage[colorlinks,citecolor=blue,urlcolor=blue]{hyperref}

\def\references{\bibliography{1-1-19}}

\newtheorem{thm}{Theorem}
\newtheorem{Lemma}{Lemma}
\newtheorem{prop}{proposition}
\newtheorem{cor}{Corollary}
\def\red{\textcolor{red}}

\newcommand{\bea}{\begin{eqnarray*}}
	\newcommand{\eea}{\end{eqnarray*}}
\newcommand{\be}{\begin{eqnarray}}
\newcommand{\ee}{\end{eqnarray}}
\newcommand{\ed}{\end{document}}

\newcommand{\btab}{\begin{tabular}}
\newcommand{\etab}{\end{tabular}}

\newcommand{\bi}{\begin{itemize}}
\newcommand{\ei}{\end{itemize}}
\newcommand{\bfi}{\begin{figure}}
\newcommand{\efi}{\end{figure}}
\newcommand{\ben}{\begin{enumerate}}
\newcommand{\een}{\end{enumerate}}
\newcommand{\bay}{\begin{array}}
\newcommand{\eay}{\end{array}}

\def\id{\hspace{.5cm}}

\def\F{Fr\'{e}chet}
\def\o{\omega}
\def\O{\Omega}
\def\C{\text{Cov}_{\Omega}}

\def\bco{\iffalse}

\def\ci{\cite}
\def\cp{\citep}
\def\eps{\varepsilon}

\newcommand{\bc}{\begin{center}}
\newcommand{\ec}{\end{center}}

\def\Cov{{\rm Cov}}

\def\Var{{\rm Var}}

\def\argmin{{\rm argmin}}
\def\arginf{{\rm arginf}}

\def\ci{\cite}
\def\cp{\citep}

\def\hs{\hspace{0.2cm}}

\def\eps{\varepsilon}

\def\o{\omega}
\def\O{\Omega}
\def\RR{\rho_{\Omega}}

\def\F{Fr\'{e}chet}

\def\fro{functional random objects}

\def\d{\rm d}

\def\d2{d_2}

\def\mbb{\mathbb}

\def\s1n{\sum_{i=1}^n}
\def\p1n{\prod_{i=1}^n}
\def\i01{\int_0^1}
\def\1d{{T_\delta}}
\def\1d{{1_\delta}}
\def\1/n{\frac{1}{n}}

\def\R{\mathbb{R}}

\vspace{-2cm}

\title[Functional Models for  Time-Varying Random Objects]{Functional Models for Time-Varying Random Objects}
\author{Paromita Dubey}
\address{University of Califonia, Davis}
\email{pdubey@ucdavis.edu}
\author[Dubey and M\"uller]{Hans-Georg M\"uller}
\address{University of Califonia, Davis}

\begin{document}

\begin{abstract} Functional data analysis provides a popular toolbox of  functional models for the analysis of  samples of random functions that  are real-valued. In recent years, samples of time-varying object data such as time-varying networks that are not in a vector space 
have been increasingly collected. These  data {can be viewed as  elements
	of a general metric space that lacks local or global linear structure
and therefore} common approaches that have been used with great success for the analysis of functional data, such as functional principal component analysis, cannot be applied.  In this paper we propose \emph{metric covariance}, a novel association measure for paired object data lying in a metric space $(\O,d)$ that we  use  to define a metric auto-covariance function for a sample of random $\O$-valued curves, where $\O$ generally will not have a  vector space or manifold structure. The proposed metric auto-covariance function is non-negative definite when the squared semimetric $d^2$ is of negative type. Then  the eigenfunctions of the linear operator with the auto-covariance function as kernel can be used as building blocks for an  
\emph{object functional principal component analysis} for   
$\O$-valued functional data, including time-varying probability distributions,  covariance matrices and time-dynamic networks.  Analogues of functional principal components for time-varying objects are obtained by applying  \F$\,$  means and  projections of distance functions of the random object trajectories   in the directions of the eigenfunctions,  leading to real-valued  \emph{\F \ scores}. Using the notion of  \emph{generalized \F \ integrals}, 
we construct \emph{object functional principal components} that lie in the metric space $\O$.   We establish asymptotic consistency of the sample based estimators for the corresponding population targets under mild metric entropy conditions on $\O$ and continuity of the $\O$-valued random curves. These concepts   are illustrated with  samples of time-varying probability distributions for human mortality, time-varying 
covariance matrices derived from trading patterns, and  time-varying networks that arise from New York taxi trips. \vspace{.3cm}

KEY WORDS:    Dynamic Networks, Fr\'echet Integral,  Functional Data Analysis, Metric Covariance, Object Data, Principal Component Analysis, Stochastic Processes.

\end{abstract}

\section{Introduction}
Time-varying data where one collects an i.i.d. sample of random functions, which take values in a general object space that does not have a linear structure,  are  increasingly common, while the statistical methodology for the analysis of such data has been lagging behind. We aim to introduce techniques that will help to fill this gap.   For the case where observations  consist of samples of random trajectories  that take values  in $\R^p$, the methodology of choice is often Functional Data Analysis (FDA)  \cp{rams:05, horv:12, mull:16:3}, where methodology 
for one-dimensional ($p=1$)  functional data is readily  available. Models for functional data that consist of vector-valued processes ($p>1$) have been studied more recently \cp{zhou:08,berr:11,chio:14, clae:14,verb:14,chio:16} as well as  the case where at each time point one records a random function, i.e. function-valued stochastic processes \cp{park:15,mull:12:3,mull:17:4}. In these models, the responses are situated in a linear space, either the Euclidean space $\R^p$ or the Hilbert space $L^2$. Functions of objects in spaces that can be locally approximated by linear spaces such as Riemannian manifolds {including} spheres have also been considered more recently \cp{lin:17,mull:18:4}.
The major objective of this paper is to overcome the global or local linearity assumptions inherent in these previous approaches. The challenge is that existing  FDA methodology relies on 
vector operations and inner products, which are no longer available. 

Functional Principal Component Analysis (FPCA)  \cp{klef:73,daux:82} has emerged as the method of choice to represent and interpret samples of random functions that take values in linear spaces.  It also provides dimension reduction by expanding an underlying random process  into the basis functions given by the eigenfunctions of the auto-covariance operator and then truncating this expansion at a finite number of expansion terms. A related tool are the modes of variation, which enable exploration of the effects of single eigendirections \cp{cast:86,jone:92} and are useful in practical applications \cp{dong:17}. 
FPCA also provides a starting point for many theoretical investigations and  FDA techniques such as functional clustering \cp{chio:07,jacq:14,suar:16} or regression and classification \cp{mull:05:4,mull:17:6}.

As we enter the era of big data, it has become increasingly common to observe more complex, often non-Euclidean, data on a time grid. Technological advances  have made it possible to record and efficiently store time courses of image, network, sensor  or other complex data. For example, neuroscientists are interested in dynamic functional connectivity, where one essentially observes samples of time-varying covariance or correlation matrices obtained from functional Magnetic Resonance Imaging (fMRI) data for each subject in a sample. 
Time-varying network data arise in various forms, e.g.  road or internet traffic networks or time-evolving social networks, and  it is of interest to extract structure and patterns from such data.  

To obtain efficient and interpretable summaries of the  information contained in samples of complex observations is a major task in modern statistics that has led for example to  the development of methods such as Geodesic Principal Component Analysis (GPCA) in the space of 
probability distributions on $\R$ \cp{bigo:17} and on more general  Hilbert spaces \cp{segu:15} that utilize  optimal transport geometry and geodesic curves under the Wasserstein metric. These approaches utilize  geodesics to connect the random distributions with the Wasserstein barycenters. We aim here at identifying dominant directions of variation in a sample of time-varying random object trajectories, where the random objects are indexed by time and are situated in a general metric space. The time-varying aspect provides for a  novel and   little explored setting, and to develop  tools that are supported by theory and are useful for the further exploration and analysis of such  data is the main motivation for this paper. 

While FPCA for samples of functions taking values in smooth Riemannian manifolds has been studied both practically and theoretically  \cp{anir:17,mull:18:4}, these approaches critically depend on the local Euclidean property of Riemannian manifolds and cannot be extended  to functional data objects that take values in more general metric spaces that do not have a tractable and relatively simple Riemannian geometry. FPCA for doubly functional data, where the observations at each time point are functions rather than scalars \cp{mull:17:4},   is based on a tensor product representation of the underlying function-valued stochastic process. The functions need to be  Hilbert-space valued, so that this approach cannot be  applied to non-Hilbertian data. 
Due to the lack of linear structure, developing a form of FPCA for random functions taking values in a metric space, which we refer to as \emph{\fro}, is a major challenge. 

Consider a totally bounded {separable} metric space $(\O,d)$ and a random sample of fully observed $\O$-valued functional data. Aiming to extend key tools of FDA to cover such data, we first revisit  the well-established  FPCA for the case of real valued functional data. The essence of FPCA is  contained in the auto-covariance  structure of the underlying random functions, i.e. their covariance at  different time points. This leads to the question how  to quantify correlation between random objects in general metric spaces that correspond to the values of the random function at different time points.  An example for such an extension of Pearson correlation to the case of   multivariate data is  the RV coefficient \cp{robe:76}, which is zero if all of the vector components are uncorrelated and strictly positive otherwise. 

In this article we introduce \emph{metric covariance}, which is a novel association measure for paired data in general metric spaces. Metric covariance differs in key aspects from  distance correlation, another measure of dependence between metric space data \cp{lyon:13,szek:17}, the latter being primarily suited to measure probabilistic independence rather than  for quantifying the strength of  `positive' or `negative' association, which is the primary goal of the former.  Unlike distance covariance, the magnitude of metric covariance quantifies the degree of association between paired data objects. The key component of FPCA is to decompose the variation in a sample of trajectories into  orthogonal directions.  An important difference between metric covariance and distance covariance, which is specifically relevant in this context, arises  when considering the associated notion of variance. In contrast to distance correlation,  metric covariance of a random object  with itself leads to  an interpretable  notion of variance for data objects, as we will demonstrate below. We also show that metric covariance is symmetric and non-negative definite whenever the squared distance $d^2$ is a semimetric of  negative type \cp{sejd:13,lyon:13,scho:38}. The notion of  \emph{metric correlation} can then be easily derived from metric covariance and random objects will be considered to be uncorrelated  if they have a  metric correlation of size 0.

In FPCA for $\R$-valued functional data, once  the auto-covariance function has been determined,   one defines a linear Hilbert-Schmidt operator whose eigenfunctions represent the orthonormal directions of variance for the  functional data in the Hilbert space $L^2$. The corresponding eigenvalues represent the fraction of variance explained by the respective functional principal components (FPCs), which are the lengths of the projections of the functional data in the direction of each eigenfunction. How can one extend these ideas to object-valued  functional data, where one does not have a linear structure or inner product? We proceed  by constructing a linear Hilbert-Schmidt operator using the proposed metric covariance as its kernel and utilize  its eigenfunctions and eigenvalues. For real valued functional data, one obtains the FPCs by the Karhunen-Lo\`eve expansion of centered functional data in the eigenbasis, where the FPCs are  the inner products of the centered functional data with respect to the eigenfunctions. Unfortunately it is not possible to `center' object functional data living in general metric spaces and one also does not have an inner product. In the case of FDA in the Hilbert space $L^2$, the inner products can be expressed as integrals. While due to the lack of linear structure there is no integral for functional random objects, the interpretation of inner products as integrals nevertheless provides a way forward that we develop in this paper. We propose two approaches for obtaining FPCs for object functional data, one in which the FPCs are scalar irrespective of the nature of the metric space in which the random objects live,  and an alternative approach  in which the FPCs themselves are random objects, i.e. $\O$ valued. 

To obtain FPCs in object space, we introduce the notion of a \emph{generalized \F \ integral} of an $\O$-valued curve with respect to a real valued function, where this integral resides in $\O$.  Generalized \F \ integrals depend on the underlying metric $d$ in $\O$ and are defined under the constraint that the real valued function in the integrand integrates to one. We draw inspiration from the covariance integral for multivariate functional data that was previously  introduced as a  \F \ integral \cp{mull:16:2}. This previous integral  is a special case of the generalized \F \ integral  introduced here; it  corresponds to the special case where  $\O$ is the space of covariance matrices and the real valued function in the integrand is the constant function one. 
We demonstrate that the  resulting Object Functional Principal Components (Object FPCs), which reside in $\O$, provide useful insights about the structure of the underlying  \fro.

For an alternative   scalar approach, we extract relatively simple characteristics from the object functional data. A first step is to define a `mean' function using the notion of \F \ means \cp{frec:48}. This mean function resides in the object function space and serves as a `central' trajectory for the object functional data. To obtain a  representative scalar FPC for a specific random object trajectory and eigenfunction, we utilize projections of the distance function between the specific  random object trajectory  and the \F \ mean trajectory on each of the eigenfunctions. The resulting  \emph{\F \ scores} encapsulate variation in the departures of \fro \ from the \F \ mean trajectory. As we illustrate in simulations and data analysis,  
plotting these  \F \ scores against each other often illustrates meaningful patterns in the sample of object functional data that are generally  hard to capture visually, due to their complexity and non-linearity.  For example, such plots can aid in detecting the presence of clusters or outliers in \fro.

In this paper, we have three major objectives. First, we lay out a framework for extending FPCA to general metric space valued functional data. The  population target parameters are the \emph{metric auto-covariance} operator, its eigenvalues and eigenfunctions and  the population \F \ mean function, which are introduced in section \ref{sec: cov}; additionally,  the  object FPCs, which are generalized \F \ integrals and the \F \ scores (section \ref{pc}).   Second, we provide sample based estimators of these population targets and establish their asymptotic properties under mild restrictions on the metric entropy of the metric space $\O$ and the continuity of the object functional data (section \ref{theory}). Proofs of all results are in section \hyperref[proofs]{S1} of the online supplement. Third, we illustrate our results through simulations (section \ref{sim}) and various data examples (section \ref{data}), which include samples of time-varying probability distributions of age at death obtained from human mortality data of 32 countries,   time-varying yellow taxi trip networks of different regions in Manhattan observed daily during the year 2016, and of  changing trade patterns between countries that can be represented as time-varying covariance matrices, followed by a brief discussion (section \ref{dis}).

\section{\textbf{Metric Covariance}}
\label{sec: cov}

\subsection{Covariance and correlation for random objects}

{We consider   a totally bounded {separable} metric space $(\O,d)$ where $d$ is a metric and an $\O$-valued stochastic process
 $X=\lbrace X(t)\rbrace_{t \in [0,1]}$ on the interval $[0,1]$.} 
 With  $P$ denoting the probability measure of the random process $X$, we are given a sample  
  $\lbrace X_i=(X_i(t))_{t \in [0,1]}: i=1,2, \hdots ,n \rbrace$  of random $\O$-valued functions on $[0,1]$ generated by $P$. The simplest case is $\O=\mbb{R}$ with the intrinsic Euclidean metric, where  $\lbrace X_1, X_2, \hdots , X_n\rbrace $ is a sample of real valued functional data. For general metric spaces $\O$, we refer to $\lbrace X_1, X_2, \hdots , X_n\rbrace $ as a sample of \fro. Inspired by the approach to FPCA for real valued functional data, our first goal  is to quantify the association between random objects $X(s)$ and $X(t)$ in $\O$, where $s$ and $t$ are two arbitrary points in the domain $[0,1]$.

For motivation, consider first the case of real random variables  $\left(U,V\right)$ with finite covariance. Imagine for a moment that we cannot add, subtract or multiply these r.v.s and  are restricted to compute their distances $d_{E}(U,V)=\lvert U-V\rvert$. As is well known,  one then  can write the variance of $U$ using an i.i.d. copy $U'$ of $U$ by
$\Var(U)=\frac{1}{2} Ed_E^2(U,U')$. 

Interestingly, this non-algebraic construction can be extended 
  to the  covariance of $U,V$:  Let $\left(U',V'\right)$ be an i.i.d. copy of $\left(U,V\right)$. We  then obtain an alternate formulation of $\Cov(U,V)$ in terms of pairwise distances as follows,
	\begin{align*}
	&\Cov(U,V)=E\left((U-E(U))(V-E(V))\right)\\ &=\frac{1}{4} E\left(d_E^2\left(U,V'\right)+d_E^2\left(U',V\right)-2d_E^2\left(U,V\right)\right).
	\end{align*}
If $(U,V)$ are $\mathbb{R}^d$-valued random variables with $d_E(\cdot,\cdot)$ denoting  the Euclidean distance in $\mathbb{R}^d$, a simple calculation shows  that in this case,
	\begin{align*}
	& \frac{1}{4} E\left(d_E^2\left(U,V'\right)+d_E^2\left(U',V\right)-2d_E^2\left(U,V\right)\right) \\ & = E\left((U-E(U))^{T}(V-E(V))\right),
	\end{align*}
	which is the inner product in the Hilbert space of $\R^d$-valued random variables with finite $E(U^TU)$. 
Next consider the case where  $(U,V)$ are $\mathcal{H}$-valued random variables, where $\mathcal{H}$ is a Hilbert space and $d_E(\cdot,\cdot)$ is replaced by $d_{\mathcal{H}}(U,V)=||U-V||_{\mathcal{H}}$, the metric that arises from the inner product $\langle \cdot,\cdot \rangle_{\mathcal{H}}$ of the Hilbert space. If the metric $d_{\mathcal{H}}(\cdot, \cdot)$ is bounded then $E(||U||^2_{\mathcal{H}}) < \infty$. One can show with some simple algebra and utilizing the Riesz representation theorem that
	\be
	\label{eq: cov_hilbert}
	 \frac{1}{4} E\left(d_{\mathcal{H}}^2\left(U,V'\right)+d_{\mathcal{H}}^2\left(U',V\right)-2d_{\mathcal{H}}^2\left(U,V\right)\right)  = E\left(\langle U-E(U),V-E(V) \rangle_{\mathcal{H}}\right),
	\ee
	which is the inner product in $L^2(\mathcal{H})$, the Hilbert space of $\mathcal{H}$-valued random variables $U$ such that $E(||U||^2_{\mathcal{H}}) < \infty$. 

What happens if $(U,V)$ are $\O$-valued random variables and we replace $d_{\mathcal{H}}$ by $d$ where $(\O,d)$ is a general metric space with no vector space structure to rely on? Or,  for which spaces does the function $\frac{1}{4} E\left(d^2\left(U,V'\right)+d^2\left(U',V\right)-2d^2\left(U,V\right)\right)$ retain desirable properties? Proposition 3 of \cite{sejd:13} implies that  whenever { $d^2$ } is a semi-metric of negative type, there exists a Hilbert space $\mathcal{H}$ and an injective map, say $f: \O \rightarrow \mathcal{H}$, with 
\be \label{map}
d^2(U,V)=||f(U)-f(V)||^2_{\mathcal{H}},
\ee
and therefore it follows from \eqref{eq: cov_hilbert} that for some `remote' Hilbert space $\mathcal{H}$ and the unknown map $f(\cdot)$,
\begin{align}
\label{eq: cov_abstract}
& \frac{1}{4} E\left(d^2\left(U,V'\right)+d^2\left(U',V\right)-2d^2\left(U,V\right)\right) \nonumber \\ & = E\left(\langle f(U)-E(f(U)),f(V)-E(f(V)) \rangle_{\mathcal{H}}\right).
\end{align}
Here a space $(\mathcal{Z},\rho)$ with a semi-metric $\rho$ is of  negative type if 
for all $n \geq 2$, $z_1, z_2, \dots, z_n \in \mathcal{Z}$ and $\alpha_1, \alpha_2, \dots, \alpha_n \in \R$ with $\sum_{i=1}^{n} \alpha_i=0$ one has
\begin{equation*}
\sum_{i=1}^{n} \sum_{j=1}^{n} \alpha_i \alpha_j \rho(z_i,z_j) \leq 0.
\end{equation*}

These considerations motivate the following definition of a  generalized version of covariance $\C(U,V)$ for paired random objects $(U,V)$ that take  values in $\O \times \O$, where $\left(\O,d\right)$ is a {separable} metric space, 
\begin{equation}
\label{eq: cov_general}
\C(U,V)=\frac{1}{4} E\left(d^2\left(U,V'\right)+d^2\left(U',V\right)-2d^2\left(U,V\right)\right),
\end{equation}
where as above  $\left(U',V'\right)$ is an  i.i.d. copy of $\left(U,V\right)$.
We refer to $\C(U,V)$ as \emph{metric covariance} of $U$ and $V$. Metric covariance is always finite if the underlying metric space is bounded and coincides with the usual notion of covariance in Euclidean spaces. 

 We also define \emph{metric correlation} between two $\O$-valued random variables as follows, 
   \begin{equation*}
   \RR(U,V)=\frac{\C(U,V)}{\sqrt{\C(U,U)\C(V,V)}}.
   \end{equation*}
   By the Cauchy-Schwarz inequality one has $-1 \leq \RR(U,V) \leq 1$.
   Metric covariance/correlation depends on the choice of the metric $d$ and different choices of $d$ might reveal different aspects of association between random objects, depending  on the underlying geometry of the metric.

\subsection{Metric auto-covariance operators}

As in the real valued Euclidean case, we define the metric  auto-covariance function $C(s,t)$ for \fro \, $\lbrace X_1, X_2, \hdots , X_n\rbrace \in \O$ as
\begin{equation*}
C(s,t)=\C\left(X(s),X(t)\right),
\end{equation*}
 for all $(s,t) \in [0,1] \times [0,1]$. 
Obviously, $C(s,t)$ is a symmetric kernel and therefore has real eigenvalues when used as the kernel of a linear Hilbert-Schmidt  operator.  The following result shows that for {separable} metric spaces $(\O,d)$ for which the squared distance function $d^2$ is of negative type,  the metric auto-covariance operator is positive semidefinite. 
 	\begin{prop}
 		\label{lma: pos_def}
 		If {$\O$ is separable} and $d^2$ is of negative type, then $C(s,t)$ is a nonnegative definite kernel.
 	\end{prop}
   {By Proposition 3 in \cite{sejd:13} and equation \eqref{eq: cov_abstract}, $\C(U,V)=0$ implies that there exists an abstract Hilbert space $\mathcal{H}$ and an injective map $f: \O \rightarrow \mathcal{H}$ such that $f(U)$ and $f(V)$ are orthogonal in $L^2(\mathcal{H})$. Note that $\Var_{\O}(U)=\C(U,U)=\frac{1}{2}E(d^2(U,U'))$, which for real valued random variables equals $\Var(U)$. 

Formally, one can define the \emph{metric auto-covariance operator} as a linear Hilbert-Schmidt integral operator $T_C$ that operates on  functions $g \in L^2([0,1])$ and utilizes the metric auto-covariance kernel, 
\begin{equation*}
(T_C g)(s)=\int_{0}^{1} C(s,t)g(t)dt.
\end{equation*}
We note that for example Theorem 4.6.4 of \ci{hsin:15} implies  the nonnegative definiteness of the kernel $C(s,t)$,  in the sense that $\langle T_Cf,f \rangle \geq 0$ for all $f$.

By Mercer's theorem there is an orthonormal basis $\lbrace\phi_i\rbrace_{i=1}^{\infty}$ of $L^2([0, 1])$ consisting of eigenfunctions of $T_C$ such that the corresponding sequence of eigenvalues $\lbrace \lambda_i\rbrace_{i=1}^{\infty}$, which are ordered in declining order,  is nonnegative, since $C(s,t)$ is positive semidefinite. The eigenfunctions corresponding to non-zero eigenvalues are continuous on $[0, 1]$ and $C$ has the representation
\begin{equation*}
{C(s,t)=\sum _{j=1}^{\infty }\lambda _{j}\phi_{j}(s)\phi_{j}(t)},
\end{equation*}
where the convergence is absolute and uniform;  see, e.g. Lemma 4.6.1 and Theorems 4.5.2, 4.6.2, 4.6.5 and 4.6.7 of \ci{hsin:15}.


We thus accomplished the first step of extending FPCA from Euclidean valued functional data to general metric space valued functional data. The eigenfunctions $\lbrace\phi_j\rbrace_{j=1}^{\infty}$ can be interpreted as principal directions of variation of the functional object process and will be ordered according to the size  of the associated eigenvalues. We can view the eigenvalues as representing a metric version of `fraction of variance explained', which is their common interpretation in the real-valued case. The only requirement for this extension is that  the squared metric $d^2$ is of negative type but this is not a severe restriction and in light of Proposition 3 of  \cite{sejd:13} is true for the following examples:

\begin{enumerate}
	\item [1.]$(\O,d)$ where $\O$ is the space of univariate probability distributions on a common compact support in $T \subset \R$. Choices of $d$ include the popular 2-Wasserstein metric or the $L^2$ metric.  
	\item[2.] $(\O,d)$ where $\O$ is the space of correlation matrices of a fixed dimension $r$, where the choice of metrics includes the Frobenius metric, log  Frobenius metric, power Frobenius metric and Procrustes metric \cp{dryd:09, pigo:14, tava:17}. 
	\item[3.] $(\O,d)$ where $\O$ is the space of networks with a fixed number, say $r$, of nodes. One can view  networks as adjacency matrices or graph Laplacians equipped with the Frobenius metric \cp{gine:17} or as resistance matrices equipped with the resistance perturbation metric \cp{monn:18}.
\end{enumerate}
We conclude that in most cases of interest the auto-covariance operator and its eigenfunctions will be well defined. 

\subsection{Interpretation of metric covariance}

When $X$ and $Y$ are real valued, classical Pearson correlation captures the strength and sign of  linear (also monotone) associations between $X$ and $Y$. 
From a geometrical perspective, Pearson correlation can be interpreted as the cosine of the angle between $X$ and $Y$. In $\mathbb{R}^d$, angles between vectors are defined using inner products, which can also be used for data in Hilbert space to characterize dependency. Specifically, for random functions in the metric space $L^2$ this idea leads to the notion of ``dynamic correlation" in functional data analysis 
\citep{mull:05:2}, which was found to be useful for data analysis in genetics \citep{opge:06} and psychology \citep{liu:16}. Dynamic correlation  turns out to be  equivalent to metric covariance  when the random objects are in the Hilbert  space $L^2([0,1])$, equipped with the usual $L^2$ metric. Metric covariance then provides a generalization beyond Hilbert spaces.  

For  general metric spaces, under the weak assumption that  the squared metric is of negative type, 
the map $f$ from object to Hilbert space in (\ref{map}) implies that  
metric covariance can be derived from the  inner product in an abstract Hilbert space, while  metric correlation is obtained by standardizing metric covariance,  and is thus tied to the notion of  an angle in an abstract space. Hence its magnitude can be interpreted as strength of association between random objects. While we use the existence of a map $f$ and an associated abstract Hilbert space, we do not require knowledge about $f$.  Metric covariance is thus  a natural extension of Pearson covariance to general metric spaces. 

In recent work \citep{mull:19:4},  Wasserstein covariance for pairs of univariate probability distributions was introduced, and was shown to have an appealing interpretation as an expected value of an inner product of optimal transport maps. More specifically, if $f_1$ and $f_2$ are the components of a random bivariate density process and  $F_1^{-1}(\cdot)$ and $F_2^{-1}(\cdot)$ the corresponding random quantile functions, the squared Wasserstein distance between $f_1$ and $f_2$ is given by
\begin{equation*}
d_W^2(f_1,f_2)= \int_{0}^{1} \{Q_1(t)-Q_2(t)\}^2 dt
\end{equation*}  and Wasserstein covariance between $f_1$ and $f_2$ was introduced as   
\begin{equation*}
\Cov_W(f_1,f_2)= E \left[ \int_{0}^{1} \left\{Q_1(t)-E\left(Q_1(t) \right) \right\} \left\{ Q_2(t)-E\left(Q_2(t) \right) \right\} dt\right].
\end{equation*} 

Wasserstein covariance is then easily seen to be  a  special case of metric covariance when the metric space-valued random objects are probability distributions and the  Wasserstein metric is used.
This
Wasserstein version of metric covariance was  found to quantify the degree of synchronization of the movement of probability mass from the marginal \F \ means of the probability distributions to the random components of a multivariate density process.  In  applications to  fMRI data, this Wasserstein version   led to new findings and insights about differences in brain connectivity of normal versus Alzheimer disease patients, a topic of special interest in neuroimaging \citep{mull:19:4}. 
The examples of dynamic correlation for Hilbert space valued random variables in functional data analysis and of Wasserstein covariance/correlation demonstrate the utility of  metric covariance/correlation in non-standard spaces and its  interpretability  in applications. This provides evidence  that 
metric covariance and metric correlation are  indeed useful tools for data analysis in general metric spaces. 

A word of caution is in order. While metric covariance covariance can be universally applied, and in the space of distributions with the Wasserstein metric has an interpretation as an inner product of transport maps, such interpretations hinge on the specific metric space  in which the random objects are located and may not be available for all spaces. In practice,  interpretations for specific scenarios can be important. The choice of the metric also matters and should be considered carefully, as it will affect the interpretation of metric covariance.

Apart from the interpretation of covariance as the expectation of an inner product, the diagonal elements  of the metric auto-covariance surface reflect a natural notion of variance of metric-space valued objects, as 
\be \label{mar} \Var_{\O}=\frac{1}{2}Ed^2(U,U'),\ee 
where $U'$  is an independent copy of $U$. This provides a variation measure that is tied to the average squared distance of objects that are independently sampled from the underlying population, which is a natural and interpretable  measure of spread that is well known to coincide with conventional variance  in the Euclidean case.   


Since it is sensible to define variance for metric-space valued random objects   as 
$$\frac{1}{2}Ed^2(U,U')=\frac{1}{2}E[d^2(U,U')-d^2(U,U)],$$ it is then natural to extend this to 
a covariance measure between random objects $(U,V)$  that reflects the difference between squared distances when sampling independently from the marginal distributions of $U$ and $V$ and when sampling from the joint distribution of $(U,V)$. This simple idea provides another avenue to suggest  $$\widetilde{\C}(U,V)=E[d^2(U,V')-d^2(U,V)].$$  
Symmetrizing this expression and adding the factor .25 to match the usual definition of covariance in the Euclidean case then leads to formula \eqref{eq: cov_general}. 
The  above arguments also lead to an interpretation of the  total variance that corresponds to  the trace of the proposed metric covariance operator  $C(s,t)$, as an integrated squared distance between the \fro $X$ and an independent copy $X'$,  
\be \label{tot} \sum_{j=1}^{\infty} \lambda_j=\int_{0}^{1} \C(X(t),X(t))dt=\frac{1}{2}  \int_{0}^{1} Ed^2(X(t),X'(t))dt, \ee
We find in our examples and applications that the eigenfunctions derived from metric covariance lead to  useful and often well-interpretable modes of variation of the time-varying metric random objects in the sense of \cite{jone:92},  
adding to the practical appeal of metric covariance for the analysis of \fro. 

To conclude this discussion, we note that metric covariance differs substantially   from distance correlation  \cp{szek:07, lyon:13}.  A distinguishing  feature of distance correlation  is  that it is  equivalent to  probabilistic independence between the distributions of $U$ and $V$ when it is 0, but we find that it is not suitable as a covariance or correlation measure for random objects in the situations that we study here. Specifically,  the auto-covariance operator it generates is not useful for our purposes.  For further details on this, see  section \hyperref[dis_cov]{S2} 
in the  supplement.

\section{\textbf{Functional Principal Components: Generalized \F \ Integrals and \F \ Scores}}
\label{pc}
\subsection{Generalized \F \ Integrals and Object Functional Principal Components}
\label{sec: object_fpc}
 FPCs in the case of real valued functional data are projections of the centered process onto  the directions of the eigenfunctions and therefore summarize how a function differs from the mean function along orthonormal eigenfunction directions. Formally the FPC of the $i^{th}$ process $X_i(t)$ and  the $k^{th}$ eigenfunction $\phi_k(t)$ is 
\begin{equation*}
\label{eq: pc_score}
\xi_{ik}=\int_{0}^{1} \left(X_i(t)-\mu(t)\right)\phi_k(t) dt,
\end{equation*}
where $\mu(\cdot)$ is the mean process. The part of the score contributing  to the variability of the functional data is $\int_{0}^{1} X_i(t) \phi_k(t) dt$, which is just a horizontal shift of the actual scores, so centering is not needed when our goal is to decompose the variability of the random processes $X$, which is fortuitous as  one  cannot `center' object data to obtain an analogue of $X(t)-\mu(t)$, as algebraic operations such as subtraction are not feasible in metric spaces. 

In the Euclidean case for any function $\phi(\cdot)$ on $[0,1]$, whenever $\int_{0}^{1} \phi(t)dt \neq 0$,  one can obtain a scaled version of the integral of $X(t)$ with respect to $\phi(t)$ as follows, 
\begin{equation*}
\int_{0}^{1}X(t)\frac{\phi(t)}{\int_{0}^{1}\phi(t)dt} dt= \quad \arginf_{\o \in \mbb{R}} \int_{0}^{1} d_E^2(\o,X(t))\frac{\phi(t)}{\int_{0}^{1}\phi(t)dt}dt.
\end{equation*}
This suggests to  define an integral of an $\O$-valued function $S(\cdot)$ with respect to a real valued function $\phi(\cdot)$ which integrates to 1. For any real valued function $\phi(\cdot)$ with $\int_{0}^{1}\phi(t)dt=1$, we  define the \emph{generalized \F \ integral} of $S(\cdot)$ with respect to $\phi(\cdot)$ as
\begin{equation}
\label{eq: frec_integral}
\int_{\oplus} S(t)\phi(t) dt= \arginf_{\o \in \O} \int_{0}^{1}d^2(\o,S(t)) \phi(t) dt, 
\end{equation}
provided that the integral $\int_{0}^{1}d^2(\o,S(t)) \phi(t) dt$ exists as a limit of Riemann sums for all $\o \in \O$ and the minimizer of the integrals over $\o \in \O$ exists and is unique. A special case of the integral in \eqref{eq: frec_integral} was  introduced as \F \ integral in \cite{mull:16:2},  where an integral for the space of covariance matrices was constructed for {$\phi \equiv 1$.}

The \F \ integrals defined here are far  more general. Generalized \F \ integrals  can  be interpreted as an extension of weighted \F \ means \cp{frec:48}. We omit the additional term  ``generalized" in the following and note that \F \ integrals can be interpreted as projections of {\fro \ onto} functions $\phi$, by weighing the elements $S(t)$ according to the value of $\phi(t)$, in direct analogy to projections in the linear function space $L^2$. This feature motivates to employ \F \ integrals to obtain object FPCs.  

 For fixed $\o \in \O$ consider the \F \ integral function
	\begin{equation*}
	I(\o)=\int_{0}^{1}d^2(\o,S(t)) \phi(t) dt,
	\end{equation*}
	which,  if it exists, is the limit of Riemann sums. A  sufficient condition for its existence is that $d^2(\o,S(t)) \phi(t)$ is a continuous function of $ t \in [0,1]$. If the metric is bounded and $S(\cdot)$ and $\phi(\cdot)$ are continuous for $t \in [0,1]$, the function $d^2(\o,S(t)) \phi(t)$ is a continuous function of $ t \in [0,1]$ and the integral $I(\o)$ exists for all $\o$. Note that for any $\o \in \O$,  $I(\o)$  is finite by the Cauchy Schwarz inequality whenever the metric space is bounded and the $L^2$ norm of the function $\phi(\cdot)$ is finite.
	
	If the integrals $I(\o)$ exist as limits of Riemann sums,   the question arises under which conditions the minimizers of the Riemann sums converge and whether the limit of the minimizers coincides with the \F \ integral $\int_{\oplus} S \phi$. Proposition \ref{lma: well_defined} addresses this question. Let $0=x_0 < x_1 < x_2 < \hdots <x_k=1$ be a partition $\mathcal{P}$ of $[0,1]$, where the $[x_j,x_{j+1}]$ are the  subintervals of the partition and the length of the $j^{th}$ subinterval is $\Delta_j=x_{j+1}-x_j$. The mesh size $\epsilon_{\mathcal{P}}$  of the partition is given by $\epsilon_{\mathcal{P}}=\max_j\Delta_j$. We select $t_0,t_1,\hdots , t_{k-1}$  such that for each $j$, $t_j \in [x_j,x_{j+1}]$.  For each $\o \in \O$, the Riemann sum $I_{\mathcal{P}}(\o)$ corresponding to the partition $\mathcal{P}$  and $t_0,t_1, \hdots ,t_{k-1}$ is given by
	\begin{equation*}
	\label{eq: Riemann sum}
	I_{\mathcal{P}}(\o)= \sum_{j=0}^{k-1}d^2(\o,S(t_j))\phi(t_j) \Delta_j
	\end{equation*}
	and the Riemann integral $I(\o)$ is obtained as a limit of Riemann sums as the partition gets finer. Formally, $I(\o)=\lim_{\epsilon_{\mathcal{P}} \rightarrow 0} I_{\mathcal{P}}(\o)$.

We will invoke  the following assumptions for the integral function $I(\o)$. For ease of notation, we suppress $t$ in $\int_{\oplus} S(t)\phi(t) dt$, writing 
$\int_{\oplus} S\phi$ in the following. 
\begin{enumerate}
	\item[(I1)] The integrand function $H(\o,t)=d^2(\o,S(t))\phi(t)$ is uniformly equicontinuous  in $t \in [0,1]$ and $\o \in \O$. 
	\item[(I2)] $\int_{\oplus} S\phi=\argmin_{\o \in \O}I(\o)$ exists and is  unique, and  $\inf_{d(\o,\int_{\oplus}S\phi) > \delta} I(\o) > I(\int_{\oplus}S\phi)$ for all $\delta > 0$.
	\item[(I3)]	There exist constants $\beta > 0, \, \nu > 0$ and $C > 0$ such that $$ I(\o)-I(\int_{\oplus} S \phi) \geq C \ d^{\beta}(\o,\int_{\oplus} S\phi)$$ whenever $d(\o,\int_{\oplus}S\phi) < \nu$.
 \end{enumerate}
 Define $\sum_{\mathcal{P},\oplus}S\phi =\argmin_{\o \in \O}I_{\mathcal{P}}(\o)$. 
\begin{prop}
	\label{lma: well_defined} 
	\begin{enumerate}
		\item Under assumption (I1), $I_{\mathcal{P}}(\o)$ converges to $I(\o)$ uniformly in $\o$ as $\epsilon_{\mathcal{P}} \rightarrow 0$. 
		\item Under assumptions (I1) and (I2), $\lim_{\epsilon_{\mathcal{P}} \rightarrow 0} d(\sum_{\mathcal{P},\oplus}S\phi,\int_{\oplus}S\phi)=0$. 
				\item If $\lim_{\epsilon_{\mathcal{P}} \rightarrow 0} h(\epsilon_{\mathcal{P}}) \sup_{\o \in \O} \lvert I_{\mathcal{P}}(\o)-I(\o)\rvert =0$ for  a function $h$  with $h(\delta) \rightarrow \infty$ as $\delta \rightarrow 0$, then under (I3), $\lim_{\epsilon_{\mathcal{P}} \rightarrow 0} h^{\frac{1}{\beta}}(\epsilon_{\mathcal{P}})d(\sum_{\mathcal{P},\oplus}S\phi,\int_{\oplus}S\phi)=0$.
	\end{enumerate}
	
\end{prop}

	As a  continuous function on a compact interval is uniformly continuous, whenever $S(\cdot)$ is continuous and $\phi(\cdot)$ is bounded and continuous,  (I1) holds since for $D=\text{diam}(\O)$,
	\begin{align*}
	& \lvert H(\o,t_1)-H(\o,t_2) \rvert \\ &= \lvert d^2(\o,S(t_1))\phi(t_1)-d^2(\o,S(t_2))\phi(t_1)+d^2(\o,S(t_2))\phi(t_1)-d^2(\o,S(t_2))\phi(t_2) \rvert \\ &\leq 2 D \ d\left(S(t_1),S(t_2)\right)\lvert \phi(t_1) \rvert +D^2 \lvert \phi(t_1)-\phi(t_2) \rvert.
	\end{align*}
	Assumption (I1) is sufficient to guarantee that the \F \ integrals are well defined, while  Assumption (I3) is a restriction on the curvature of the function $I(\o)$ near its minimizer, implying convergence rates of the approximations of the \F \ integrals. A few examples of spaces that satisfy assumptions (I2) and (I3) are as follows.  
	\begin{enumerate}
		\item[1.] Let $(\O,d_W)$ be the space of univariate probability distributions on a common support $T \subset R$. For any $\o \in \O$,  denote the corresponding random distribution and quantile functions by $Q(\o)$.  The squared 2-Wasserstein metric between distributions $\o_1$ and $\o_2$ is 
		\begin{equation*}
		d^2_W(\o_1,\o_2)=d^2_{L^2}(Q(\o_1),Q(\o_2))=\int_{0}^{1} (Q(\o_1)(u)-Q(\o_2)(u))^2du.
		\end{equation*}
		For any $S(t)$ taking values in $\O$, where we view $Q(S(t))$ as the quantile function of the distribution at time $t \in [0,1]$, writing $Q(S(t))(u)$ for  the $u^{th}$ quantile of the distribution at time $t$,  define $Q^*(u)={\int_{0}^{1}Q(S(t))(u)\phi(t)dt}$. Since $\int_{0}^{1}\phi(t)dt=1$, a simple calculation shows that  for any $\o \in \O$,
		\begin{equation*}
		\arginf_{\o \in \O} I(\o)=\arginf_{\o \in \O} d^2_{L^2}(Q(\o),Q^*),
		\end{equation*}
		therefore  the minimizer exists and is unique by the convexity of the space of univariate quantile functions.  By the  orthogonal projection theorem the minimizer  $\tilde{\o}$  is uniquely characterized by
		\begin{equation*}
	{	\langle Q^*-Q(\tilde{\o}),Q(\o)-Q(\tilde{\o}) \rangle_{L^2} \leq 0,}
		\end{equation*}
		for all $\o \in \O$ and therefore it is enough to choose $\nu=C=1$ and $\beta=2$ in (I3). 
		\item [2.] 
		Consider the space of graph Laplacians or graph adjacency matrices of connected, undirected and simple graphs with a fixed number
		 $r$ of nodes $(\O,d_F)$, equipped with  the Frobenius metric $d_F$. For any $\o \in \O$, 
		\begin{equation*}
		d^2_F(\o_1,\o_2)=\sum_{j=1}^{r} \sum_{k=1}^{r}(\o_{1,jk}-\o_{2,jk})^2.
		\end{equation*}
		For any $S(t)$ taking values in $\O$, let $S_{jk}(t)$ be the $(j,k)^{th}$ entry of the graph Laplacian or the graph adjacency matrix. Define $S^*_{jk}={\int_{0}^{1}S_{jk}(t)\phi(t)dt}$. Since $\int_{0}^{1}\phi(t)dt=1$, it can be easily seen that for any $\o \in \O$,
		\begin{equation*}
		\arginf_{\o \in \O} I(\o)=\arginf_{\o \in \O} d^2_{F}(\o,S^*),
		\end{equation*}
		and so the minimizer exists and is unique by the convexity of the space of graph Laplacians \cp{gine:17} and the space of graph adjacency matrices. 
		Again, by the orthogonal projection theorem, the minimizer $\tilde{\o}$ is uniquely characterized by
		\begin{equation*}
	{	\sum_{j=1}^{r}\sum_{k=1}^{r} (\tilde{\o}_{jk}-S^*_{jk})(\tilde{\o}_{jk}-\o_{jk})\leq0,}
		\end{equation*}
		for all $\o \in \O$ and therefore it is enough to choose $\nu=C=1$ and $\beta=2$ in (I3). 
	\item [3.]The same arguments also imply  that $(\O,d_F)$ satisfies assumptions (I2) and (I3) when $\O$ is the space of correlation matrices of a fixed dimension $r$. 
	\end{enumerate}

 As we have seen, for general metric spaces $\O$, under mild assumptions on the boundedness of the metric and continuity of the functions $S(\cdot)$ and $\phi(\cdot)$, the \F \ integral has nice properties if it exists and is unique. Moreover, when $\O$ is bounded and $\int_{0}^{1}\lvert \phi(t) \rvert dt < \infty$, 
\begin{equation*}
\lvert I(\o_1)-I(\o_2) \rvert \leq 2D \ d(\o_1,\o_2) \int_{0}^{1}\lvert \phi(t) \rvert dt,
\end{equation*} 
and therefore $I(\o)$ is a continuous function of $\o \in \O$. This ensures that the \F \ integral always exists when $\O$ is compact.

We now define the FPCs corresponding to the bounded continuous eigenfunctions $\phi_k$  of the metric auto-covariance operator in the object space using \F \ integrals. For this, we assume  that  all  trajectories $\lbrace X_i(t)\rbrace_{t \in [0,1]}$ have continuous sample paths almost surely and the metric space $\O$ is bounded, and furthermore that 
\begin{enumerate}
	\item[(A1)] $\int_{0}^{1}\phi_k(t)dt \neq 0$
	\item [(A2)] $\int_{\oplus} X_i\phi^*_k$ exists and is unique almost surely for all $i=1,\hdots,n$, where $\phi^*_k(t)={\phi(t)}/{\int_{0}^{1}\phi(t)dt}$.
\end{enumerate}
 Then   {\it object functional principal components} (object FPCs).  
  for $X_i$ and   $\phi_k$ are defined as the \F \ integrals 
\begin{equation}
\label{eq: principal_object}
\psi^{ik}_{\oplus}=\int_{\oplus} X_i\phi^*_k,
\end{equation}
which are random objects in $\O$.  Similar to ordinary FPCA one can choose a number of basis functions aiming to explain a desired percentage of variation in the data utilizing the eigenvalues of the metric auto-covariance operator. If $\O=\R$, the object FPCs correspond to a location and scale shifted version of the ordinary FPCs. 

\subsection{\F \ Scores}
\label{sec: scalar_fpc}
Exploratory data analysis such as checking for clusters  or  outliers often benefits  from plotting the FPCs against each other for the case of real valued functional data. FPCs defined using \F \ integral live in the object space $\O$ and therefore  visualizing them is non-trivial.  One approach is to obtain  their projections to a lower-dimensional real space using multi-dimensional scaling or its variants \cp{krus:64,belk:02} and then visualizing the projections. Here we propose another approach for obtaining a scalar version of object FPCs. The resulting scalar FPCs are interpretable and can be plotted against each other and are thus useful  for exploratory data analysis.

In the real valued case, one obtains projections of the deviations of the observed random curves from the mean curve onto dominant eigenfunctions. While the concept of a mean function can be generalized to object functional data using  \F \ means \cp{frec:48},  one cannot `center' object data and does not have directional information. Nevertheless, it is possible to  study how distances of sample curves from the mean curve project onto a few dominant eigenfunctions, in analogy to the  real valued case. 
Formally, given a random object process $\lbrace X(t) \rbrace_{t \in [0,1]}$,  the population \F \ mean function is 
\begin{equation*}
\label{eq: mean}
\mu_{\oplus}(t)=\argmin_{\o \in \O} E(d^2(\o,X(t))),
\end{equation*}
where we assume existence and uniqueness of the minimizer. 
For real valued functional data under the Euclidean metric the \F \ mean function coincides with the usual pointwise mean function. Defining distance functions 
\begin{equation*}
D_i(t)=d\left(X_i(t),\mu_{\oplus}(t)\right)
\end{equation*}
for sample trajectories $X_i$, 
we represent the scalar functions $D_i$  in the eigenbasis of the metric auto-covariance operator, obtaining the coefficients 
\begin{equation}
\label{eq: principal_score}
\beta_{ik}= \int_{0}^{1} D_i(t) \phi_k(t) dt= \int_{0}^{1} d\left(X_i(t),\mu_{\oplus}(t)\right) \phi_k(t) dt.
\end{equation}
We refer to the scalars $\beta_{ik}$ as the \emph{\F \ scores}.

The \F \ scores can be interpreted as decomposition of the departures of the sample elements from the `central' \F \ mean curve in predominant directions of variation. They can be plotted against each other and have the potential to provide interesting insights, as we will illustrate in the data applications. 
Considering the existence of the \F \ scores, with $D$ denoting as before the diameter of the totally bounded metric space $\O$, continuity of the \F \ mean function implies that for  any $t_1,t_2 \in [0,1]$, 
\begin{align*} 
& \lvert d^2(X_i(t_1),\mu_{\oplus}(t_1))-d^2(X_i(t_2),\mu_{\oplus}(t_2))\rvert \\ & =\lvert d^2(X_i(t_1),\mu_{\oplus}(t_1))-d^2(X_i(t_1),\mu_{\oplus}(t_2))+d^2(X_i(t_1),\mu_{\oplus}(t_2))- d^2(X_i(t_2),\mu_{\oplus}(t_2))\rvert\\ & \leq 2D \ \lbrace  d(\mu_{\oplus}(t_{1}),\mu_{\oplus}(t_2))+ {d}(X_i(t_1),X_i(t_2))\rbrace.
\end{align*}
Thus for continuous eigenfunction $\phi_k(\cdot)$, the function $d^2(X_i(t),\mu_{\oplus}(t))\phi_k(t)$ is a continuous function of $t \in [0,1]$ almost surely and therefore the \F \ scores will exist.  Proposition  \ref{lma: mean_cont} shows that under assumption (A3)  the \F \ mean function is indeed continuous.
\begin{enumerate}
	\item[(A3)] For each  $t \in [0,1]$, the pointwise \F \ mean $\mu_{\oplus}(t)$ exists and is unique, and   $$\inf_{d(\o,\mu_{\oplus}(t))  > \gamma} E(d^2(\o,X(t))) > E(d^2(\mu_{\oplus}(t),X(t)))$$ for any $\gamma > 0$.
\end{enumerate}
\begin{prop}
	\label{lma: mean_cont}
	If the  random object process $\lbrace X(t) \rbrace_{t \in [0,1]}$ has almost surely continuous paths, then $\mu_{\oplus}(\cdot)$ is continuous under (A3).\end{prop}

Assumption (A3) is satisfied for the space $(\O,d_W)$  of univariate probability distributions with the 2-Wasserstein metric and also for  the space $(\O,d_F)$, where $\O$ is the space of covariance matrices or  alternatively graph Laplacians of fixed dimension with the Frobenius metric $d_F$ \citep{mull:18:5, mull:19:3}.

\section{Estimation and Theory}
\label{theory}
Having defined suitable population targets, our goal now is to construct appropriate estimators, starting with a sample of \fro. An empirical estimator of the metric auto-covariance operator $C(s,t)$ as defined in section \ref{sec: cov} is given by
\begin{equation}
\label{eq: cov}
\hat{C}(s,t)=\frac{1}{4n(n-1)} \sum_{i \neq j} f_{s,t}(X_i,X_j), 
\end{equation}
where
\begin{equation*}
f_{s,t}(X_i,X_j)=d^2(X_i(s),X_j(t))+d^2(X_j(s),X_i(t))-d^2(X_i(s),X_i(t))-d^2(X_j(s),X_j(t)).
\end{equation*}
Observe that for each $s,t \in [0,1]$, $\hat{C}(s,t)$ is a $U$-statistic and the class $\lbrace \hat{C}(s,t): s,t \in [0,1]\rbrace $ is a family of $U$-statistics.

Noting that $\hat{C}(s,t)$ can be viewed  as a stochastic process indexed by the function class $\mathcal{F}=\lbrace f_{s,t}(\cdot,\cdot): s,t \in [0,1]\rbrace$, where $f_{s,t}(x,y)=d^2(x(s),y(t))+d^2(y(s),x(t))-d^2(x(s),x(t))-d^2(y(s),y(t))$ allows us to apply the theory of $U$-processes \cp{nola:87,nola:88,arco:93} for weak convergence \cp{bill:68,well:96}. For the uniform convergence of $\lbrace \hat{C}(s,t): s,t \in [0,1]\rbrace $,  we need an assumption on the rate of continuity of the \fro. 
\begin{itemize}
	\item[(A4)] The process $X(\cdot)$ is almost surely $\alpha$-H\"{o}lder continuous for some $0 < \alpha \leq 1$, where the H\"{o}lder constant has a finite second moment, i.e. for some non negative function $G(X)$ one has 
	\begin{equation*}
	d(X(s),X(t)) \leq G(X) |s-t|^\alpha,
	\end{equation*}	
	where $E(G(X))^2 < \infty.$
\end{itemize}

\begin{thm}
	\label{lma: C_limit}
	Under assumption (A4), the sequence of stochastic processes 
		\begin{equation*}
	U_n(s,t)=\sqrt{n} (\hat{C}(s,t)-C(s,t))
	\end{equation*}
	converges weakly to a Gaussian process with mean 0 and covariance function 
	\begin{equation*}
	{R_{(s,t), (u,v)}= \Cov\left(f_{s,t}(X,X'),f_{u,v}(X,X')\right),}
	\end{equation*}
	{where $X'$ is an i.i.d copy of $X$.}
\end{thm}

  Writing $\hat{\lambda}_{j}$ and $\hat{\phi_{j}}$ for the eigenvalues and eigenfunctions of $\hat{C}(s,t)$, uniform convergence and rates of convergence of these estimates of the eigenvalues and eigenfunctions of the metric auto-covariance operator to their targets are obtained  as a direct consequence of Proposition  \ref{lma: C_limit} under the following assumption on the spacings of the eigenvalues. 

\begin{enumerate}
	\item[(A5)] For each $j \geq 1,$  the eigenvalue $\lambda_j$ has multiplicity 1, i.e it holds that  $\delta_j > 0$,  where $\delta_j=\min_{1\leq l\leq j}(\lambda_l-\lambda_{l+1})$.
\end{enumerate}
\begin{cor}(\cite{bosq:00})
	\label{cor: eigen}
	Under assumptions (A4) and (A5),
	\begin{equation*}
	|\hat{\lambda}_j-\lambda_j|=O_P(1/\sqrt{n}).
	\end{equation*}
	\begin{equation*}
	\sup_{s \in [0,1]} \left| \hat{\phi}_j(s)-\phi_j(s)\right| =O_P(1/{\delta_j \sqrt{n}}).
	\end{equation*}
\end{cor}
 
{As in classical FDA, the eigenfunctions $\phi_j$  are uniquely identifiable  only up to a sign change. 
For  theoretical considerations such as the convergence in Corollary \ref{cor: eigen}, we may always assume that true and estimated eigenfunctions are aligned in the sense that  $\langle \hat{\phi}_j, \phi_j \rangle \geq 0$}. Our next objective is to  obtain sample estimators for  the object FPCs (\ref{eq: principal_object})  defined in section \ref{sec: object_fpc}. For each $j$, consider the following estimators of $\phi^*_j(t)$, 
\begin{equation*}
\hat{\phi}_j^*(t)=\frac{\hat{\phi}_j(t)}{\int_{0}^{1}\hat{\phi}_j(t)dt}.
\end{equation*}
A natural estimator for the  \F \ integral ${\psi}^{ik}_{\oplus}$ is then 
\begin{equation}
\label{eq: int}
\hat{\psi}^{ik}_{\oplus}=\int_{\oplus} X_i \hat{\phi}^*_j=\argmin_{\o \in \O} \int_{0}^{1} d^2(\o,X_i(t)) \hat{\phi}^*_j(t) dt.
\end{equation}
To obtain convergence of $\hat{\psi}^{ik}_{\oplus}$ to its population target, we make the following assumptions.
\begin{itemize}
	\item[(A6)] For every $i$ and $k$, ${\psi}^{ik}_{\oplus}$ and $\hat{\psi}^{ik}_{\oplus}$ exist and are unique almost surely. Moreover for any $\epsilon > 0$, $c_{\eps}=\inf_{d(\o,\psi^{ik}_{\oplus})> \epsilon}\left(\int_{0}^{1} d^2(\o,X_i(t))  \phi^*(t) dt-\int_{0}^{1} d^2(\psi^{ik}_{\oplus},X_i(t))  \phi^*(t) dt\right) > 0$ almost surely.
	\item[(A7)] There exist constants $\beta_1 > 1, \, \nu' > 0$ and $C' > 0$ such that  almost surely,
	\begin{equation*}
	\left(\int_{0}^{1} d^2(\o,X_i(t))  \phi^*(t) dt-\int_{0}^{1} d^2(\psi^{ik}_{\oplus},X_i(t))  \phi^*(t) dt\right) \geq C' d^{\beta_1}(\o,\psi^{ik}_{\oplus}), 
	\end{equation*} whenever $d(\o,\psi^{ik}_{\oplus}) < \nu'$.
\end{itemize}
Assumption (A6) on the existence and uniqueness of the \F \ integrals is used to establish consistency. Assumption (A7) is a restriction on the local  behavior of the integrals around the minimizer and determines the  rate of convergence.
\begin{thm}
	\label{lma: fpc}
	Under assumptions (A1), (A2), (A4) and (A6), $$d(\hat{\psi}^{ik}_{\oplus},\psi^{ik}_{\oplus})=o_P(1).$$ If additionally (A7) holds, then $$d(\hat{\psi}^{ik}_{\oplus},\psi^{ik}_{\oplus})=O_P(n^{-1/(2\beta_1)}).$$
\end{thm}
Here we choose $\hat{\phi}_j$ to be such that $\langle \hat{\phi}_j, \phi_j \rangle \geq 0$ which ensures matching signs for the true and estimated eigenfunctions in the computation of $\hat{\psi}^{ik}_{\oplus}$ and $\psi^{ik}_{\oplus}$. 

Next we provide estimates of the \F \ scores and study their asymptotics. The starting point is the  following estimator of the population \F \ mean function,
\begin{equation} \label{mean}
 \hat{\mu}_\oplus(t)=\argmin_{\o \in \O} \frac{1}{n} \sum_{i=1}^{n}d^2(X_i(t),\o).
\end{equation} We need the following assumptions: \begin{itemize}
	\item[(A8)] The \F \ mean function estimate $\hat{\mu}_\oplus(t)$ exists and is unique almost surely for all $t \in [0,1]$. Additionally, for every $\eps > 0$, there exists $\tau(\eps) > 0$ such that
	\begin{equation*}
	\lim_{n \rightarrow \infty}P \left(\inf_{s \in [0,1]} \inf_{d(\o,\hat{\mu}_\oplus(s)) > \epsilon} \frac{1}{n} \sum_{l=1}^{n} \lbrace d^2(X_i(s),\o)-d^2(X_i(s),\hat{\mu}_\oplus(s))\rbrace \geq \tau(\eps) \right) = 1.
	\end{equation*}
{\item[(A9)] There exists a sufficiently small $\delta > 0$  and constants $0< \nu_\delta \leq 1$ and $H_\delta > 0$, such that for all $\O$-valued functions $\o(\cdot)$ with $d_\infty(\o,\mu_\oplus) < \delta$,  where $d_\infty(\o,\mu_\oplus)=\sup_{s \in [0,1]} d_\infty(\o(s),\mu_\oplus(s))$, the functions $\o(\cdot)$ are $\nu_\delta$-H\"{o}lder continuous with H\"{o}lder constant bounded above by $H_\delta$, i.e.
	\begin{equation*}
	d(\o(s),\o(t)) \leq H_\delta |s-t|^{\nu_{\delta}}.
	\end{equation*}}
	\item[(A10)] For $I(\delta)=\int_{0}^{1} \sup_{s \in [0,1]} \sqrt{\log N(A{\eps\delta},B_{\delta}(\mu_\oplus(s)),d)} d\eps $, it holds that  $I(\delta)=O(1)$ as $\delta \rightarrow 0$ for all sufficiently small $\delta > 0$ and for any constant $A>0$.  Here $B_{\delta}(\mu_\oplus(s))=\{\o \in \O: d(\o,\mu_\oplus(s)) < \delta\}$ is the $\delta$-ball around $\mu_\oplus(s)$ and $N(\eps,B_{\delta}(\mu_\oplus(s)),d)$ is the covering number, i.e. the minimum number of balls of radius $\eps$ required to cover $B_{\delta}(\mu(_\oplus(s))$ 
	\cp{well:96}. 
	\item[(A11)]  There exist $\alpha> 0, \ D > 0$ and $\beta_2 > 1$ such that 
	\begin{equation*}
	\inf_{s \in [0,1]} \inf_{d(\o,\mu_\oplus(s)) < \alpha} \lbrace E(d^2(X(s),\o))-E(d^2(X(s),\mu_\oplus(s))) -D d^{\beta_2}(\o,\mu_\oplus(s))\rbrace \geq 0.
	\end{equation*}
\end{itemize}
\begin{prop} 
	\label{lma: mean_consistency}
	Under assumptions (A3) and (A8),
	\begin{equation*}
	\sup_{t \in [0,1]} d(\hat{\mu}_\oplus(t),\mu_\oplus(t))=o_P(1).
	\end{equation*}
\end{prop}
Assumptions (A4) and (A9)-(A11) are required to obtain an entropy condition for the space of \fro \  (Lemma \ref{lma: entropy} below),  which is used to  establish the rate of convergence of the sample \F \ mean function. We note that (A9), where we  assume that in a sufficiently close neighborhood of the true \F \ mean function $\mu_\oplus(t)$ all object functions have a common rate of H\"{o}lder continuity and a 
common H\"{o}lder constant, is weaker than assumptions that have been required in  classical FDA \cite[see e.g.][]{mull:06:11}, where one deals with real-valued random functions.   
Assumption (A10) is a bound on the covering number of the object metric space and is satisfied by common instances for random objects that include the  examples discussed at the end of section \ref{sec: scalar_fpc}.   

We write $\o(\cdot)$ for $\O$-valued functions $[0,1]\rightarrow \O$ and define
\begin{equation*}
V_n(\o,s)= \frac{1}{n} \sum_{i=1}^n \lbrace d^2(X_i(s),\o(s))- d^2(X_i(s),\mu_\oplus(s)) \rbrace,
\end{equation*}
\begin{equation*}
V(\o,s)= E \lbrace d^2(X(s),\o(s))- d^2(X(s),\mu_\oplus(s)) \rbrace.
\end{equation*}
Here $\hat{\mu}_\oplus(\cdot)$ is the minimizer of $V_n(\o,s)$ and $\mu_\oplus(\cdot)$ is the minimizer of $V(\o,s)$. We refer to $\hat{\mu}_\oplus(\cdot), \mu_\oplus(\cdot)$ and $\o(\cdot)$ as $\hat{\mu}_\oplus,\mu_\oplus$ and $\o$ in the following.  
To derive the rate of convergence of $\hat{\mu}_\oplus$, we first obtain a  bound  for  $E\left(\sup_{s \in [0,1]} \sup_{d_{\infty}(\o,\mu_\oplus) < \delta} \left|V_n(\o,s)-V(\o,s)\right|\right)$ for small $\delta > 0$, where $d_{\infty}(\o,\mu_\oplus)=\sup_{s \in [0,1]} d(\o(s),\mu_\oplus(s))$. For this, we define function classes
\begin{equation}
\label{func_class}
\mathcal{F}_{\delta}=\lbrace  f_{\o,s}(x)= d^2(x(s),\o(s))-d^2(x(s),\mu_\oplus(s)): s \in [0,1], \, d_{\infty}(\o,\mu_\oplus) < \delta \rbrace.
\end{equation}

It is easy to see that an envelope function for this class is the constant function $F(x)=2M\delta$, where $M$ is the diameter of $\O$. The $L^2$-norm of this envelope function is $||F||_{2}=2M\delta$. By Theorem 2.14.2 of \cite{well:96} we have
\begin{equation}
\label{eq: tail_bound}
E\left(\sup_{s \in [0,1]} \sup_{d_{\infty}(\o,\mu_\oplus) < \delta} \left|V_n(\o,s)-V(\o,s)\right|\right) \leq \frac{2M\delta \ J_{[]}(1,\mathcal{F}_{\delta},L^{2}(P))}{\sqrt{n}},
\end{equation}
where $J_{[]}(1,\mathcal{F}_{\delta},L^{2}(P))$ is the bracketing integral of the function class $\mathcal{F}_\delta$, 
\begin{equation*}
J_{[]}(1,\mathcal{F}_{\delta},L^{2}(P))= \int_{0}^{1} \sqrt{1+\log N(\eps ||F||_{2},\mathcal{F}_{\delta},L^2(P))} d\eps.
\end{equation*}
Here $N(\eps ||F||_{2},\mathcal{F}_{\delta},L^2(P))$ is  the minimum number of balls of radius $\eps ||F||_{2}$ required to cover the function class $\mathcal{F}_{\delta}$ under the $L^2(P)$ norm. Lemma \ref{lma: entropy} provides the behavior of the bracketing integral of the function class $\mathcal{F}_\delta$, a key step  for the proof of Theorem \ref{lma: rate}.
\begin{Lemma} 
	\label{lma: entropy}
	Under assumptions (A4),(A9) and (A10), it holds  for the function class $\mathcal{F}_{\delta}$ as defined in \eqref{func_class} that  $J_{[]}(1,\mathcal{F}_{\delta},L^{2}(P)) =O(\sqrt{\log{1/\delta}})$ as $\delta \rightarrow 0$.
\end{Lemma}
\begin{thm}
	\label{lma: rate}
	Under assumptions (A3)-(A4) and (A8)-(A11),
	\begin{equation*}
	\sup_{s \in [0,1]} d(\hat{\mu}_\oplus(s),\mu_\oplus(s))=O_P\left(\left(\frac{\sqrt{\log{n}}}{n}\right)^{1/\beta_2}\right).
	\end{equation*}
\end{thm}
Setting $\hat{D}_i(t)=d(X_i(t),\hat{\mu}_\oplus(t))$, an application is  the convergence of the estimated \F \ scores 
\begin{equation}
\label{eq: score}
\hat{\beta}_{ik}=\int_{0}^{1} \hat{D}_i(t)\hat{\phi}_k(t)dt.
\end{equation}
\vspace{-0.85cm}
\begin{cor}
	\label{cor: scalar_fpc}
	Under assumptions (A3)-(A5) and (A8)-(A11),
	\begin{equation*}
	\left|\hat{\beta}_{ik}-\beta_{ik}\right| =O_P\left(n^{-1/2}+\left(\frac{\sqrt{\log{n}}}{n}\right)^{1/\beta_2}\right).
	\end{equation*}
\end{cor}
Following widely adopted convention, we assume throughout that true and estimated eigenfunctions are aligned in the sense that  $\langle \hat{\phi}_j, \phi_j \rangle \geq 0$,  as the scores are identifiable only up to a sign change.

\section{Simulations}
\label{sim}
We illustrate the utility of the proposed  methods through simulations for two settings: In the first setting,  the space $\O$ consists of univariate probability distributions equipped with the 2-Wasserstein metric and in the second setting,   $\O$ consists of networks with fixed number of nodes, represented as graph adjacency matrices and equipped with the Frobenius metric.
\subsection{Time-varying probability distributions}
We generated random samples of sizes $n=25, 50$ and $100$ of `distribution'-valued curves on the domain $[0,1]$, where for each $t \in [0,1]$,  $X_i(t)$ is a normal distribution with mean $\mu_i(t)$ and variance $\sigma^2_i(t)$ with 
\bea
\mu_i(t)&=&1+U_i \phi_1(t)+ V_i \phi_3(t), \quad U_i \sim N(0,12), \quad  V_i \sim N(0,1),\\
{ \sigma_i(t)}&=&3+ W_i \phi_2(t)+Z_i \phi_3(t), \quad  W_i \sim \sqrt{72}U(0,1), \quad Z_i \sim \sqrt{9}U(0,1),  \eea
with $\phi_1(t)=(t^2-0.5)/0.3416$,  $\phi_2(t)=\sqrt{3}t$, $\phi_3(t)=(t^3-0.3571t^2-0.6t+0.1786)/0.0895$ where  $\phi_1, \phi_2$ and $\phi_3$ are orthonormal on $[0,1]$.  We use the 2-Wasserstein metric for the distribution space $\O$.  For these specifications, the  metric auto-covariance function is 
\begin{equation}
C(s,t)=12\phi_1(s)\phi_1(t)+6\phi_2(s)\phi_2(t)+1.75\phi_3(s)\phi_3(t),  \end{equation} 
and $\phi_1(\cdot), \phi_2(\cdot)$ and $\phi_3(\cdot)$ are the first 3 eigenfunctions.

        We applied the proposed method to estimate the metric auto-covariance operator for the simulated data and obtained its eigenvalues and eigenfunctions. Denoting the  estimated metric auto-covariance surface and the estimated $j$-th eigenvalue and eigenfunction 
 obtained at the $k^{th}$ simulation run by $\hat{C}_k(s,t)$, respectively  $\hat{\lambda}_{j,k}$ and $\hat\phi_{j,k}$,   we computed mean integrated squared errors (MISE)  
\bea
\text{MISE}(C)&=& \frac{1}{100} \sum_{k=1}^{100} \int_{0}^{1} \int_{0}^{1} \left( \hat{C}_k(s,t)-C(s,t)\right)^2 ds dt,\\
\text{MISE}(\phi_j)&=& \frac{1}{100} \sum_{k=1}^{100} \int_{0}^{1} \left( \hat{\phi}_{j,k}(s)-\phi_j(s) \right)^2 ds, \quad
\text{MISE}(\lambda_j)=\frac{1}{100} \sum_{k=1}^{100} \left( \hat{\lambda}_{j,k}-\lambda_j \right)^2. 
\eea
Figure \ref{fig: f1} shows the true and estimated metric auto-covariance surfaces and their  eigenfunctions for one randomly chosen simulation run for $n=25$ and $n=100$. We find that the proposed method has negligible bias as sample size increases. 
The MISEs are reported in Table \ref{tab:B-1} and seen to decrease with increasing sample sizes.
\captionof{table}{Mean Squared Errors for the estimators of  the metric auto-covariance kernel $C$ and the eigenfunctions $\phi_1, \phi_2$ in dependence on sample size when the \fro \ are distributions.} 
\label{tab:B-1} 
\begin{center}
	\begin{tabular}{ |c|c|c|c|c|c|c|c|} 
		\hline
		$n$ & $C$ & $\phi_1$ & $\phi_2$&  $\phi_3$ &$\lambda_1$ & $\lambda_2$ & $\lambda_3$ \\  \hline
		25 & 12.2709
		& 1.2841    & 1.5351  & 1.5202&10.8634 & 7.8471& 3.5798\\ 
		50 & 8.6598
		 & 0.0504& 0.0201  & 0.0030& 0.9748 & 0.6482 & 0.3680\\ 
		100 & 4.0697& 0.0158 &  0.0084  & 0.0047& 0.1239& 0.0607 & 0.0314\\
		\hline
	\end{tabular}
	\label{tab:B-1}
\end{center} 
\begin{figure}
	\centering \vspace{-.265cm}
	\includegraphics[scale = .38]{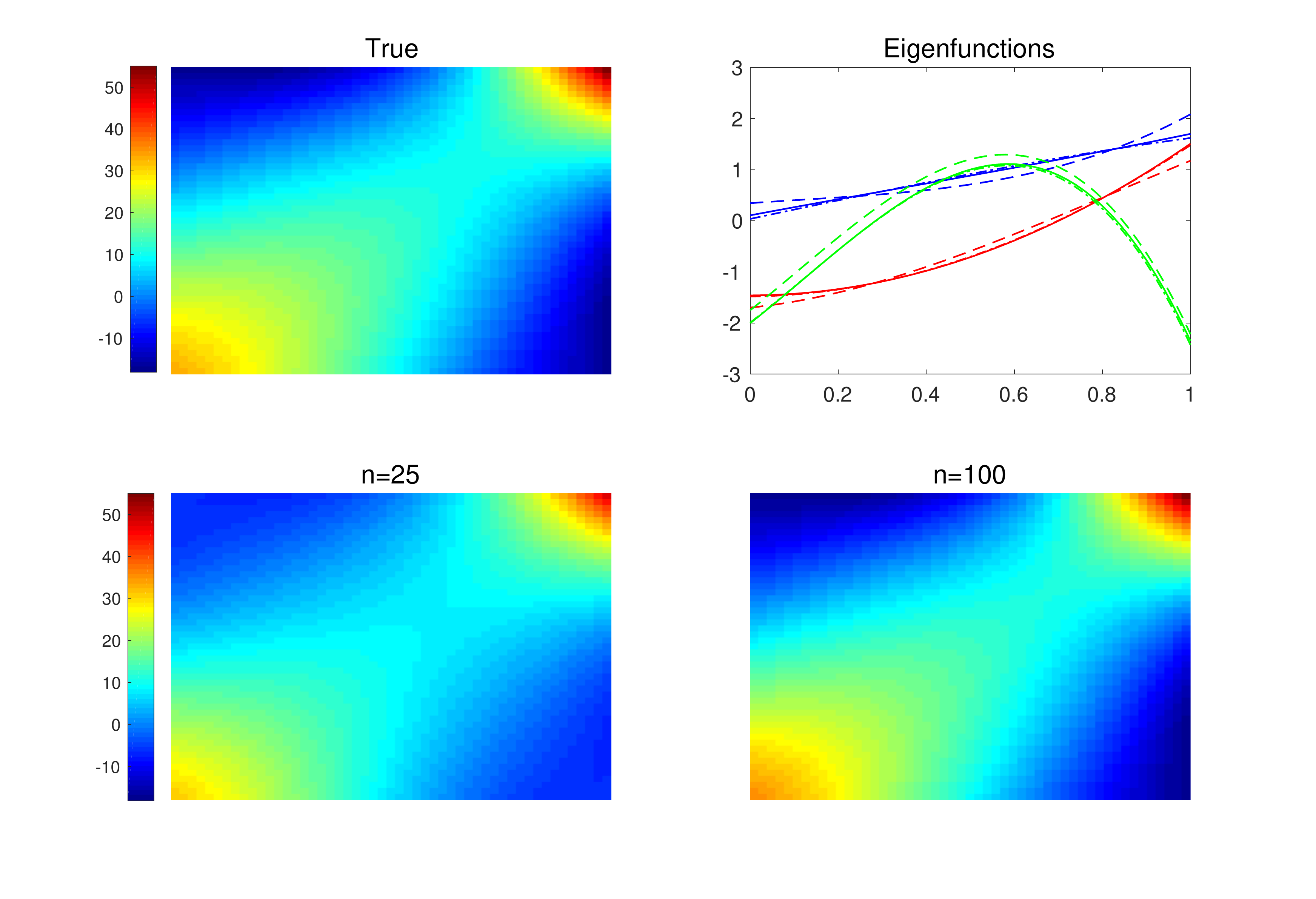} \vspace{-.95cm}
	 \caption{{True (top left)  and estimated (bottom panels) metric auto-covariance contour plots \eqref{eq: cov} for simulation with distributions as \fro. The red curves in the top right panel correspond to the first, blue curves to the second and green curves to the third
	 eigenfunction, depicting true (solid) and estimated (dashed for $n=25$ and dot-dashed for $n=100$) eigenfunctions.}}
	\label{fig: f1}
\end{figure} 
To illustrate the nature of the simulated random density trajectories, four density-valued random functions that are part of a sample of density-valued random functions as generated in one Monte Carlo run  are displayed in Figure \ref{fig: f24}, reflecting  variation in means and variances of the Gaussian  distributions as a function of time for the four selected subjects. The estimated  object FPCs, i.e. the \F \ integrals of the object curves along the first two eigenfunctions, from one Monte Carlo run are in Figure \ref{fig: f2} for  sample size 50. 
 Here the  first object FPCs  reflect  variation in location of the distributions and the second object FPCs variation in the variance of the distributions, which is what we expect  in view of  how these data were generated.  The  object FPCs are found  to be  useful for discovering the underlying  modes of variation for distributions as \fro.

\subsection{Time-varying networks}
\label{sec: sim_net}
In each iteration, we generated random samples of sizes $n=25, 50$ and $100$ of time varying random networks with 10 nodes each in the time interval $[0,1]$. For generating the edge weights, we followed the model described below. We assumed that the network has two communities, the first five nodes belonging to one community and the second five nodes to the other one. For each fixed time $t$, the edge weights within each community and also those  between the communities are the same, where the latter are smaller than the within community edge weights. Formally, if $p_{1,i}(t), p_{2,i}(t)$ and $p_{12,i}(t)$ denote the edge weight at time $t \in [0,1]$ for the first community, the second community and between communities,  for the $i^{th}$ network valued curve we generated 
\begin{equation}
\label{net_gen}
p_{1,i}(t)=0.5+U_i \phi_1(t) +V_i \phi_3(t), \quad p_{2,i}(t)= 0.5+W_i \phi_2(t) + Z_i \phi_3(t), \quad p_{12,i}(t)= 0.1. 
\end{equation}
Here the $U_i$, $V_i$, $W_i$ and $Z_i$ were generated from the uniform distributions $U(0,0.4)$, $U(0,0.1)$, $U(0,0.3)$ and $U(0,0.1)$, respectively. The functions $\phi_1(t)$, $\phi_2(t)$ and $\phi_3(t)$  are orthonormal polynomials derived from Jacobi polynomials $P_n^{(\alpha,\beta)}(x)$ \cp{toti:05}, which are classical orthogonal polynomials for $\alpha, \beta > 1$. They  are orthogonal with respect to the basis $(1+x)^{\beta}(1-x)^\alpha$ on $[-1,1]$. With a suitable change of basis, 
one can obtain a version of the Jacobi polynomials on $[0,1]$ which are orthonormal with respect to the weight function $x^\beta(1-x)^\alpha$ on $[0,1]$. We selected $\phi_1(t)$, $\phi_2(t)$ and $\phi_3(t)$  as
\begin{equation*}
 \phi_j(t)=\frac{(P_{2j}^{(4,3)}(2t-1))t^{1.5} (1-t)^{2}}{[\int_{0}^{1}(P_{2j}^{(4,3)}(2t-1))^2t^{1.5} (1-t)^2dt]^{1/2}} \quad \text{for} \ j=1,2,3.
\end{equation*}}
The weighted networks are represented as graph adjacency matrices with the Frobenius metric. Here the true metric auto-covariance function is 
\begin{equation}
C(s,t)=0.266 \phi_1(s)\phi_1(t)+0.15 \phi_2(s)\phi_2(t)+0.0417 \phi_3(s)\phi_3(t),
\end{equation}
and $\phi_1(\cdot), \phi_2(\cdot)$ and $\phi_3(\cdot)$ are the first three eigenfunctions.

 \begin{figure}
 	\centering
 	\includegraphics[scale = .42]{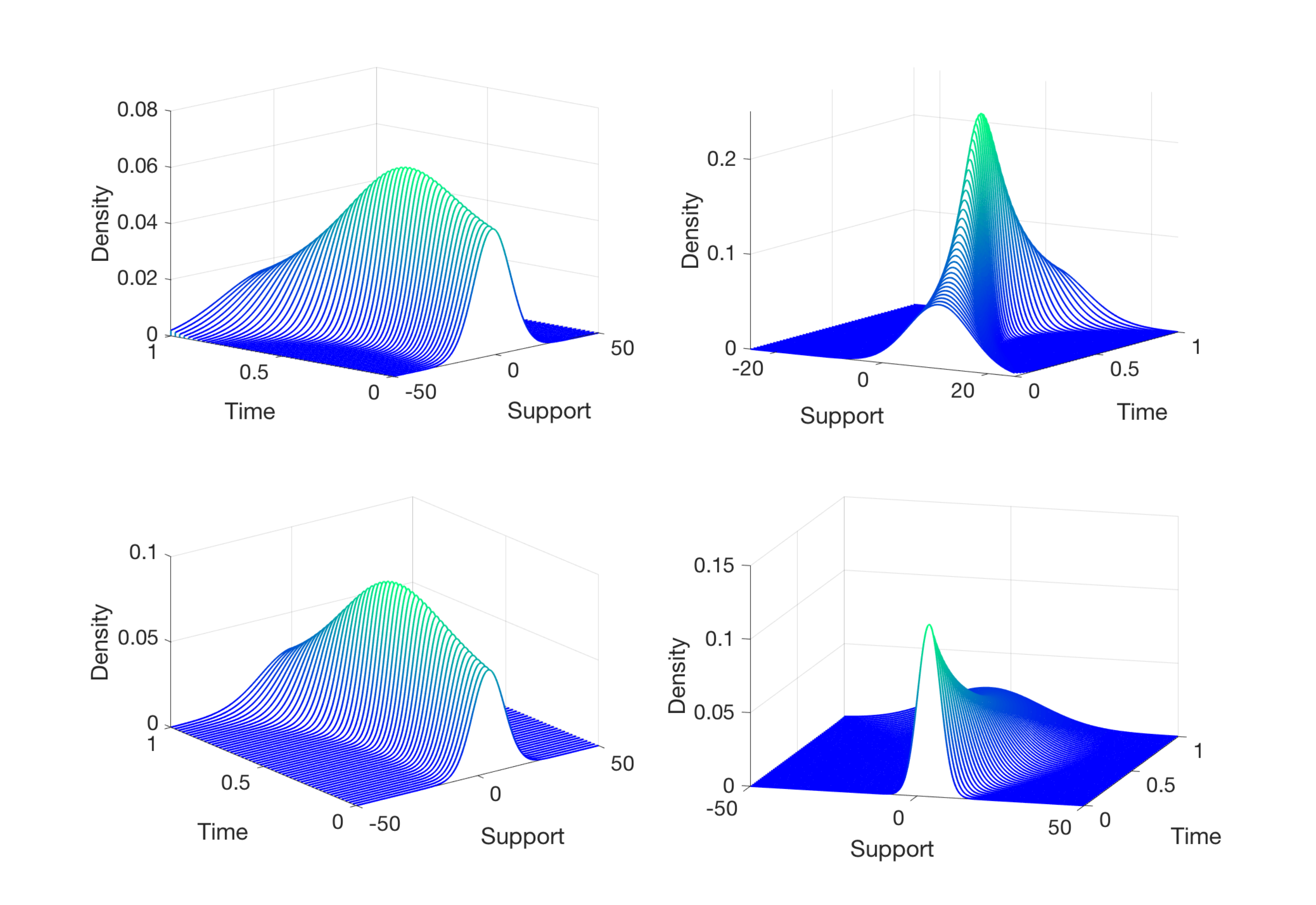}
 	\caption{Four randomly chosen observations of density-valued trajectories, selected from the sample of distributions generated by  one of the Monte Carlo runs. The densities are plotted as a function of time.}
 	\label{fig: f24}
 \end{figure}
\begin{figure}
	\centering
	\includegraphics[scale = .25]{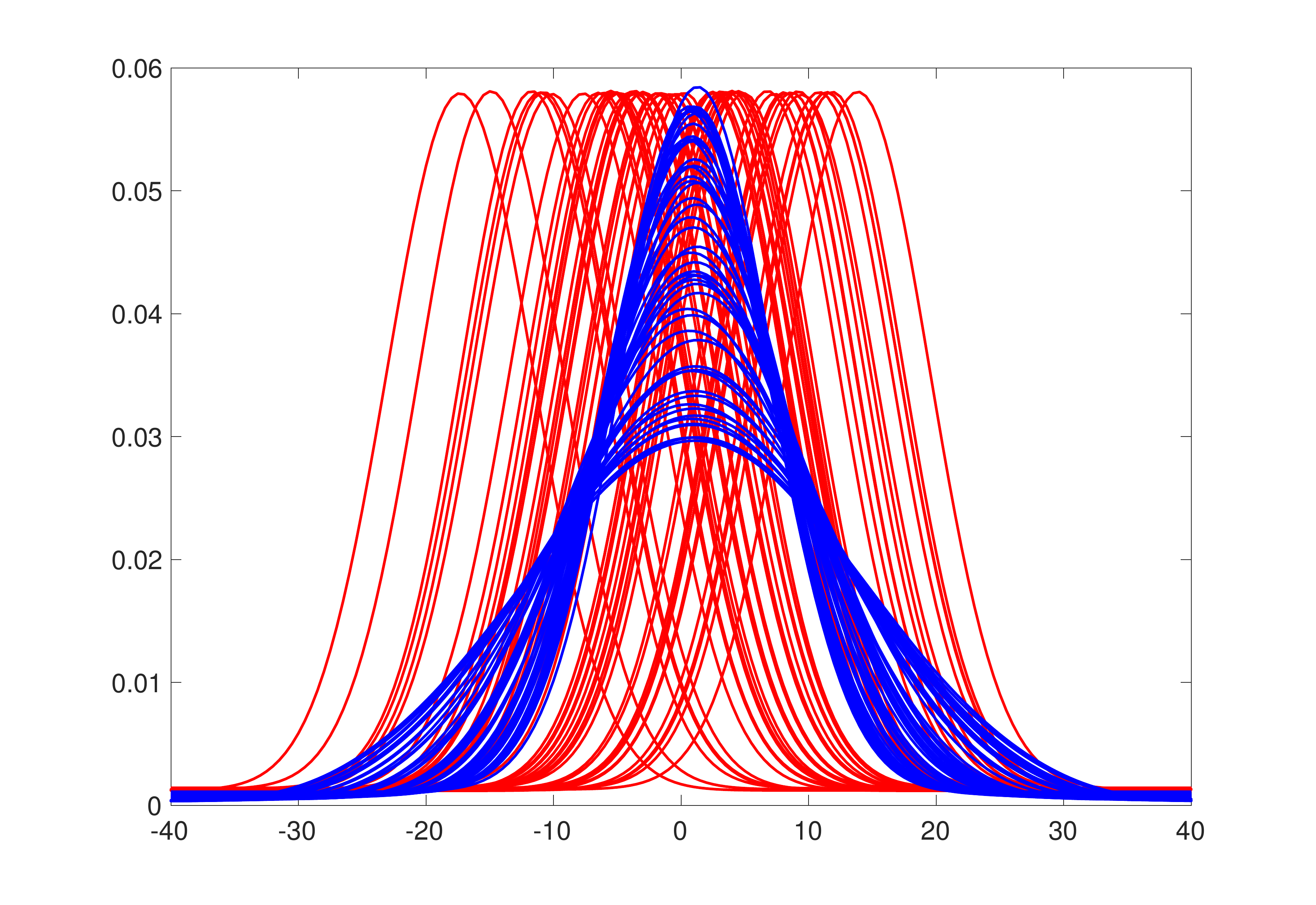}
	\caption{Estimated \F \ integrals \eqref{eq: int} for the first (red) and second (blue) eigenfunction for the sample elements,  for simulated time-varying probability distributions.}
	\label{fig: f2}
\end{figure}

\begin{figure}
	\centering  \vspace{-.85cm}
	\includegraphics[scale = .4]{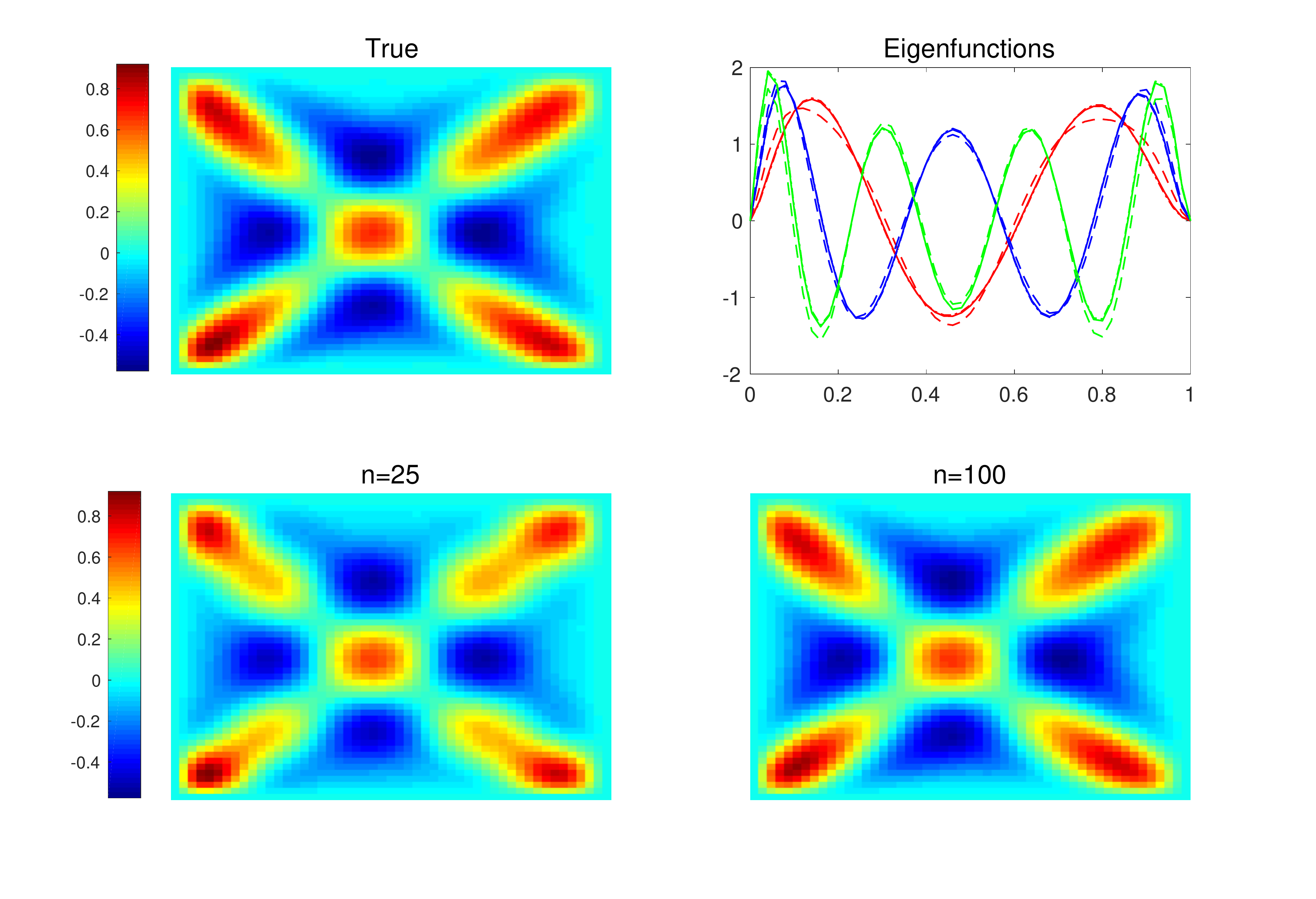} \vspace{-1cm}
	\caption{True (top left)  and estimated (bottom panels) metric auto-covariance contour plots \eqref{eq: cov} for simulation with networks as \fro. The red curves in the top right panel correspond to the first, blue curves to the second and green curves to the third
	 eigenfunction, depicting true (solid) and estimated (dashed for $n=25$ and dot-dashed for $n=100$) eigenfunctions.}	\label{fig: f3}
\end{figure}
We estimated the metric auto-covariance operator from the simulated data and obtained its eigenfunctions for different sample sizes. Figure \ref{fig: f3} displays  the true and estimated metric auto-covariance surfaces and corresponding eigenfunctions for one randomly chosen simulation run for $n=25$ and $n=100$.  The MISEs were computed as described for the previous simulation setting and are reported in Table \ref{tab: 2}. They decrease with increasing sample sizes. The proposed method is seen to work  very  well.
\captionof{table}{Mean Integrated Squared Errors for the estimators of the metric auto-covariance kernel $C$ and eigenfunctions/eigenvalues  $\phi_j, \lambda_j, \, j=1,2,3$, in dependence on sample size for samples of  \fro \ that correspond to time-varying networks.}\label{tab: 2}
\begin{center}
	\begin{tabular}{ |c|c|c|c|c|c|c|c|c|} 
		\hline
		$n$ & $C$ & $\phi_1$ & $\phi_2$ & $\phi_3$&$\lambda_1$ & $\lambda_2$ & $\lambda_3$ \\  \hline
		25 & 0.0039
	& 0.0039  & 0.0017& 0.0007 & 0.0025 &   0.0010  &  0.0007\\ 
		50 & 	0.0017
		 & 0.0093   & 0.0046 &   0.0021 & 0.0007 &   0.0003  &  0.0001 \\ 
		100 & 0.0010&0.0130  &  0.0063  &  0.0028 & 0.0001 &   0.0001  &  0.0001\\
		\hline
	\end{tabular}
\end{center}

The object FPCs were obtained using \F \ integrals \eqref{eq: int}. For visualization they are presented as a  ``networks.mov" in the supplementary materials. In the movie the leftmost plot corresponds to \F \ integrals for the first eigenfunction which, as expected due to the true model, shows variation only in the edge weights of the first community. The middle plot corresponds to \F \ integrals for the second eigenfunction and indicates variation only in the edge weights of the second community.  The rightmost plot corresponds to \F \ integrals for the third eigenfunction, where variation in both the first and the second community edge weights can be discerned. 

\section{Data Applications}
\label{data}
\subsection{Mortality data}
\label{mort}
\begin{figure}[t]
	\centering
	\includegraphics[scale = .46]{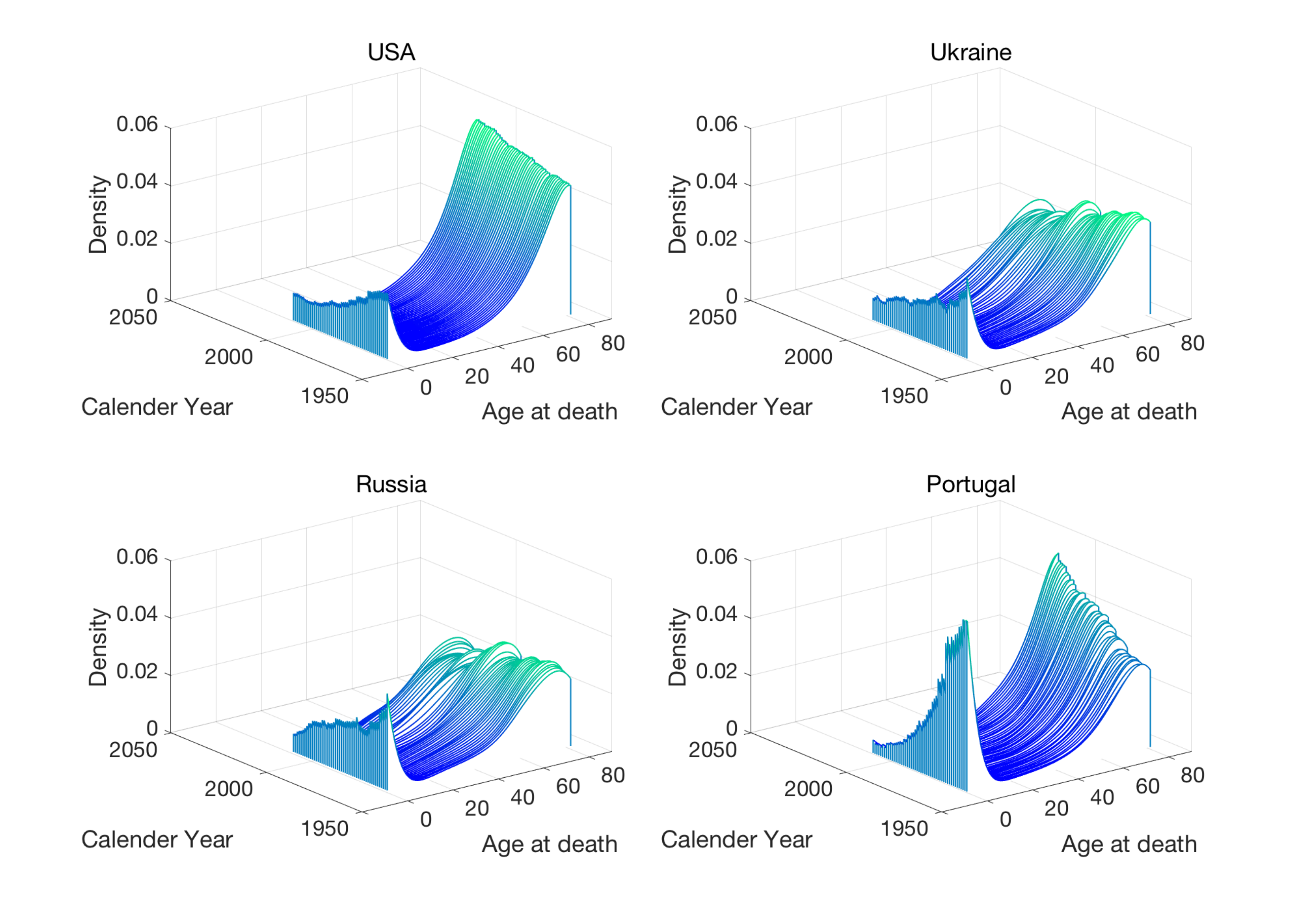}
	\caption{Time-varying age at death density functions for the age interval $[0,80]$ for males in USA, Ukraine, Russia and Portugal.}
	\label{fig: f25}
\end{figure}
\begin{figure}
	\centering 
	\includegraphics[scale = .46]{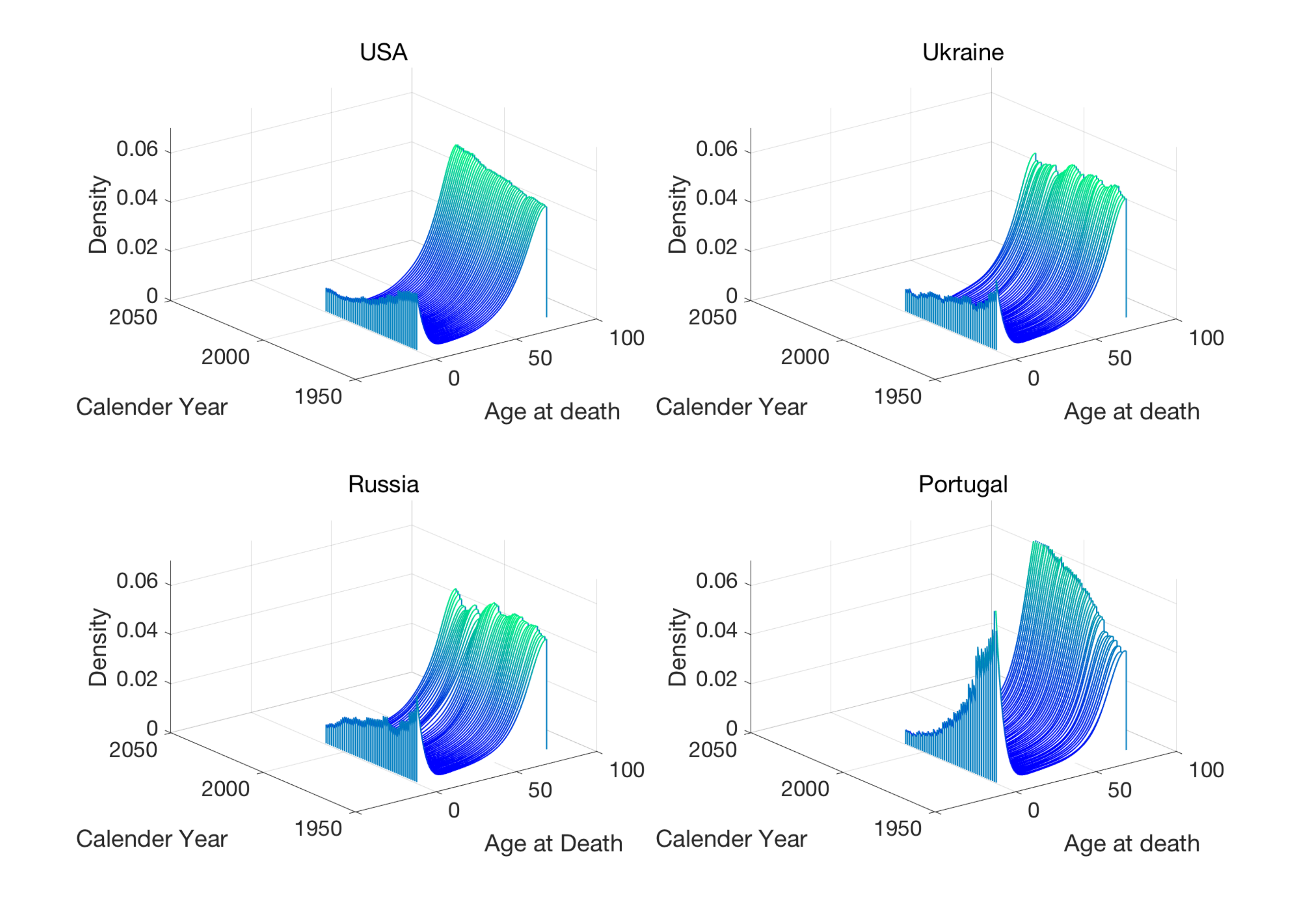} 
	\caption{Time-varying age at death density functions for the age interval $[0,80]$ for females in  USA, Ukraine, Russia and Portugal.}
	\label{fig: f26}
\end{figure}

The Human Mortality Database provides life table data differentiated by gender and is available at \url{www.mortality.org}. Currently the mortality database contains life table data for 37 countries spanning over 5 decades. One can obtain histograms from life tables and smooth these with  local least squares to obtain estimated probability density functions for age at death. We carried this out for  the age interval $[0,80]$. The mortality data can then be viewed as  samples of time varying univariate probability distributions, for a sample of 32 countries,  where the time axis corresponds to calendar years between 1960 and 2009 and the observation made at each  calendar year  for each country corresponds to the age at death distribution for that year.  We included the 32 countries which had complete records over the entire calendar period. For each country and year, we used the Hades package available at \url{https://stat.ucdavis.edu/hades/} for smoothing the histograms and used bandwidth$=2$ to obtain the age-at-death densities. For illustration, the time-varying age at death distributions represented as density functions for the age interval $[0,80]$ and indexed by calendar year are displayed  for four selected countries, USA, Ukraine, Russia and Portugal,  for males in Figure \ref{fig: f25}  and for females in Figure \ref{fig: f26}.

\begin{figure}
	\centering
	\includegraphics[width=6.3in]{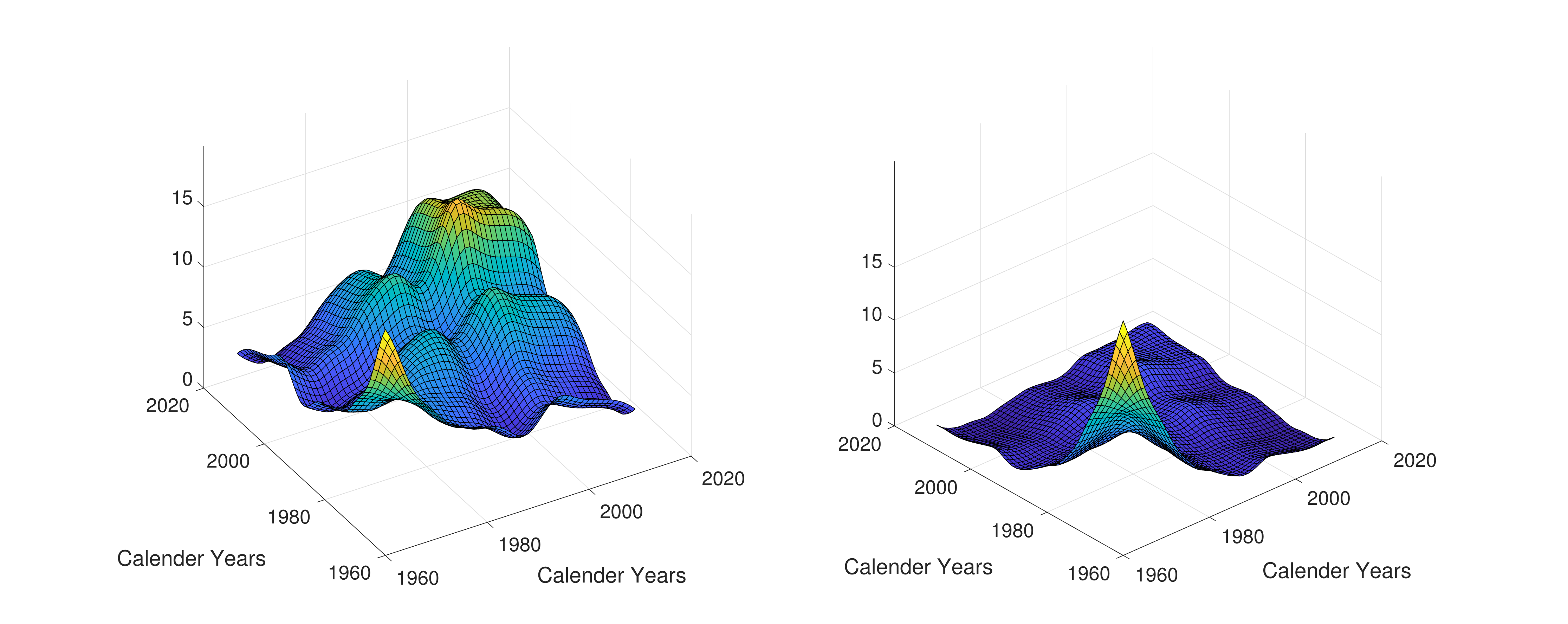}
	\caption{Estimated metric auto-covariance surfaces \eqref{eq: cov} for males (left) and females (right) for densities as \fro, as obtained for the mortality data.}
	\label{fig: f4}
\end{figure}

\begin{figure}
		\centering
		\includegraphics[scale = .39]{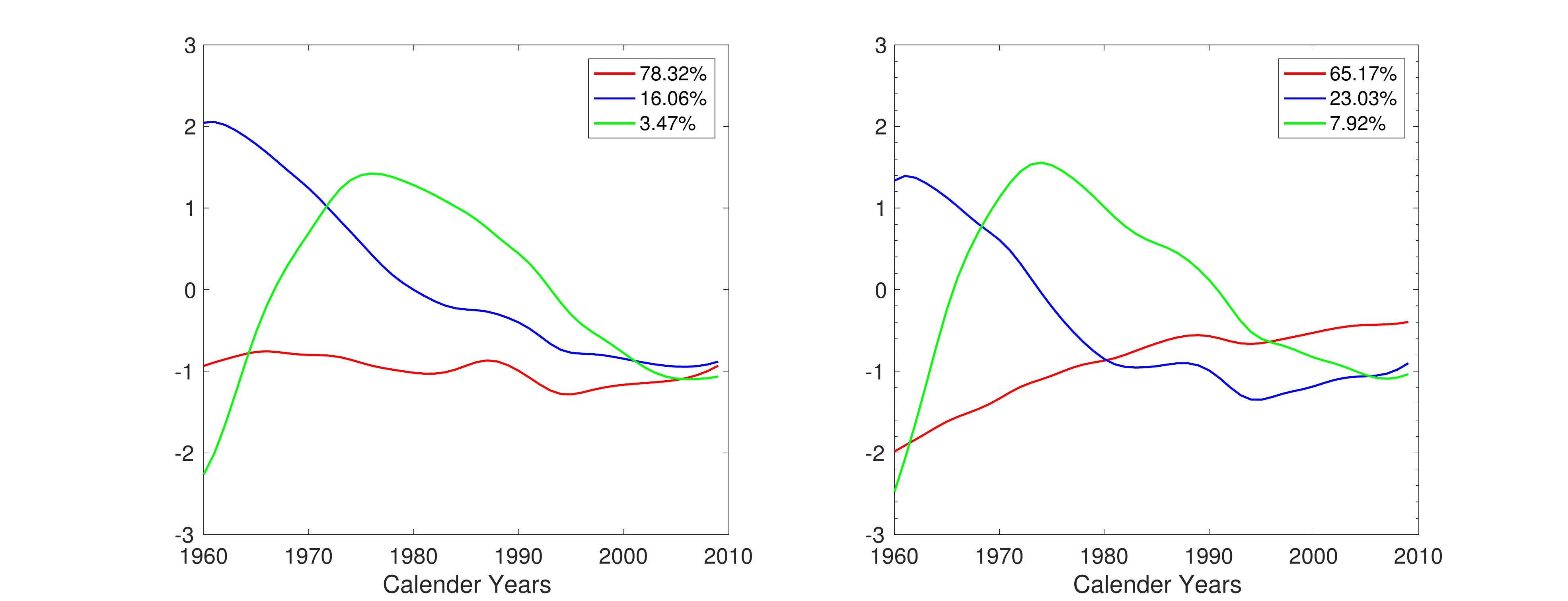}
		\caption{The eigenfunctions of the estimated metric auto-covariance surface for males (left) and for females (right) for the mortality data.}
		\label{fig: f5}
	\end{figure}

\id Choosing the 2-Wasserstein metric for the probability distributions space, the estimated metric auto-covariance surfaces for males and females can be inspected in Figure \ref{fig: f4} and the eigenfunctions of the corresponding auto-covariance operators in  Figure \ref{fig: f5}. The auto-covariance functions and eigenfunctions indicate that there are systematic differences  between males and females.

\begin{figure}
 	\centering  \vspace{-.3cm}
 	\includegraphics[scale = .44]{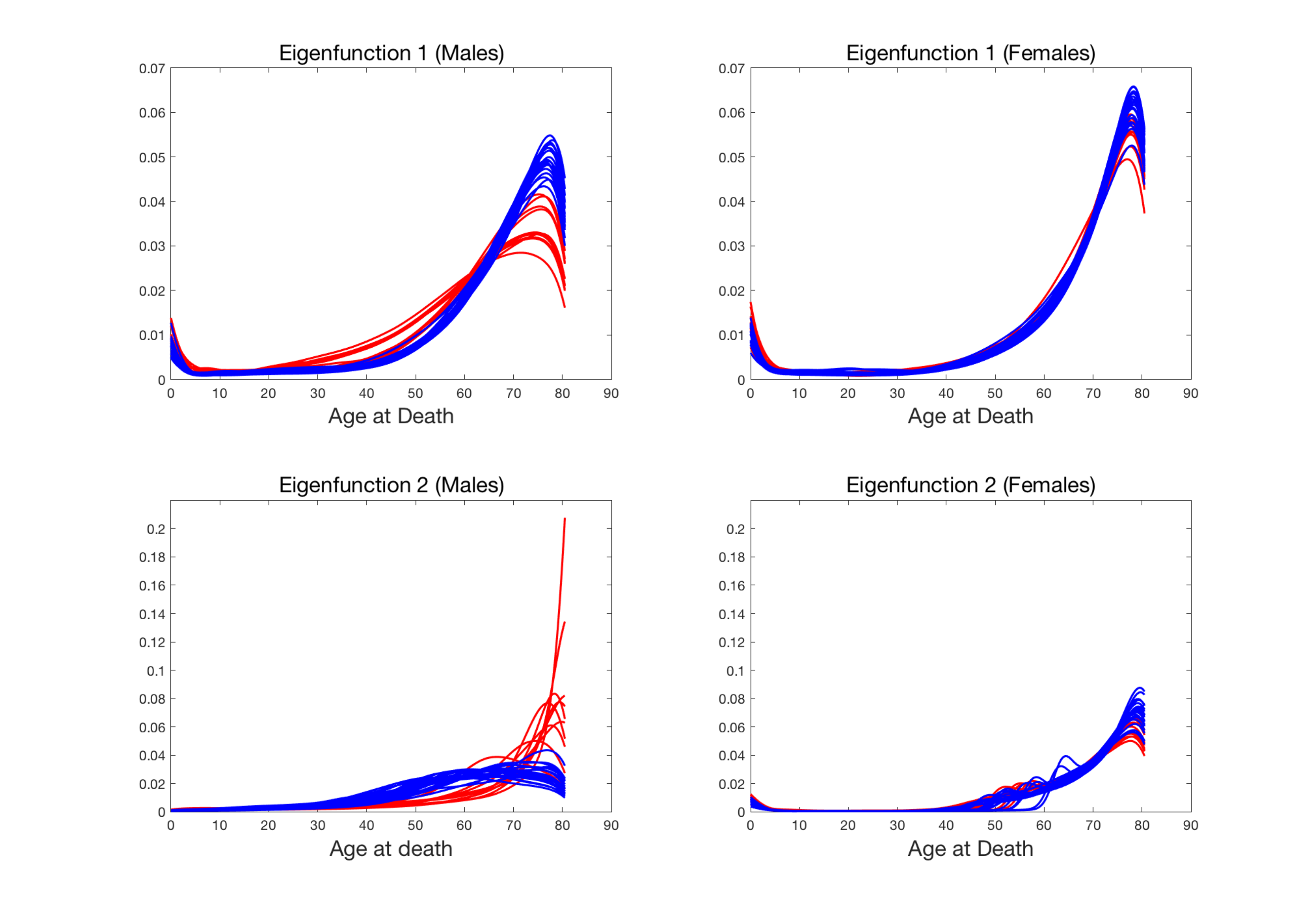} \vspace{-1cm}
 	\caption{ {\F \ integrals \eqref{eq: int} for the first eigenfunction for males (top left) and females (top right) and for the second eigenfunction for males (bottom left) and females (bottom right) for the mortality data. The red curves correspond to the Eastern European countries, namely, Belarus, Bulgaria, Czech Republic, Hungary, Latvia, Lithuania, Poland, Slovakia, Ukraine, Russia and Estonia and the blue curves to the other countries.}}
 	\label{fig: f6}
 \end{figure}

\id The resulting  object FPCs, i.e. the \F \ integrals, are illustrated in Figure \ref{fig: f6} for the first two eigenfunctions. The object FPCS are distributions that are represented as densities for males and females. The  Eastern European countries included in the data base, namely, Belarus, Bulgaria, Czech Republic, Hungary, Latvia, Lithuania, Poland, Slovakia, Ukraine, Russia and Estonia underwent major political upheaval due to the end of Communist rule in these regions during the period between the late 1980s and early 1990. This is reflected in  clear distinctions between the Eastern European countries (red) and the rest (blue) in the \F \ integrals for the males but much less so for the females, which indicates that particularly male mortality was affected by the political upheavals. 


 \id The sample \F \ mean function at a particular calendar year  corresponds to the sample average of the quantile functions of the different countries at that calendar year and is illustrated in the  movies ``mean\_males.mov" and ``mean\_females.mov" in the supplementary materials. Figure \ref{fig: f7} illustrates  the scalar FPCs, i.e. the \F \ scores for the second eigenfunction plotted against the \F \ scores for the first eigenfunction for males and females. Russia is an outlier for the first eigenfunction for males and Portugal is an outlier for the second eigenfunction, even though it does not belong to the above list of Eastern European countries. One could speculate that this might be related to the fact Portugal in 1974 moved  to a democratic government  after four decades of authoritarian dictatorship. Figures \ref{fig: f25} and \ref{fig: f26} suggest higher infant mortality for both males and females in Portugal during the earlier era. Another interesting observation is that the order of outliers is reversed for females, as Russia turns out to be an outlier for females for the second eigenfunction and Portugal for the first. The plots of the \F \ scores against each {other} indicate that there are clear distinctions between the two groups of countries and Portugal. 
 
\begin{figure}
	\centering
	\includegraphics[scale = .4]{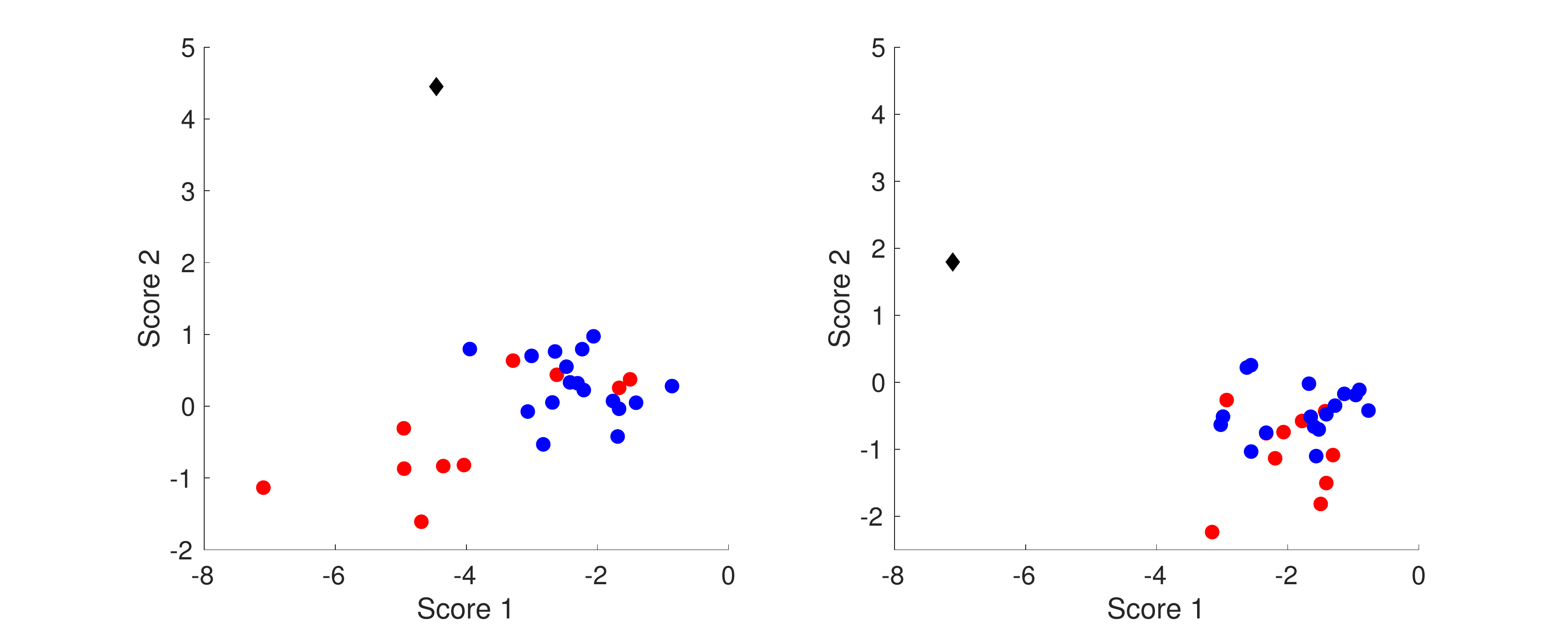}
	\caption{\F \ scores \eqref{eq: score} for the first and the second eigenfunctions plotted against each other for males (left) and for females (right) for the mortality data. The black diamond corresponds to Portugal, the red dots to the Eastern European countries, namely, Belarus, Bulgaria, Czech Republic, Hungary, Latvia, Lithuania, Poland, Slovakia, Ukraine, Russia and Estonia and the blue dots to the other countries.}
	\label{fig: f7}
\end{figure}
\subsection{Time-varying networks for New York taxi data}
\label{taxi}
New York City Taxi and Limousine Commission (NYC TLC) provides records on pick-up and drop-off dates/times, pick-up and drop-off locations, trip distances, itemized fares, rate types, payment types, and driver-reported passenger counts for yellow and green taxis available at \url{http://www.nyc.gov/html/tlc/html/about/trip_record_data.shtml}. The time resolution of these data is in the order of  seconds. Of interest are networks which represent how many people traveled between places of interest and the evolution of these networks during a typical day. To study this, we constructed samples of time-varying networks where the sample elements  are the recordings for each day in the year 2016. Three days (23 and 24 Jan and 13 March) were excluded from the study due to incomplete records. 
	
\id We focus on the Manhattan area, which has the highest traffic and  split the area according to the provided location shape files into 10 zones,  which form the regions of interest. Details about the  zones are  in section \hyperref[data_des]{S6.1} of the online supplement. Yellow taxis provide the  predominant taxi service in Manhattan. We divided each day into 20 minute intervals, and for each interval constructed a network with nodes corresponding to the 10 selected zones and edge weights representing the number of people who traveled between the zones connecting the edges within the 20 minute interval. The edge weights were normalized by the maximum edge weight for each day so that they lie in $[0,1]$. We thus have a  time-varying network for each  of the  363 days in 2016 for which complete records are available, where the  time points {at which} the network-valued functions are evaluated correspond to the 20 minute intervals of a 24 hour day. The observations at each time point correspond to a 10 dimensional graph adjacency matrix which characterizes the network between the 10 zones of Manhattan for the particular 20 minute interval.

\id We choose the Frobenius metric as metric between the graph adjacency matrices.  The sample \F \ mean function at a particular time point therefore corresponds to the sample average of the graph adjacency matrices of 363 networks corresponding to different days for that time point. It is  illustrated in the  movie ``mean\_NY.mov" in the supplementary materials. Figure \ref{fig: f10} illustrates the estimated auto-covariance function and associated eigenfunctions.
The plots of the \F \ scores  for second, third and fourth  eigenfunction against the scores for the first eigenfunction can be found in 
Figure  \ref{fig: f11}, where
 the blue dots correspond to Mondays to Thursdays, the green dots to Fridays and the red dots to Saturdays and Sundays.
  Several interesting patterns emerge: Weekdays and weekends form clearly distinguishable clusters. Special holidays 
show similar patterns as weekends. Several outliers can be identified using the projection scores for the  eigenfunctions, which turn out to be special days: For the first eigenfunction, the outliers correspond to New Year day and November 6, 2016, which is the day when daylight saving ends. March 13, 2016 is the day {on which} the daylight saving begins but was excluded as it did not have complete records. For the second eigenfunction, an outlying point is  Independence Day, 4 July 2016, and  for the third eigenfunction, 14 February 2016 which is Valentine's day.   Another day that stands out is 18 September 2016. On further investigating it was found that between 17 to 19 September 2016 three bombs exploded and several unexploded ones were found in the New York metropolitan area (\url{https://en.wikipedia.org/wiki/2016_New_York_and_New_Jersey_bombings}). 
\begin{figure}
	\centering
	\begin{subfigure}{7cm}
		\centering\includegraphics[width=8cm]{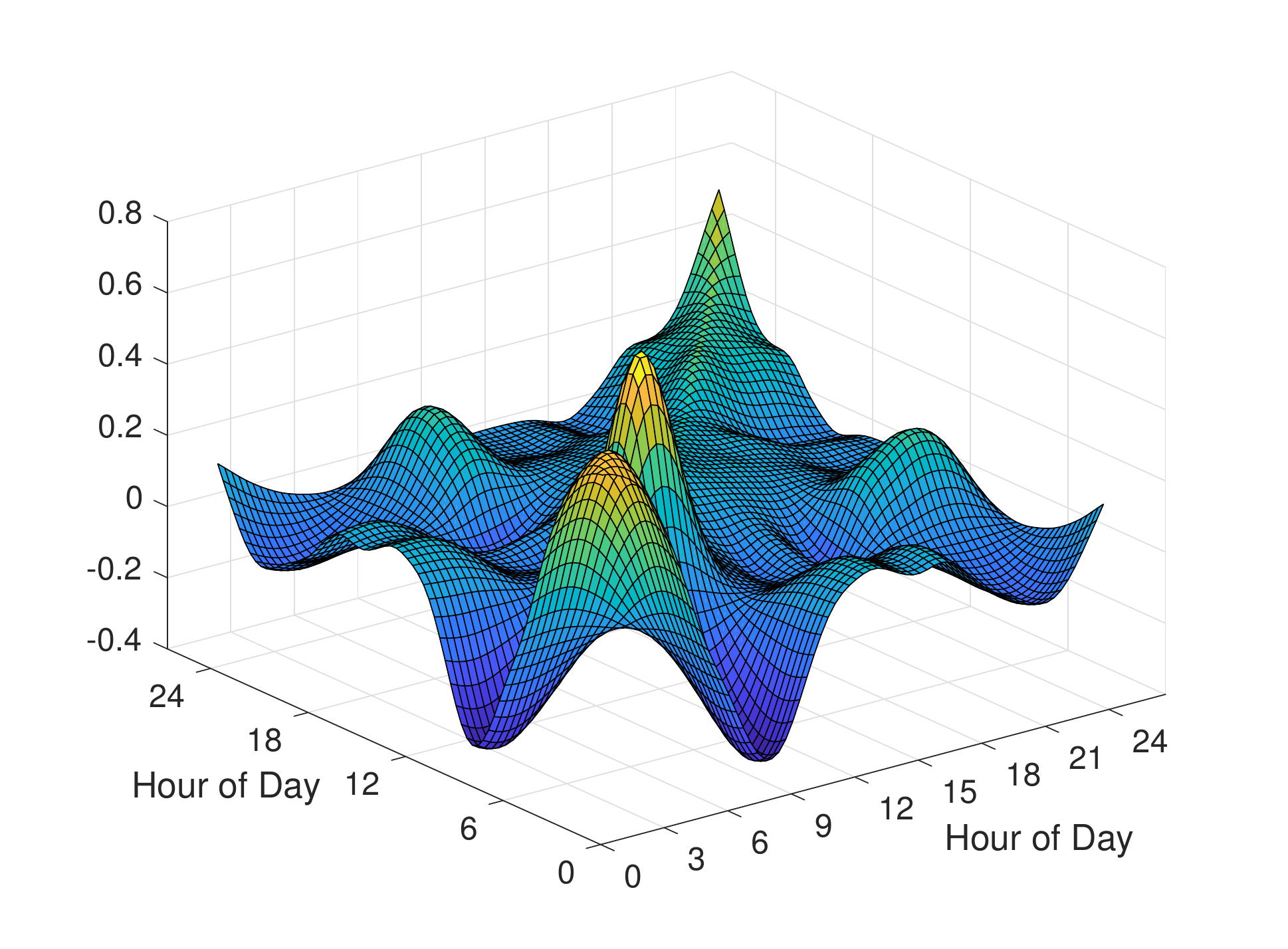}
		\label{fig: f8}
	\end{subfigure}
	\hspace{0.2 in}
	\begin{subfigure}{7cm}
		\centering\includegraphics[width=7cm]{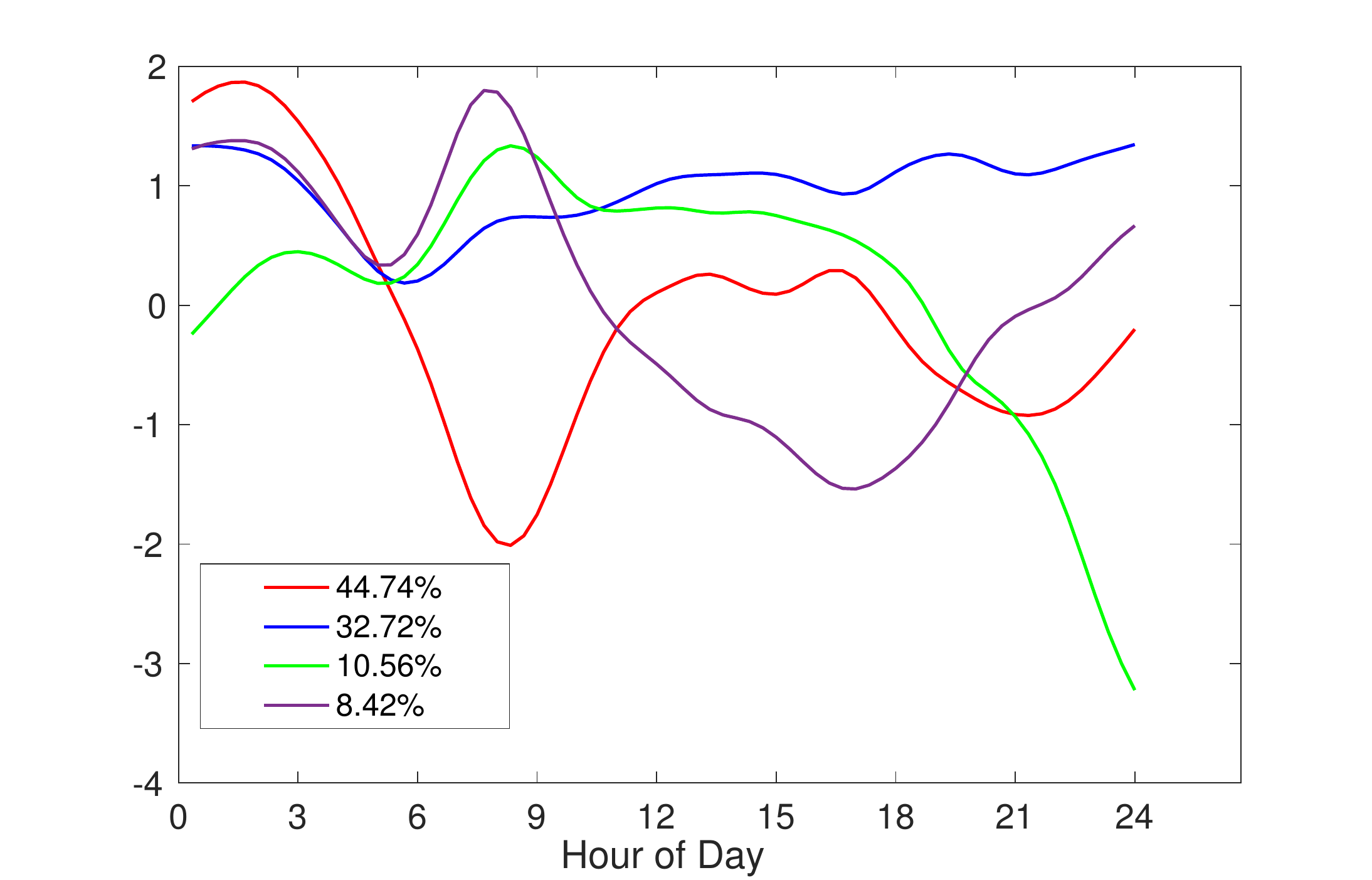}
		\label{fig: f9}
	\end{subfigure}
	\caption{Estimated metric auto-covariance surface \eqref{eq: cov} (left) and the corresponding  eigenfunctions (right) for the New York taxi data, viewed as time-varying networks.} 
	\label{fig: f10}
\end{figure}

 \begin{figure}
   	\hspace{-1.5cm} \includegraphics[scale = .39]{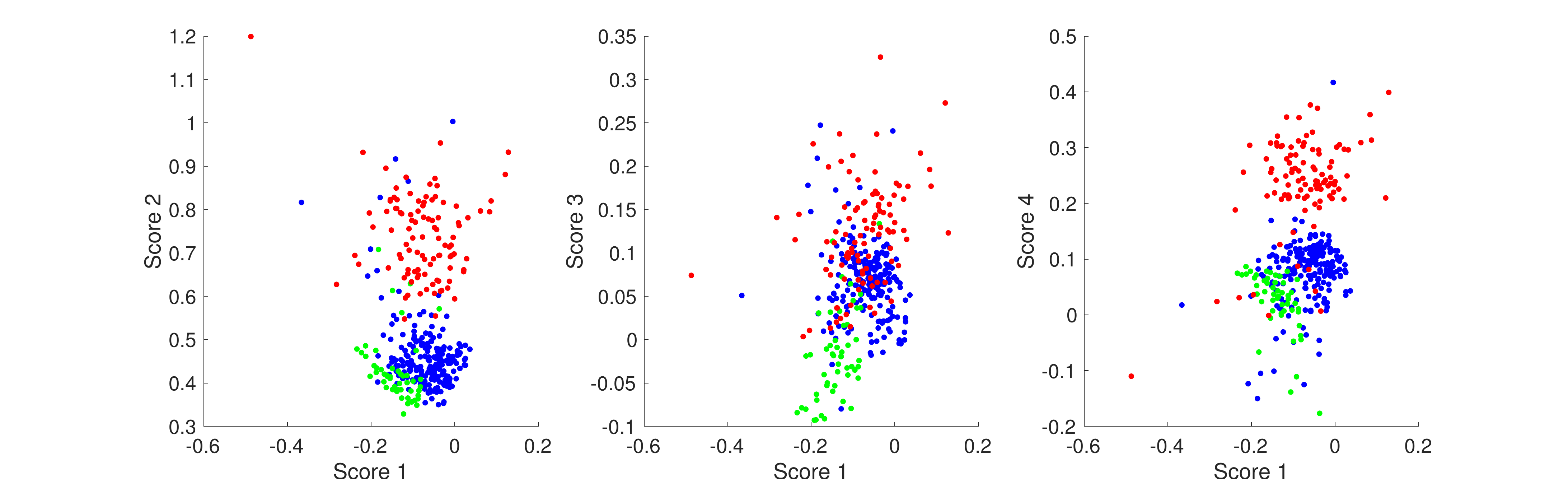}
   	\caption{\F \ scores \eqref{eq: score} for the second (left), third (middle) and fourth (right) eigenfunctions in the $y$-axis plotted against \F \ scores for the first eigenfunction in the $x$-axis, for the New York taxi data. The blue dots correspond to Mondays to Thursdays, the green dots to Fridays and the red dots to Saturdays and Sundays.}
   	\label{fig: f11}
   \end{figure}

\id We then repeated the analysis separately for three groups of days, namely the weekdays Monday-Thursday (group 1),  Fridays and weekends (group 2) and holidays (group 3).   We present the results in Figure 17 (Supplement) and in several movies whose descriptions can be found in sections \hyperref[NY_groups]{S3} and \hyperref[mov]{S5} of the online supplement. 

\subsection{World trade data}
\label{trade}
The Center for International Data at University of California, Davis (\url{http://cid.econ.ucdavis.edu/nberus.html}) provides detailed documentation of United Nations trade data for the years 1962-2000. The dataset, publicly available at {\url{https://cid.econ.ucdavis.edu/nberus.html}}, contains bilateral trade data during this time period for several commodities and countries. We studied  the time period 1970 to 1999 for  46 actively trading countries and the 26 most common types of commodities. The list of chosen countries and commodities can be found in the section \hyperref[data_des]{S6.2} of the online supplement.   For each country, commodity and year, we represent current trade as the ratio of the amount of total trade, i.e. import export value (in thousands of US dollars), to the amount of total trade recorded for the same commodity and country in the year 2000, yielding a 26-dimensional vector of trade ratios. 

\id Viewing  the countries as sampling units,  we obtain for each country and calendar year $t$  a  $26 \times 26$-dimensional raw covariance matrix of commodities trade ratios as $\tilde{\Sigma}(t)=(Q(t) -\bar{Q}(t))(Q(t) -\bar{Q}(t))^T$, where $Q(t)$ is the country-specific 26-dimensional vector of commodities trade ratios  for year $t$ and the mean vector $\bar{Q}$ is  obtained as cross-sectional average over all 46 countries. These raw time-varying raw covariances were then smoothed using local \F \ regression with a Gaussian kernel \cp{mull:19:3,mull:19:2} to obtain samples of smooth time-varying 26-dimensional covariance matrices between the components of  commodities trade for each of the \red{46} countries over the time period 1970 to 2000,  yielding  time-varying covariance matrices  over the time period 1970 to 2000 as \fro. 

\begin{figure}
	\centering
	\begin{subfigure}{6cm}
		\centering\includegraphics[width=6cm]{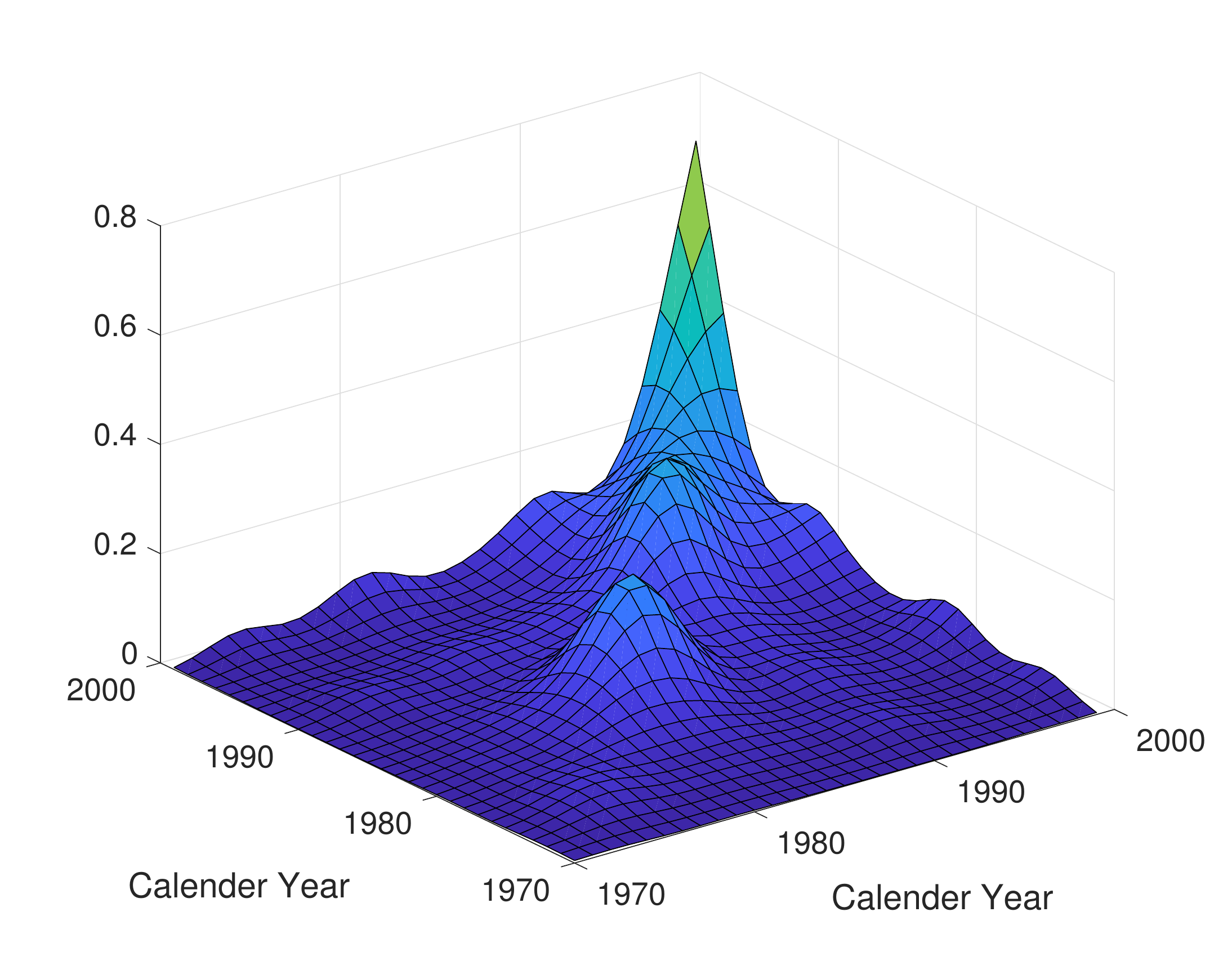}
		\caption{Estimated metric auto-covariance.}
		\label{fig: f20}
	\end{subfigure}
	\hspace{0.2 in}
	\begin{subfigure}{6cm}
		\centering\includegraphics[width=7.5 cm]{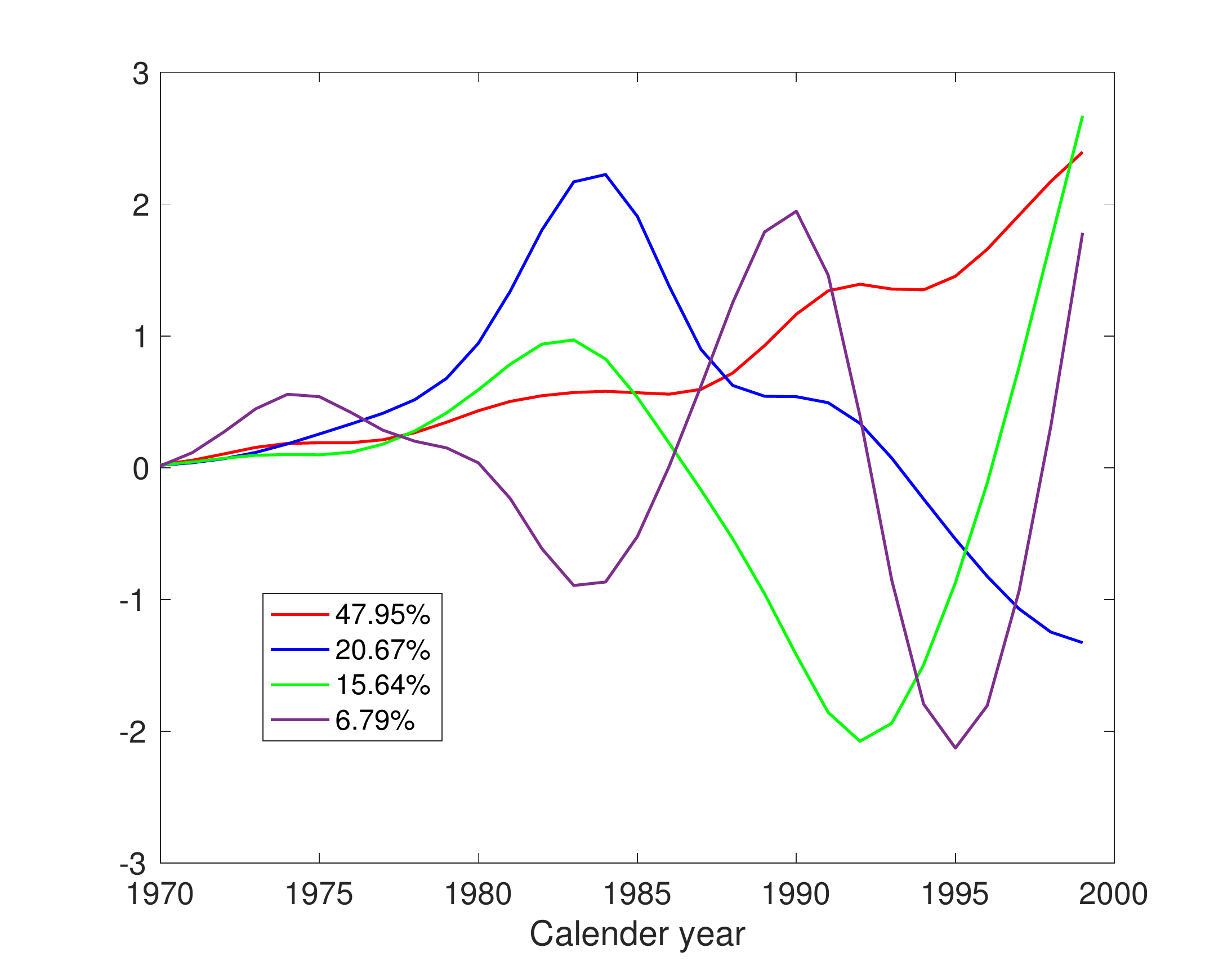}
		\caption{Eigenfunctions.}
		\label{fig: f21}
	\end{subfigure}
	\caption{Estimated metric auto-covariance surface \eqref{eq: cov} (left) and corresponding first four eigenfunctions (right) for the trade data.}
	\label{fig: f19}
\end{figure}

\id When adopting  the Frobenius metric, the sample \F \ mean function at calendar year $t$ corresponds to the sample average of the smoothed covariance across 46 countries for year $t$.  Figure \ref{fig: f20} illustrates the estimated metric auto-covariance function and Figure \ref{fig: f21} its eigenfunctions. The metric auto-covariance and its eigenfunctions provide  insights  about world trade patterns over the time period 1970 to 1999. The first eigenfunction  represents increased variability due to overall expansion in world trade over the years from 1970 to 1999. The slope of the first eigenfunction is more gradual before 1985 but increases sharply starting 1985, stagnates a little around 1990 and then again picks up. This can be connected to the boom in world trade towards the last decade of the new millennium. The second eigenfunction corresponds to a contrast before 1990 and after 1990. The peak in the second eigenfunction between 1980-1985 could be related to a major economic downturn caused by recession affecting several countries in the dataset during the early 1980s. The recession began in USA in 1981 and continued \red{until} 1982 and affected many of the developed Western countries. The third eigenfunction captures effects of the early 1990s recession, which compared to the 1980s recession was  much milder. 
\begin{figure}
	\centering
	\includegraphics[scale = .6]{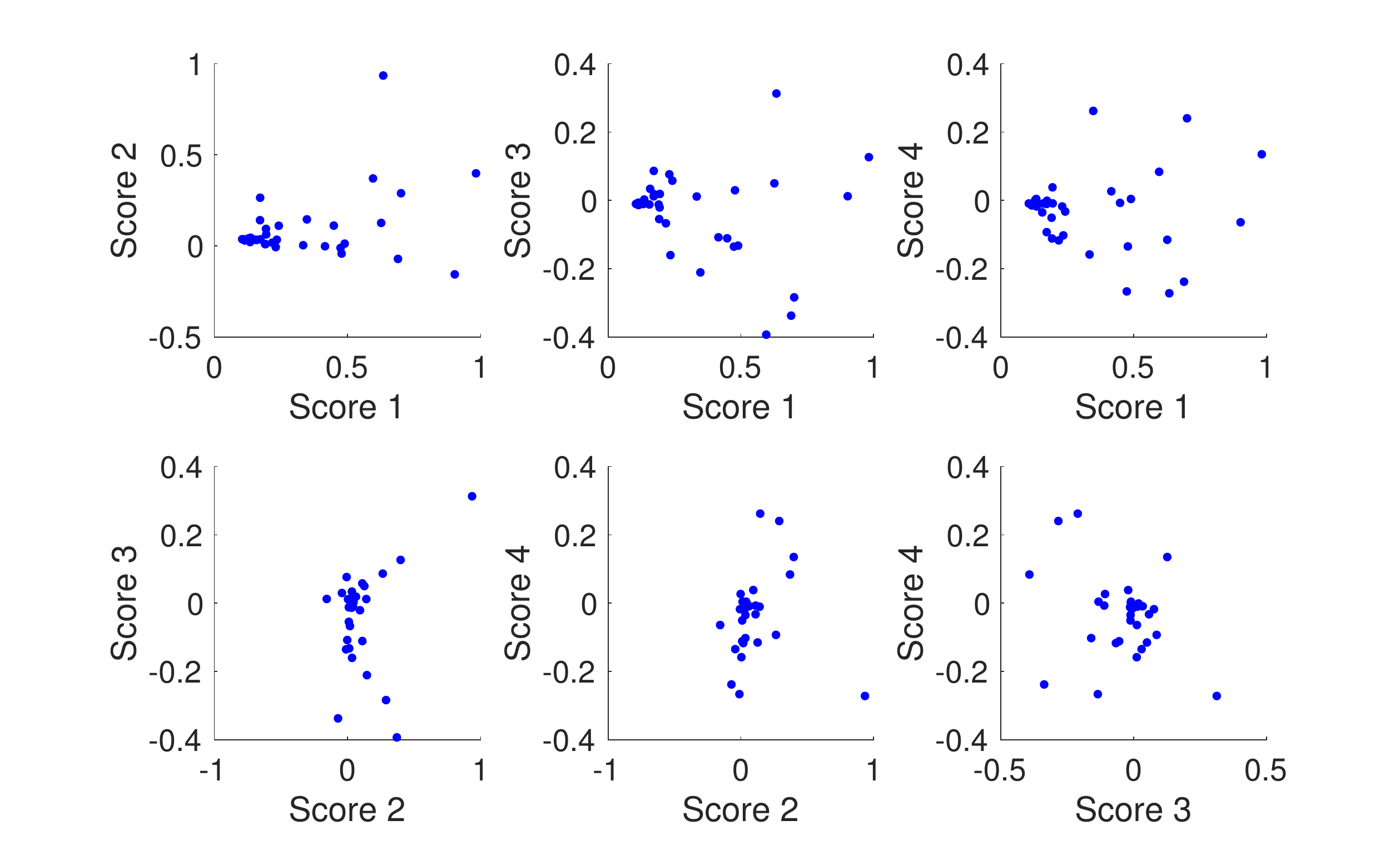}
	\caption{\F \ scores for different eigenfunctions plotted against each other for the trade data. }
	\label{fig: f22}
\end{figure}

\id In Figure \ref{fig: f22}, the \F \ scores for the first four eigenfunctions are plotted against each other.  Thailand and Egypt have high \F \ scores for the first eigenfunction and Saudi Arabia ranks the highest for the second eigenfunction. Chile, Israel, Hong Kong and Bulgaria turn out to figure prominently in the third eigenfunction.  Further visualization can be found in section \hyperref[trade_vis]{S4} of the online supplement, including a movie described in section \hyperref[mov]{S5} of the online supplement that demonstrates the object FPCs.

\section{Discussion}
\label{dis}
We propose an extension of functional data methods to the case of \fro. The basis of our approach is  metric covariance, a novel covariance measure for paired metric space valued data.  Eigenfunctions of the metric covariance operator for time-varying object data aid in creating a version of object FPCA, where the object FPCs in the metric space $\O$  are obtained as \F \ integrals, a general and versatile concept. Alternatively,  components of variation can be quantified by  \F \ scores, which are real numbers. For the precursor problem, where one has non-functional time-varying object data, i.e.  one has observations for just one random object function over time, methods for metric-space valued regression have  been considered previously \cp{stei:10,fara:14,mull:19:3},
often  under the special assumption that the regression responses are on a Riemannian manifold \cp{shi:09,flet:13,hink:12, su:12, yuan:12,corn:17}. However the more general object  function case, which is characterized by samples of random functions that are object-valued, is considerably more  challenging, as the absence of a linear structure in the object space both globally and locally imposes serious limitations on the methods that can be applied. 

\id The tools we propose here for \fro,  namely metric covariance,  the metric auto-covariance operator and its eigenfunctions,  the \F \ integrals and the \F \ scores make it possible to obtain compact summaries,  visualizations and interpretations of the observed samples of time-varying object data that in themselves are highly complex and difficult to  quantify. These tools can provide  insights into the 
patterns of variability of the object trajectories, as we demonstrated in the simulations and data examples. The quantification of {\fro \ can} also be used for other tasks. For  example the object FPCs that we introduce reside in the object space and can serve as responses for a regression model, where predictors are Euclidean vectors and responses are random object trajectories, which are  summarized by these object FPCs. Implementing such a regression approach is analogous to  the principal component approach for function-to-function regression  \cp{mull:05:5}. Various regression models can then be 
implemented  through \F \ regression \cp{mull:19:3}. For the case where {\fro \ feature} as predictors in a regression setting,  one can employ the vector of  \F \ scores that summarize each random object trajectory  as predictors. The ensuing regression, classification and clustering models will be interesting topics for future research.

\id A core challenge that one faces when modeling and analyzing samples of random object trajectories is that in  contrast to the situation for real-valued processes, one cannot expect to represent object-valued 
processes in terms of an analogue to the Karhunen-Lo\`eve expansion, due to the lack of a linear structure in the object space $\O$. In some special cases such expansions are possible, for example through a transformation method, whenever random objects  can be transformed to a linear space,  as exemplified for the case of objects that are probability distributions \cp{mull:16:1} or for the case of Riemannian manifold-valued objects \cp{mull:18:4}. Apart from such special cases, it is an open problem whether more general useful representations of \fro can be found. Another open problem is inference for such data, e.g., comparing two groups or testing for structural features of auto-covariance.   Here the metric auto-covariance operator that we introduce in this paper and also the Fr\'echet mean function could prove useful for the extension of
tests that have been considered for real-valued functional data \cp[for some recent examples, see][]{asto:17,cons:17,chen:17,choi:18}.
These and many other open problems in this area indicate that there is ample potential for future research.  

\section*{Acknowledgements}

We wish to thank the reviewers for their comprehensive and helpful comments that led to numerous improvements and new perspectives.  
This research was supported by NSF grant DMS-1712864 and NIH grant 5UH3OD023313.
\references

\newpage 
\section*{Online Supplement}\label{app}

\subsection*{S1. \hs Proofs}
\label{proofs}
\subsubsection*{Proof  of Proposition~\ref{lma: pos_def}}	It is clear that $C(s,t)$ is a symmetric function. To prove that it is nonnegative definite we need to show that for any positive integer $k$, 
	\begin{equation*}
	\sum_{i=1}^{k}\sum_{j=1}^{k}a_ia_jC(s_i,s_j) \geq 0,
	\end{equation*}
	for any $a_1,a_2,\dots ,a_k$ in $\mbb{R}$ and $s_1,s_2, \dots ,s_k$ in $[0,1]$. Since $d^2$ is a semimetric of negative type, by Proposition 3 in \cite{sejd:13} there exists a Hilbert space $\mathcal{H}$ and an injective map $f: \O \rightarrow H $ such that $d^2(\o_1,\o_2)= \lVert f(\o_1)-f(\o_2)\rVert^2_{\mathcal{H}}$. We therefore have that for $x,x',y,y' \in \O$,
	\begin{align*}
	& d^2(x,y')+d^2(x',y)-2d^2(x,y) 
	\\ = &\lVert f(x)-f(y')\rVert^2_{\mathcal{H}}+\lVert f(x')-f(y)\rVert^2_{\mathcal{H}}-2\lVert f(x)-f(y)\rVert^2_{\mathcal{H}} \\ = & \lVert f(y')\rVert^2_{\mathcal{H}}+ \lVert f(x')\rVert^2_{\mathcal{H}}-\lVert f(y)\rVert^2_{\mathcal{H}}-\lVert f(x)\rVert^2_{\mathcal{H}}+ 4 \left\langle  f(x),f(y) \right\rangle _{\mathcal{H}}\\ & -2\left\langle  f(x),f(y') \right\rangle _{\mathcal{H}}-2\left\langle  f(x'),f(y) \right\rangle _{\mathcal{H}},
	\end{align*}
	which implies that for i.i.d copies $(X,Y)$ and $(X',Y')$ 
	\begin{align*}
	C(X,Y)=\frac{1}{4}E\left(4\left\langle  f(X),f(Y) \right\rangle _{\mathcal{H}}-2 \left\langle  f(X),f(Y') \right\rangle _{\mathcal{H}}-2
	\left\langle  f(X'),f(Y) \right\rangle _{\mathcal{H}}\right).
	\end{align*} 
	Let $H_i=f(X(s_i))$ and $H_i'=f(X'(s_i))$ for $i=1,2, \hdots , k$ where $X'$ is an i.i.d copy of $X$. Then
	\begin{equation*}
	C(s_i,s_j)=\frac{1}{4}E\left(4\left\langle  H_i,H_j \right\rangle _{\mathcal{H}}-2 \left\langle  H_i,H_j' \right\rangle _{\mathcal{H}}-2
	\left\langle  H_i',H_j \right\rangle _{\mathcal{H}}\right),
	\end{equation*}
	which leads to
	\begin{align*}
	&	\sum_{i=1}^{k}\sum_{j=1}^{k}a_ia_jC(s_i,s_j) \\ &= \frac{1}{4} E\left(4\left\langle  \sum_{i=1}^k a_iH_i,\sum_{j=1}^k a_jH_j \right\rangle _{\mathcal{H}}-2 \left\langle  \sum_{i=1}^k a_iH_i,\sum_{j=1}^k a_jH_j' \right\rangle _{\mathcal{H}}-2
	\left\langle  \sum_{i=1}^k a_iH_i', \sum_{i=1}^k a_jH_j \right\rangle _{\mathcal{H}}\right) \\ & \geq \frac{1}{4} E\left(4 \lVert \sum_{i=1}^k a_iH_i\rVert^2_{\mathcal{H}}\right) - 2 E\left(\lVert \sum_{i=1}^k a_iH_i\rVert_{\mathcal{H}}\right) E\left(\lVert \sum_{j=1}^k a_jH_j'\rVert_{\mathcal{H}}\right)\\&-  2 E\left(\lVert \sum_{i=1}^k a_iH_i'\rVert_{\mathcal{H}}\right) E\left(\lVert \sum_{j=1}^k a_jH_j\rVert_{\mathcal{H}}\right)
	\\ &= \Var\left(\lVert \sum_{i=1}^k a_iH_i\rVert_{\mathcal{H}}\right).
	\end{align*}
	The last step follows from the Cauchy-Schwarz inequality. This completes the proof.

\subsubsection*{Proof  of Proposition~\ref{lma: well_defined}}
	\begin{enumerate}
		\item  For any $\gamma > 0$, by (I1) there exists $\delta > 0$ such that, whenever $\lvert t_1-t_2 \rvert < \delta$, 
		\begin{align*}
		\sup_{\o \in \O}\lvert H(\o,t_1)-H(\o,t_2) \rvert < \gamma.
		\end{align*}
		For any partition $\mathcal{P}$ as described above such that $\epsilon_{\mathcal{P}} < \delta$ we find
		\begin{align*}
		&\sup_{\o \in \O}\lvert I_{\mathcal{P}}(\o)-I(\o)\rvert \\ & =\sup_{\o \in \O}\lvert \sum_{j=0}^{k-1}H(\o,t_j)\Delta_j-\int_{0}^{1}H(\o,t)dt\rvert \\ & =\sup_{\o \in \O}\lvert \sum_{j=0}^{k-1} \int_{x_j}^{x_{j+1}}\lbrace H(\o,t_j)-H(\o,t) \rbrace dt\rvert \\ & = \sum_{j=0}^{k-1} \int_{x_j}^{x_{j+1}} \sup_{\o \in \O}\lvert H(\o,t_j)-H(\o,t) \rvert dt  \quad < \, \gamma,
		\end{align*}
		which completes the proof for part (a).
		\item Observe that
		\begin{align*}
		& \lvert I(\sum_{\mathcal{P},\oplus}S\phi)-I(\int_{\oplus}S\phi)\rvert \\ & \leq \lvert I(\sum_{\mathcal{P},\oplus}S\phi)-I_{\mathcal{P}}(\sum_{\mathcal{P},\oplus}S\phi)+I_{\mathcal{P}}(\sum_{\mathcal{P},\oplus}S\phi)-I(\int_{\oplus}S\phi)\rvert \\ &\leq \sup_{\o \in \O} \lvert I(\o)-I_{\mathcal{P}}(\o)\rvert + \lvert \min_{\o \in \O} I_{\mathcal{P}}(\o)-\min_{\o \in \O} I(\o)\rvert \\ &\leq  2 \sup_{\o \in \O} \lvert I(\o)-I_{\mathcal{P}}(\o)\rvert,
		\end{align*}
		and therefore by part (a), $\lvert I(\sum_{\mathcal{P},\oplus}S\phi)-I(\int_{\oplus}S\phi)\rvert \rightarrow 0$ as $\epsilon_{\mathcal{P}} \rightarrow 0$. 
		
		Now assume that $\lim_{\epsilon_{\mathcal{P}} \rightarrow 0} d(\sum_{\mathcal{P},\oplus}S\phi,\int_{\oplus}S\phi) \neq 0$. Then there must exist a sequence of partitions $\lbrace\mathcal{P}_n\rbrace$ and a $\gamma > 0$ such that $\epsilon_{\mathcal{P}_n} \rightarrow 0$ but $d(\sum_{\mathcal{P}_n,\oplus}S\phi,\int_{\oplus}S\phi)\geq \gamma$. For this sequence of partitions we observe that,
		\begin{equation*}
		\lvert I(\sum_{\mathcal{P}_n,\oplus}S\phi)-I(\int_{\oplus}S\phi)\rvert \geq \lvert \inf_{d(\o,\int_{\oplus}S\phi) > \gamma} I(\o)-I(\int_{\oplus}S\phi)\rvert > 0
		\end{equation*}
		and therefore $\lim_{\epsilon_{\mathcal{P}_n} \rightarrow 0}\lvert I(\sum_{\mathcal{P}_n,\oplus}S\phi)-I(\int_{\oplus}S\phi)\rvert \geq \lvert \inf_{d(\o,\int_{\oplus}S\phi) > \gamma} I(\o)-I(\int_{\oplus}S\phi)\rvert >0 $ by (I2), which is a contradiction to part (a). Therefore the assumption that \\$\lim_{\epsilon_{\mathcal{P}} \rightarrow 0} d(\sum_{\mathcal{P},\oplus}S\phi,\int_{\oplus}S\phi) \neq 0$ cannot be true, which completes the proof for part(b).
		\item  Let $\delta > 0$ be such that whenever $\epsilon_{\mathcal{P}} < \delta$,  it holds that $d(\sum_{\mathcal{P},\oplus}S\phi,\int_{\oplus}S\phi) < \nu$. Assume that $\lim_{\epsilon_{\mathcal{P}} \rightarrow 0} h^{1/\beta}(\epsilon_{\mathcal{P}})d(\sum_{\mathcal{P},\oplus}S\phi,\int_{\oplus}S\phi) \neq 0$. Then there exists a sequence of partitions $\lbrace\mathcal{P}_n\rbrace$ and a $\gamma > 0$ such 
		that $\epsilon_{\mathcal{P}_n} < \delta$, while $d(\sum_{\mathcal{P}_n,\oplus}S\phi,\int_{\oplus}S\phi)\geq \frac{\gamma}{h^{1/\beta}(\epsilon_{\mathcal{P}})}$. For this sequence of partitions we observe that
		\begin{align*}
		& h(\epsilon_{\mathcal{P}_n}) \lvert I(\sum_{\mathcal{P}_n,\oplus}S\phi)-I(\int_{\oplus}S\phi)\rvert \\ & \geq h(\epsilon_{\mathcal{P}_n}) \lvert \inf_{\frac{\gamma}{h^{1/\beta}(\epsilon_{\mathcal{P}_n})} \leq d(\o,\int_{\oplus}S\phi) < \nu} I(\o)-I(\int_{\oplus}S\phi)\rvert \\ & \geq \frac{C h(\epsilon_{\mathcal{P}_n}) \gamma^{\beta}}{h(\epsilon_{\mathcal{P}_n})}
		\end{align*}
		by (I3). Therefore $\lim_{\epsilon_{\mathcal{P}_n} \rightarrow 0}h(\epsilon_{\mathcal{P}_n})\lvert I(\sum_{\mathcal{P}_n,\oplus}S\phi)-I(\int_{\oplus}S\phi)\rvert \geq C \gamma^{\beta} >0$,  which results in a contradiction and completes the proof  for part (c).
	\end{enumerate}

\subsubsection*{Proof  of Lemma~\ref{lma: mean_cont}}
	Consider $t \in [0,1]$ and a sequence $\lbrace t_n \rbrace \in [0,1]$ such that $t_n \rightarrow t$. We aim to show that $d(\mu_{\oplus}(t_n),\mu_{\oplus}(t)) \rightarrow 0$.
	
  Observe that almost surely continuous sample curves on the compact interval $[0,1]$ are uniformly continuous and since $\O$ is bounded, by bounded convergence, for all $t \in [0,1]$ and sequences $\lbrace t_n \rbrace \in [0,1]$ such that $t_n \rightarrow t$,  there exists a $\delta > 0$ for every $\epsilon > 0$ such that whenever $\lvert t_n-t\rvert < \delta$,  it holds that $E(d(X(t_n),X(t)) < \epsilon$ for all but finitely many $n$. For  processes $E(d^2(\o,X(t)))$, 
	\begin{equation*}
	\lvert E(d^2(\o,X(t_n))) - E(d^2(\o,X(t)))\rvert \leq 2D \ E(d(X(t_n),X(t)),
	\end{equation*}
	and therefore, given any $\epsilon > 0$,  there exists a $\delta>0$ such that whenever $\lvert t_n-t \rvert < \delta$,  one has  $\sup_{\o \in \O} \lvert E(d^2(\o,X(t_n))) - E(d^2(\o,X(t)))\rvert <  \eps$. This implies that
	\begin{align*}
	& \lvert E(d^2(\mu_{\oplus}(t_n),X(t))) -  E(d^2(\mu_{\oplus}(t),X(t)))\rvert \\ & \le \lvert E(d^2(\mu_{\oplus}(t_n),X(t))) -E(d^2(\mu_{\oplus}(t_n),X(t_n))) \\&\hspace{1cm} +E(d^2(\mu_{\oplus}(t_n),X(t_n)))- E(d^2(\mu_{\oplus}(t),X(t)))\rvert \\ & \leq \sup_{\o \in \O} \lvert E(d^2(\o,X(t_n))) - E(d^2(\o,X(t)))\rvert+ \lvert \min_{\o \in \O} E(d^2(\o,X(t_n))) - \min_{\o \in \O}  E(d^2(\o,X(t)))\rvert \\ & \leq 2\sup_{\o \in \O} \lvert E(d^2(\o,X(t_n))) - E(d^2(\o,X(t)))\rvert \quad < \, 2\epsilon.
	\end{align*}
	
	Assume $d(\mu_{\oplus}(t_n),\mu_{\oplus}(t)) \nrightarrow 0$. Then one can find  an $\eta >0 $ such that for any $\delta > 0$, there exists a subsequence $\lbrace t_{n_k}\rbrace$ of $\lbrace t_{n}\rbrace$ for which $\lvert t_{n_k}-t \rvert < \delta$ but $d(\mu_{\oplus}(t_{n_k}),\mu_{\oplus}(t)) \geq \eta$. Then by (A3),
	\begin{align*}
	& \lvert E(d^2(\mu_{\oplus}(t_{n_k}),X(t))) - E(d^2(\mu_{\oplus}(t),X(t)))\rvert \\ & \geq  \lvert \inf_{d(\o,\mu_{\oplus}(t))  > \eta} E(d^2(\o,X(t))) - E(d^2(\mu_{\oplus}(t),X(t)))\rvert \\ & >0.
	\end{align*}
	This leads to a contradiction for $\epsilon={\lvert \inf_{d(\o,\mu_{\oplus}(t))  > \eta} E(d^2(\o,X(t))) - E(d^2(\mu_{\oplus}(t),X(t)))\rvert }/{2}$, thus completing  the proof.

\subsubsection*{Proof  of Theorem~\ref{lma: C_limit}} 
	Denoting  the usual $L^2$ norm by  $||\cdot||_2$, observe that for $s_1,t_1,s_2,t_2 \in [0,1]$,  one has 
	\begin{align*}
	& |f_{s_1,t_1}(x,y)-f_{s_2,t_2}(x,y)|\\ & \leq 4M \lbrace d(x(s_1),x(s_2))+d(x(t_1),x(t_2))+d(y(s_1),y(s_2))+d(y(t_1),y(t_2))\rbrace \\ & \leq  4M  \left(G(x)+G(y)\right) (|s_1-s_2|^\alpha+|t_1-t_2|^\alpha),
	\end{align*}	
	implying 
	\begin{equation*}
	||f_{s_1,t_1}-f_{s_2,t_2}||_2 \leq 8M ||G||_2(|s_1-s_2|^\alpha+|t_1-t_2|^\alpha).
	\end{equation*}
	Observe that for any $0<u<1$, if we take $|s_1-s_2|<\left(\frac{u}{16}\right)^{\frac{1}{\alpha}}$ and $|t_1-t_2|<\left(\frac{u}{16}\right)^{\frac{1}{\alpha}}$, then $||f_{s_1,t_1}(X)-f_{s_2,t_2}(Y)||_2\leq Mu ||G||_2$. Therefore, with  $s_1,s_2, \hdots, s_K$ and $t_1,t_2,\hdots, t_L$ forming  $\left(\frac{u}{4}\right)^{\frac{1}{\alpha}}$-nets for $[0,1]$ with metric $|\cdot|$,  the brackets $[f_{s_i,t_j}\pm Mu ||G||_2]$ cover the function class $\mathcal{F}$  \cp{well:96} and are of length $2Mu ||G||_2$. This implies 
	\begin{equation*}
	N_{[]}(2Mu ||G||_2,\mathcal{F},L^2(P \otimes P)) \leq N\left(\left(\frac{u}{4}\right)^{\frac{1}{\alpha}},[0,1],|\cdot|\right)^2,
	\end{equation*}
	where $N_{[]}(\eps,\mathcal{F},L^2(P))$ is the bracketing number, which is the minimum number of $\eps$-brackets needed to cover $\mathcal{F}$, where an $\eps$-bracket is composed of pairs of functions $[l,u]$ such that $||l-u||_2 < \eps$, and $N$ is the covering number. Hence for any $\epsilon > 0$,  for some constant $K > 0$,
	\begin{equation*}
	N_{[]}(\epsilon,\mathcal{F},L^2(P \otimes P)) \leq K \epsilon^{-2/\alpha} < \infty.
	\end{equation*}
	The result now follows from Theorem 4.10 of \cite{arco:93}, observing 
	\begin{align*}
	& \int_{0}^{1} \sqrt{\log  N_{[]}(\epsilon,\mathcal{F},L^2(P \otimes P))} d\eps \\ & \leq \eps \sqrt{\log K} + \int_{0}^{1} \sqrt{-\frac{2}{\alpha}\log\eps} d\eps \\ & = \eps \sqrt{\log K}+ \sqrt{\frac{2}{\alpha}} \Gamma\left(\frac{3}{2}\right) < \infty.
	\end{align*}

\subsubsection*{Proof  of Corollary~\ref{cor: eigen}}
	For any fixed $j$, Lemma 4.3 in \cite{bosq:00} gives $|\hat{\lambda}_j-\lambda_j| \leq \sup_{s,t \in [0,1]}\left|\hat{C}(s,t)-C(s,t)\right|$. Uniform mapping then implies 
		\begin{equation*}
	|\hat{\lambda}_j-\lambda_j|=O_P(1/\sqrt{n}).
	\end{equation*}
	Under assumption (A5), $\sup_{s \in [0,1]} \left| \hat{\phi}_j(s)-\phi_j(s)\right| \leq 2\sqrt{2}\delta_j^{-1}\sup_{s,t \in [0,1]}\left|\hat{C}(s,t)-C(s,t)\right|$, and therefore 
	\begin{equation*}
	\sup_{s \in [0,1]} \left| \hat{\phi}_j(s)-\phi_j(s)\right| =O_P(1/\sqrt{n}),
	\end{equation*}
which completes the proof. 

\subsubsection*{Proof  of Theorem~\ref{lma: fpc}}
Since 
	\begin{align}
	\left|\int_{0}^{1}\hat{\phi}(t)dt-\int_{0}^{1}\phi(t)dt\right| \leq \sup_{s \in [0,1]} \left| \hat{\phi}(s)-\phi(s)\right| =O_P(1/\sqrt{n}),
	\end{align}  for 
all sufficiently large $n$, $\left|\int_{0}^{1}\hat{\phi}(t)dt\right| \geq {\left|\int_{0}^{1}{\phi}(t)dt\right|}/{2}$, and since $\left|\int_{0}^{1}{\phi}(t)dt\right| \neq 0$,
	\begin{align*}
	&\sup_{s \in [0,1]} |\hat{\phi}^*(s)-\phi^*(s)| \\ & \leq 2 \frac{\sup_{s \in [0,1]}\left|\hat{\phi(s)}\right| \left|\int_{0}^{1}\hat{\phi}(t)dt-\int_{0}^{1}\phi(t)dt\right| + \left|\int_{0}^{1}\hat{\phi}(t)dt\right| \sup_{s \in [0,1]} \left| \hat{\phi}(s)-\phi(s)\right|}{\left|\int_{0}^{1}{\phi}(t)dt\right|^2}\\ & = O_P(1/\sqrt{n}).
	\end{align*}
	As a direct consequence,
	\begin{align}
	&\sup_{\o \in \O} \left|\int_{0}^{1} d^2(\o,X_i(t)) \hat{\phi}^*(t) dt-\int_{0}^{1} d^2(\o,X_i(t))  \phi^*(t) dt\right| \nonumber \\ & \leq M^2 \sup_{s \in [0,1]} |\hat{\phi}^*(s)-\phi^*(s)| =O_P(1/\sqrt{n}), \label{eq: integrand}
	\end{align}
	whence
	\begin{align}
	&P\left(d(\hat{\psi}^{ik}_{\oplus},\psi^{ik}_{\oplus})>\epsilon\right) \nonumber \\ & \leq P\left(\left|\int_{0}^{1} d^2(\hat{\psi}^{ik}_{\oplus},X_i(t))  \phi^*(t) dt-\int_{0}^{1} d^2(\psi^{ik}_{\oplus},X_i(t))  \phi^*(t) dt\right| > c_{\eps}\right) \label{eq: eq1}\\ & \leq P\left(\sup_{\o \in \O} \left|\int_{0}^{1} d^2(\o,X_i(t))\hat{\phi}^*(t) dt-\int_{0}^{1} d^2(\o,X_i(t))  \phi^*(t) dt\right| > \frac{c_{\eps}}{2}\right) \label{eq: eq2}, 
	\end{align}
	which implies that $d(\hat{\psi}^{ik}_{\oplus},\psi^{ik}_{\oplus})=o_P(1)$ by equation \eqref{eq: integrand}. Here  \eqref{eq: eq2} follows from \eqref{eq: eq1} using the fact that
	\begin{align*}
	& \left|\int_{0}^{1} d^2(\hat{\psi}^{ik}_{\oplus},X_i(t))  \phi^*(t) dt-\int_{0}^{1} d^2(\psi^{ik}_{\oplus},X_i(t))  \phi^*(t) dt\right| \\ & \leq \left|\int_{0}^{1} d^2(\hat{\psi}^{ik}_{\oplus},X_i(t))  \left(\phi^*(t) -\hat{\phi}^*(t)\right) dt\right| \\& +\left|\inf_{\o \in \O} \int_{0}^{1} d^2(\o,X_i(t))  \hat{\phi}^*(t) dt-\inf_{\o \in \O}\int_{0}^{1} d^2(\o,X_i(t))  \phi^*(t) dt\right| \\ & \leq 2 \sup_{\o \in \O} \left|\int_{0}^{1} d^2(\o,X_i(t)) \hat{\phi}^*(t) dt-\int_{0}^{1} d^2(\o,X_i(t))  \phi^*(t) dt\right| .
	\end{align*}
	From assumption (A7), 
	\begin{align*}
	& P\left(n^{1/(2\beta_1)}d(\hat{\psi}^{ik}_{\oplus},\psi^{ik}_{\oplus})>2^L\right)\\ & \leq  P\left(\frac{2^L}{n^{1/(2\beta_1)}} < d(\hat{\psi}^{ik}_{\oplus},\psi^{ik}_{\oplus}) < \nu'\right) + P\left(d(\hat{\psi}^{ik}_{\oplus},\psi^{ik}_{\oplus})\geq \nu'\right) \\ &= P\left(\sup_{\o \in \O} \left|\int_{0}^{1} d^2(\o,X_i(t))\hat{\phi}^*(t) dt-\int_{0}^{1} d^2(\o,X_i(t))  \phi^*(t) dt\right| > \frac{2^{\beta_1L}}{n^{1/2}}\right)  + P\left(d(\hat{\psi}^{ik}_{\oplus},\psi^{ik}_{\oplus})\geq \nu'\right) \\ \leq & P\left(\sqrt{n} \sup_{s \in [0,1]} |\hat{\phi}^*(s)-\phi^*(s)| > \frac{2^{\beta_1L}}{M^2}\right)  + P\left(d(\hat{\psi}^{ik}_{\oplus},\psi^{ik}_{\oplus})\geq \nu'\right).
	\end{align*}
	Therefore  $P\left(n^{1/2\beta_1}d(\hat{\psi}^{ik}_{\oplus},\psi^{ik}_{\oplus})>2^L\right)$ can be made as small as possible by choosing sufficiently large $n$ and  $L$,  using equation \eqref{eq: integrand} and the fact that $d(\hat{\psi}^{ik}_{\oplus},\psi^{ik}_{\oplus})=o_P(1)$, thus completing the proof.

\subsubsection*{Proof  of Proposition~\ref{lma: mean_consistency}}
	By Theorem 1.5.4 in \cite{well:96}, it suffices to show asymptotic equicontinuity of the processes $Z_n(s)=d(\hat{\mu}_\oplus(s),\mu_\oplus(s))$, i.e.  for any $\theta > 0$,
	\begin{equation}
	\label{eq: equicont}
	\lim_{\delta \rightarrow 0} \limsup_{n \rightarrow \infty} P\left(\sup_{|s-t|<\delta} \left|Z_n(s)-Z_n(t)\right| > 2\theta\right) = 0, 
	\end{equation}
	in addition to the pointwise convergence of $Z_n(s)$, i.e. for all $s \in [0,1]$ it holds that 
	\begin{equation}
	\label{eq: ptwise}
	Z_n(s)=o_P(1).
	\end{equation}
	We observe that equation \eqref{eq: ptwise} follows from Theorem 1 in \cite{mull:19:3}. To establish equation \eqref{eq: equicont}, by Lemma \ref{lma: mean_cont} and as $\left|Z_n(s)-Z_n(t)\right| \leq d(\mu_\oplus(s),\mu_\oplus(t))+d(\hat{\mu}_\oplus(s),\hat{\mu}_\oplus(t))$,  it suffices to show that 
	\begin{equation}
	\label{eq: equicontinuity}
	\lim_{\delta \rightarrow 0} \limsup_{n \rightarrow \infty} P\left(\sup_{|s-t|< \delta}  d(\hat{\mu}_\oplus((s),\hat{\mu}_\oplus((t)) > \theta\right) = 0.
	\end{equation}

	To show equation \eqref{eq: equicontinuity}, 
	suppose that $d(\hat{\mu}_\oplus(s),\hat{\mu}_\oplus(t)) > \theta$ with $s,t \in [0,1]$.
	\begin{enumerate}
		\item[Step 1.] Since the functions are continuous and the domain is compact, it holds that  almost surely $\sup_{|s-t|< \delta} d(X(t),X(s)) \rightarrow 0$ as $\delta \rightarrow 0$. By the boundedness of the metric and  dominated convergence, 
		\begin{equation}
		\label{eq: dct}
		\lim_{\delta \rightarrow 0} E( \sup_{|s-t|< \delta} d(X(t)),X(s)) ) \rightarrow 0. 
		\end{equation}
		Now  \eqref{eq: dct} implies  that for any $a > 0$,
		\begin{align*}
		&  P\left(\sup_{|s-t|< \delta} \sup_{\o \in \O} \left|\frac{1}{n}\sum_{i=1}^n d^2(X_i(t),\o)-\frac{1}{n}\sum_{i=1}^n d^2(X_i(s),\o)\right| > a \right) \\ & \leq  \frac{2M \ E\left(\sup_{|s-t|< \delta} \frac{1}{n}\sum_{i=1}^n d(X_i(t),X_i(s))\right)}{a} \\ & \leq  \frac{2M \ E( \sup_{|s-t|< \delta} d(X(t)),X(s)))}{a} \rightarrow 0 \quad \text{as} \ \delta \rightarrow 0, 
		\end{align*}
		where $M=\text{diam}(\O)$.
		
		\item[Step 2.] We observe that 
		\begin{align*}
		& \left|\frac{1}{n}\sum_{i=1}^n d^2(X_i(t),\hat{\mu}_\oplus(s))-\frac{1}{n}\sum_{i=1}^n d^2(X_i(t),\hat{\mu}_\oplus(t))\right| \\ & \leq \left|\frac{1}{n}\sum_{i=1}^n d^2(X_i(t),\hat{\mu}_\oplus(s))-\frac{1}{n}\sum_{i=1}^n d^2(X_i(s),\hat{\mu}_\oplus(s))\right| \\ & \hspace{1cm} +\left|\frac{1}{n}\sum_{i=1}^n d^2(X_i(s),\hat{\mu}_\oplus(s))-\frac{1}{n}\sum_{i=1}^n d^2(X_i(t),\hat{\mu}_\oplus(t))\right| \\ & \leq 2 \sup_{\o \in \O} \left|\frac{1}{n}\sum_{i=1}^n d^2(X_i(t),\o)-\frac{1}{n}\sum_{i=1}^n d^2(X_i(s),\o)\right|. 
		\end{align*}
		\item[Step 3.] Now we find 
		\begin{align}
		& P\left(\sup_{|s-t|< \delta}  d(\hat{\mu}_\oplus(s),\hat{\mu}_\oplus(t)) > \theta\right) \nonumber\\ & \leq  P \left( B \cap A_n \right) + P\left(B \cap A_n^C\right) \label{eq: eqA1}
		\\ & \leq  P \left( \sup_{|s-t|< \delta} 2 \sup_{\o \in \O} \left|\frac{1}{n}\sum_{i=1}^n d^2(X_i(t),\o)-\frac{1}{n}\sum_{i=1}^n d^2(X_i(s),\o)\right|  \geq \tau(S)\right) + P\left(B \cap A_n^C\right). \nonumber
		\end{align}
	\end{enumerate}
	where $\ A_n=\lbrace\sup_{|s-t|< \delta} \left|\frac{1}{n}\sum_{i=1}^n d^2(X_i(t),\hat{\mu}_\oplus(s))-\frac{1}{n}\sum_{i=1}^n d^2(X_i(t),\hat{\mu}_\oplus(t))\right| \geq \tau(S)\rbrace$, $B= \lbrace\sup_{|s-t|< \delta}  d(\hat{\mu}_\oplus(s),\hat{\mu}_\oplus(t)) > S\rbrace$ in \eqref{eq: eqA1} with $\tau(S)$ as defined in (A8). The  last step follows from \eqref{eq: eqA1} using Step 2. From Step 1, choosing $a=\frac{\tau(S)}{2}$,
	\begin{equation*}
	\lim_{\delta \rightarrow 0} P \left( \sup_{|s-t|< \delta} 2 \sup_{\o \in \O} \left|\frac{1}{n}\sum_{i=1}^n d^2(X_i(t),\o)-\frac{1}{n}\sum_{i=1}^n d^2(X_i(s),\o)\right|  \geq \tau(S)\right)=0
	\end{equation*} 
	for any $n$, and from (A1),  $\lim_{n \rightarrow \infty} P\left(B \cap A_n^C\right)=0$. This completes the proof.

	\subsubsection*{Proof  of Lemma~\ref{lma: entropy}}
	Define functions 
	$f_{\o,s}(x)=d^2(x(s),\o(s))-d^2(x(s),\mu_\oplus(s))$. We find 
	\begin{align*}
	& \left| f_{\o_1,s_1}(x)-f_{\o_2,s_2}(x)\right|  \\ & \leq  \left| f_{\o_1,s_1}(x)-f_{\o_1,s_2}(x)\right|+ \left| f_{\o_1,s_2}(x)-f_{\o_2,s_2}(x)\right| \\ & \leq 2M \left(2 d(x(s_1),x(s_2))+d(\mu_\oplus(s_1),\mu_\oplus(s_2))+d(\o_1(s_1),\o_1(s_2))+d(\o_1(s_2),\o_2(s_2))\right).
	\end{align*}
	Note that $d(\o_1(s_2),\o_2(s_2)) \leq d(\o_1(s_2),\mu_2(s_2))+d(\o_1(s_2),\mu_2(s_2))$. By assumptions (A4) and (A9), it holds that  almost surely,
	\begin{align*}
	& \left| f_{\o_1,s_1}(X)-f_{\o_2,s_2}(X)\right|   \leq 4M \left[G(X) |s_1-s_2|^\alpha+H_\delta|s_1-s_2|^{\nu_\delta}  \right]\\ & +4M \left[d(\o_1(s_2),\mu_2(s_2))+d(\o_1(s_2),\mu_2(s_2))\right],
	\end{align*}
	which implies that
	\bea
	&&||f_{\o_1,s_1}-f_{\o_2,s_2}||_{2}\\
	 && \quad \leq 4M \left[||G||_{2} |s_1-s_2|^\alpha+H_\delta|s_1-s_2|^{\nu_\delta}+d(\o_1(s_2),\mu_2(s_2))+d(\o_1(s_2),\mu_2(s_2))\right].
	\eea
	It follows that for some $0<u<1$, if we take $|s_1-s_2|<\left(\frac{u}{4}\right)^{\frac{1}{V}}$ with $V=\min(\alpha,\nu_\delta)$ and $\o_1,\o_2$ such that $d_\infty(\o_1,\mu)< \frac{u}{8}$ and $d_\infty(\o_2,\mu)< \frac{u}{8}$, then $||f_{\o_1,s_1}-f_{\o_2,s_2}||_{2} < Mu(||G||_2+H_\delta+1)$. Therefore if $s_1,s_2, \hdots, s_K$ is a  $\left(\frac{u}{4}\right)^{\frac{1}{V}}$-net for $[0,1]$ with metric $|\cdot|$ and $\o_1,\o_2,\hdots, \o_L$ is a  $\frac{u}{8}$-net for $B_{\delta}(\mu(\cdot))$ with metric $d_\infty$,  the brackets $[f_{s_i,\o_j}\pm Mu(||G||_2+H_\delta+1)]$ cover the function class $\mathcal{F}_\delta$ and are of length $2Mu(||G||_2+H_\delta+1)$. We conclude that 
	\begin{equation*}
	N_{[]}(2Mu(||G||_2+H_\delta+1),\mathcal{F}_{\delta},L^2(P)) \leq N\left(\left(\frac{u}{4}\right)^{\frac{1}{V}},[0,1],|\cdot|\right) N\left(\frac{u}{8},B_{\delta}(\mu_\oplus(\cdot)),d_\infty \right).
	\end{equation*}
	Applying  \ci{well:96} (page 84),  for any function class $\mathcal{F}$ and for any $r$,
	\begin{equation*}
	N(\eps,\mathcal{F},L^r(P)) \leq 	N_{[]}(2\eps,\mathcal{F},L^r(P)), 
	\end{equation*}
	so that  for appropriate constants $K_1,K_2,C > 0$,
	\begin{align*}
	&\log N(2M\delta\eps,\mathcal{F}_{\delta},L^2(P)) \\ & \leq \log N\left(K_1(\eps\delta)^{1/V},[0,1],|\cdot|\right) + \log N\left(K_2{\eps\delta},B_{\delta}(\mu_\oplus(\cdot)),d_\infty \right) \\ & \leq \log \left(C\left(\frac{1}{\eps\delta}\right)^{1/V}\right)+ \log N\left(K_2{\eps\delta},B_{\delta}(\mu_\oplus(\cdot)),d_\infty \right).
	\end{align*}
	Observe that $\log N\left(K_2{\eps\delta},B_{\delta}(\mu_\oplus(\cdot)),d_\infty \right) \leq \sup_{s \in [0,1]} \log N\left(K_2{\eps\delta},B_{\delta}(\mu_\oplus(s)),d \right)$ because $d_\infty(\o_1,\o_2)=\sup_{s \in [0,1]} d(\o_1(s),\o_2(s))$ and $d(\o_1(s),\o_2(s))$ is a uniformly continuous function in $s$ so that  the supremum is attained. Therefore, $d_\infty(\o_1,\o_2)= d(\o_1(s^*),\o_2(s^*))$ for some $s^* \in [0,1]$. Finally we observe that
	\begin{align*}
	& \int_{0}^{1} \sqrt{1+\log N(\eps ||F||_{2},\mathcal{F}_{\delta},L^2(P))} d\eps \\ & = \int_{0}^{1} \sqrt{1+\log N(2M\delta\eps,\mathcal{F}_{\delta},L^2(P))} d\eps \\ & \leq \sqrt{\log(C)}+ \int_{0}^{1} \sqrt{-\frac{1}{V}\log(\eps\delta)}d\eps+\int_{0}^{1}\sup_{s \in [0,1]} \sqrt{\log N\left(K_2{\eps\delta},B_{\delta}(\mu_\oplus(s)),d \right)}d\eps \\ & \leq \sqrt{\log(C)}+\frac{1}{\sqrt{V}}  \int_{0}^{1} \sqrt{-\log(\eps\delta)}d\eps+ \int_{0}^{1}\sup_{s \in [0,1]} \sqrt{\log N\left(K_2{\eps\delta},B_{\delta}(\mu_\oplus(s)),d \right)}d\eps. 
	\end{align*}
	Assumption (A10) then implies $ J_{[]}(1,\mathcal{F}_{\delta},L^{2}(P))= O(\sqrt{-\log \delta})$  as $\delta \rightarrow 0$, which completes the proof.

\subsubsection*{Proof  of Theorem~\ref{lma: rate}}
	For a sequence $\{q_n\}$  define the sets 
	\begin{equation*}
	S_{j,n}(x)= \lbrace \o(\cdot) : 2^{j-1} < q_n d^{\beta_2/2}_\infty(\o,\mu_\oplus) \leq 2^j\rbrace .
	\end{equation*}
	Choose $\alpha > 0$ to satisfy (A11) and also small enough such that (A3) and (A4) hold for all $\delta < \alpha$ and choose $\tilde{\alpha}=\alpha^{\beta_2/2}$. For any integer $L$,
	\begin{align}
	& P\left(q_n d^{\beta_2/2}_{\infty}(\hat{\mu}_\oplus,\mu_\oplus)> 2^L \right) \nonumber \\ & \leq 
	P\left(d_{\infty}(\hat{\mu}_\oplus,\mu_\oplus) \geq \alpha\right)+ \sum_{j \geq L, 2^j \leq q_n \tilde{\alpha}} P\left(\hat{\mu}_\oplus \in S_{j,n}\right) \nonumber \\ & \leq 
	P\left(d_{\infty}(\hat{\mu}_\oplus,\mu_\oplus) \geq \alpha\right)+ \sum_{j \geq L, 2^j \leq q_n \tilde{\alpha}} P\left( \sup_{\o \in S_{j,n}} \left|V_n(\o,s)-V(\o,s)\right| \geq D \frac{2^{2(j-1)}}{q^2_n},\right) \label{eq: onion}
	\end{align}
	where \eqref{eq: onion} follows  by observing 
	\begin{equation*}
	\sup_{\o \in S_{j,n}} \left|V_n(\o,s)-V(\o,s)\right| \geq  \left|\inf_{\o \in S_{j,n}} V_n(\o,s)- \inf_{\o \in S_{j,n}} V(\o,s)\right| \geq D \frac{2^{2(j-1)}}{q^2_n}.
	\end{equation*}
	The first term in \eqref{eq: onion} goes to zero by Proposition \ref{lma: mean_consistency} and for each $j$ in the second term it holds that  $d_\infty(\o,\mu_\oplus) \leq \alpha$.  By Lemma \ref{lma: entropy}, $ J_{[]}(1,\mathcal{F}_{\delta},L^{2}(P))=O(\sqrt{\log {1/\delta}})$, and therefore is bounded above by $J\sqrt{\log {1/\delta}}$ for all small enough $\delta > 0$,  where $J > 0$ is a constant. Using equation \eqref{eq: tail_bound}, Lemma \ref{lma: entropy} and the  Markov inequality,  the second term is upper bounded up to a constant by
	\begin{equation}
	\label{eq: upp}
	\sum_{j \geq L, 2^j \leq q_n \tilde{\alpha}} \frac{2MJ 2^j}{n} \sqrt{\log {n/2^{j+1}}}\frac{q_n^2}{D 2^{2(j-1)}}.
	\end{equation}
	Since $\sqrt{\log {n/2^{j+1}}}$ is dominated by $\sqrt{\log {n}/2}$, setting $q_n=\frac{\sqrt{n}}{(\log{{n}})^{1/4}}$, the series in \eqref{eq: upp} is upper bounded by $\frac{8MJ}{D} \sum_{j \geq L, 2^j \leq q_n \tilde{\alpha}} \frac{1}{2^j}$ , which converges and can be made sufficiently small by choosing $L$ and $n$ large. This proves the desired result that $d_\infty(\hat{\mu}_\oplus,\mu_\oplus)=O_P(q_n^{-2/\beta_2})=O_P\left(\left(\frac{n}{\sqrt{\log{n}}}\right)^{-1/\beta_2}\right)$.

\subsubsection*{Proof  of Corollary~\ref{cor: scalar_fpc}} Observing that 
	\begin{align*}
	& \left|\hat{\beta}_{ik}-\beta_{ik}\right| \\ & \leq \left|\int_{0}^{1} d(X_i(t),\hat{\mu}_{\oplus}(t)) \left(\hat{\phi}_k(t)-\phi_k(t)\right) dt\right|+ \left|\int_{0}^{1} \phi(t) \left(d(X_i(t),\hat{\mu}_{\oplus}(t))-d(X_i(t),\mu_{\oplus}(t))\right)dt\right| \\ & \leq M \sup_{s \in [0,1]} \left|\hat{\phi}_k(s)-\phi_k(s)\right|+ \int_{0}^{1}|\phi(t)|dt \sup_{s \in [0,1]} d(\hat{\mu}_{\oplus}(s),\mu_{\oplus}(s)), 
	\end{align*}
	the result follows from Corollary \ref{cor: eigen} and Theorem \ref{lma: rate}.

\subsection*{S2. \hs  Comparison of Metric Covariance with Distance Covariance}
\label{dis_cov}
For two random variables $X$ and $Y$ with marginal probability measures $P_X$ and $P_Y$ and joint probability measure $P_{XY}$, testing for probabilistic dependence corresponds to testing 
\begin{equation}
\label{eq: test}
H_0:\,  P_{XY}=P_X P_Y \quad \text{versus} \quad H_1:\,  P_{XY} \neq P_X P_Y.
\end{equation}

Implementation of these tests is usually based on  a metric in the space of probability measures. As shown in \cite{lyon:13}, distance correlation \cp{szek:07,szek:17}  provides a suitable metric for this purpose,  provided $X$ and $Y$ take values in metric spaces which are of ``strong negative type" \citep{lyon:13} and  include Euclidean spaces and separable Hilbert spaces. Then independence of $X$ and $Y$ is equivalent to the distance correlation being 0. 
While it is often of interest to determine whether distance correlation is zero, which then implies independence of $X$ and $Y$,  the magnitude of distance correlation if not zero is hard to interpret.  This fact is emphasized for example in \cite{jako:17}  (page 61) where distance covariance is characterized to be useful  to test for independence \eqref{eq: test} but much less so to measure  degree of dependence between random variables $X$ and $Y$ in general metric spaces. 


\id As a concrete example, we compare distance correlation/covariance with metric correlation/covariance  
for the case of distribution spaces with the Wasserstein metric. Writing $Q_1, Q_2$ for the quantile function functions corresponding to distributions $F_1, F_2$, the distance covariance $\text{dCov}(F_1,F_2)$ between $F_1$ and $F_2$ is found to correspond to 
\bea 
\text{dCov}(F_1,F_2)&=&E \left[ {\int_{0}^{1} \{Q_1(t)-Q'_1(t)\}^2 dt}  {\int_{0}^{1} \{Q_2(t)-Q'_2(t)\}^2 dt}\right]^{1/2} \\ && + \, E \left[ {\int_{0}^{1} \{Q_1(t)-Q'_1(t)\}^2 dt} \right]^{1/2}  E \left[ {\int_{0}^{1} \{Q_2(t)-Q'_2(t)\}^2 dt}\right]^{1/2}  \\ && -  \, 2 E \left[{\int_{0}^{1} \{Q_1(t)-Q'_1(t)\}^2 dt}  {\int_{0}^{1} \{Q_2(t)-{\tilde{Q}}_2(t)\}^2 dt}\right]^{1/2},
\eea
where $F'_1$ is an independent copy of $F_1$ and $F'_2,\, \tilde{F}_2$ are two independent copies of $F_2$. This expression for distance covariance is rather unintuitive, and it is hard to interpret as a measure for the strength of covariation between $F_1$ and $F_2$. 

\id A numerical comparison provides further  illumination. We implemented distance covariance as an alternative covariance/correlation for \fro \  in  a simulation study,  where we compared the utility of the proposed metric covariance with that of distance covariance \citep{szek:17,lyon:13} for carrying out FPCA of regular scalar-valued functional data. We consider a simple setting as in classical FDA where the time-varying random objects $X(s)$ are real valued for $s \in [0,1]$. In the following, we take $U_i$, $V_i$ and $Y_i$ to be distributed as $N(0,3)$, $N(0,1)$ and $N(0,0.25)$, respectively. As described in the simulation setting for time varying networks in Section \ref{sec: sim_net}, we take $\phi_1(t)$, $\phi_2(t)$ and $\phi_3(t)$  to be orthonormal polynomials derived from the Jacobi polynomials $P_n^{(\alpha,\beta)}(x)$ \cp{toti:05}, which are classical orthogonal polynomials for $\alpha, \beta > 1$.  The expressions for $\phi_1(t)$, $\phi_2(t)$ and $\phi_3(t)$  are
\bea
 \phi_1(t)&=&\frac{(P_3^{(1,0.5)}(2t-1))t^{0.25} (1-t)^{0.5}}{[\int_{0}^{1}(P_3^{(1,0.5)}(2t-1))^2t^{0.5} (1-t)dt]^{1/2}} \\  \phi_2(t)&=&\frac{(P_4^{(1,0.5)}(2t-1))t^{0.25} (1-t)^{0.5}}{[\int_{0}^{1}(P_4^{(1,0.5)}(2t-1))^2t^{0.5} (1-t)dt]^{1/2}}\\\phi_3(t)&=&\frac{P_5^{(1,0.5)}(2t-1))t^{0.25} (1-t)^{0.5}}{[\int_{0}^{1}(P_5^{(1,0.5)}(2t-1))^2t^{0.5} (1-t)dt]^{1/2}}.
\eea

\id We generated 1000 i.i.d. realizations $X_i(s)$ as follows on a fine grid of $[0,1]$, 
\begin{equation}
\label{eq2}
X_i(s)=U_i \phi_1(t)+V_i \phi_2(t)+ Y_i \phi_3(t).
\end{equation}
\begin{figure}
	\begin{center}
	\includegraphics[scale = .35]{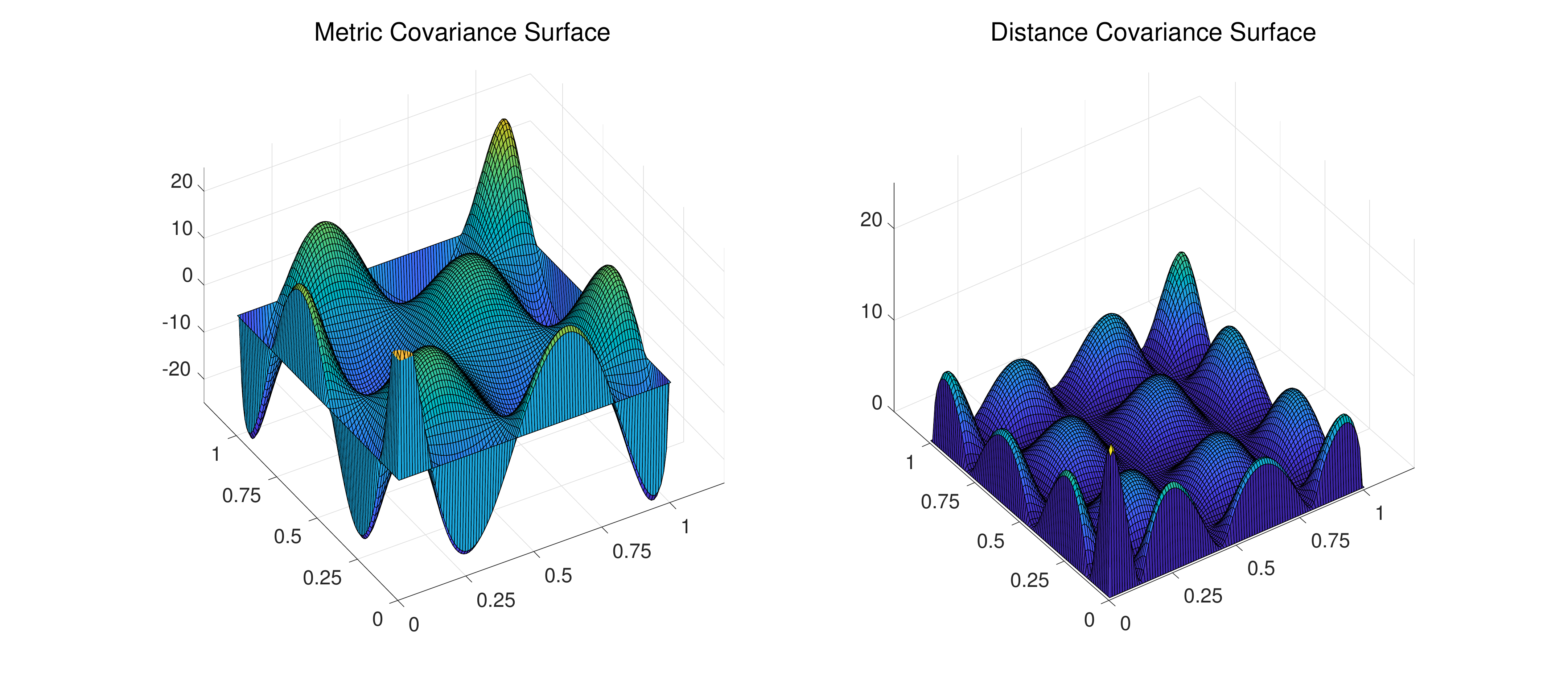}
	\end{center}
	\caption{Metric covariance surface (left) and distance covariance surface (right) for functional data generated according to model in \eqref{eq2}.}
	\label{fig: fa1}	
\end{figure}
\begin{figure}
	\begin{center}
	\includegraphics[scale = .33]{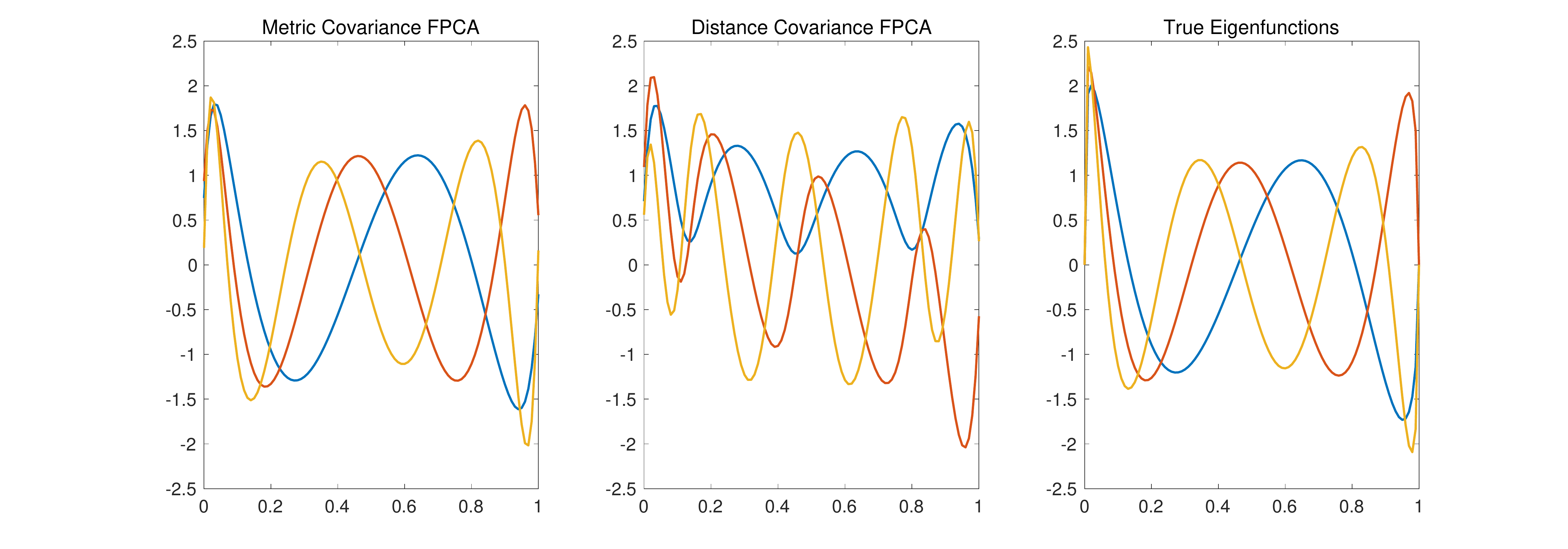}
	\end{center}
	\caption{Eigenfunctions obtained from using metric covariance based  (left panel) and distance covariance based (middle panel)  kernel for simulated functional data, generated according to model  \eqref{eq2}. Also shown are  the true underlying eigenfunctions (right panel).  The blue curves correspond to the first, the red curves to the second and the yellow curves to the third eigenfunction.}
	\label{fig: fa2}
\end{figure}
It is clear from the construction of $X_i$ that in model \eqref{eq2}, the first, second and third eigenfunctions are given by $\phi_1(t), \phi_2(t)$ and $\phi_3(t)$ respectively. We evaluated the estimated metric covariance and distance covariance surfaces on a fine grid, which led to the surfaces depicted in Figure \ref{fig: fa1},  and then obtained the first three eigenfunctions using these surfaces as covariance kernels. The resulting eigenfunctions are presented in Figure \ref{fig: fa2}.  As can be seen from Figures \ref{fig: fa1} and \ref{fig: fa2}, metric covariance delivers the eigenfunctions that one would expect for functional principal component analysis, while  distance covariance as an alternate notion of covariance leads to seemingly arbitrary and  
uninterpretable eigenfunctions, so is clearly not suitable in this context. 

\id In classical FPCA one aims to identify dominant modes of variation of functional data that are derived from the  eigenfunctions of the auto-covariance operator. The interpretation of these modes of variation provides valuable insights in many applications, and this is why interpretability of the eigenfunctions is important. 
Another major goal is to decompose the variation of functional data in a parsimonious and interpretable way into orthogonal directions. 
As illustrated in the simulation above, the lack of clear interpretation of the eigenfunctions associated with the distance covariance operator is a big hurdle for this program.  When using distance correlation, the corresponding distance variance and also the total variation have an unintuitive and complex form that  makes distance covariance rather unsuitable for our purposes. In contrast, the proposed metric covariance (\ref{mar}) works well for quantifying the variation and co-variation of random objects.  It gives rise to the total variation measure 
(\ref{tot}) for \fro \ and emerges as a bona fide extension of the proven and successful FPCA for scalar-valued functional data.

\subsection*{S3. \hs  Metric auto-covariance surfaces and eigenfunctions for New York taxi data}
\label{NY_groups}

\begin{figure}[H]
	\centering
	\begin{subfigure}{7cm}
		\centering\includegraphics[width=7cm]{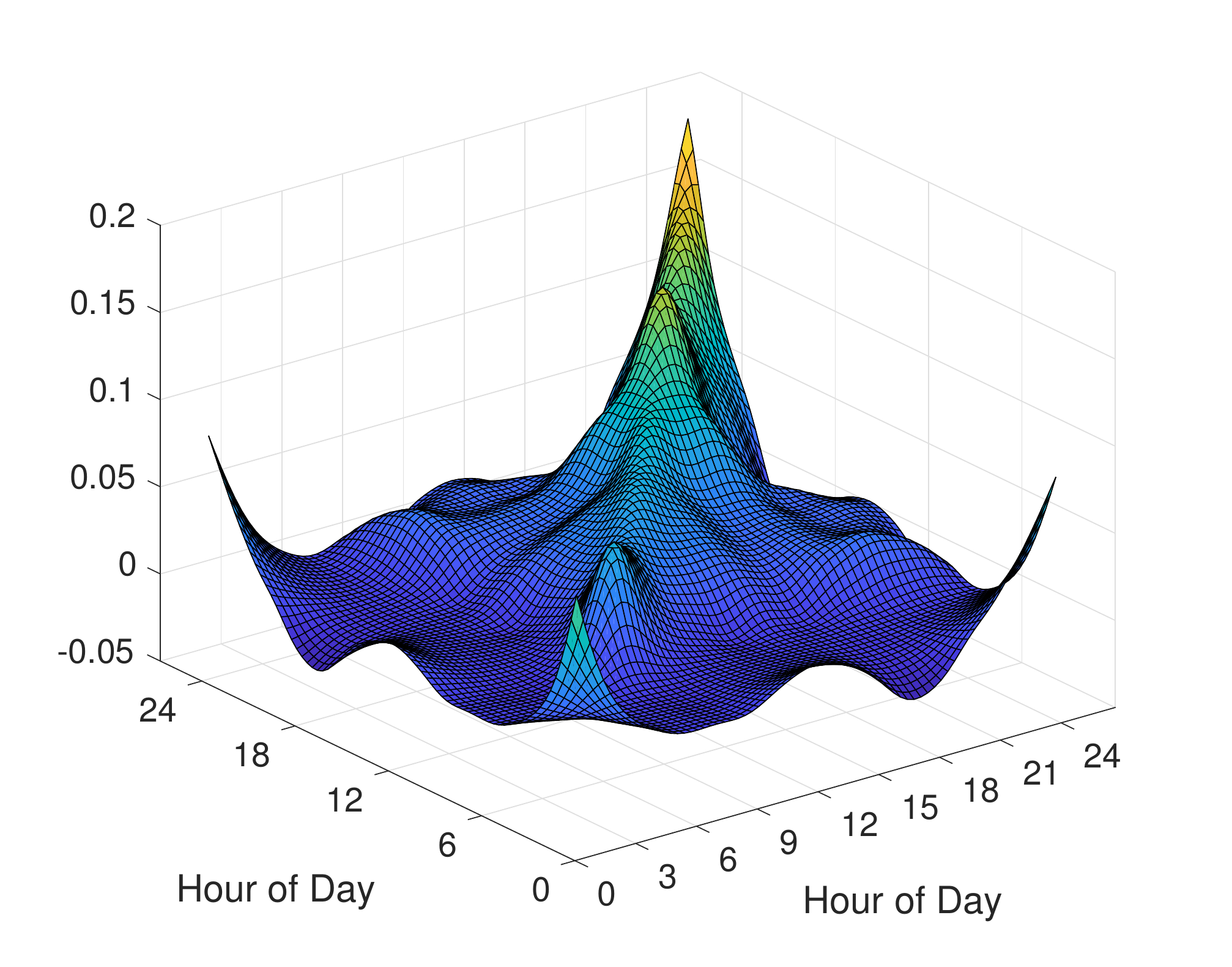}
		\caption{Mondays-Thursdays}
		\label{fig: f13}
	\end{subfigure}
	\hspace{0.2 in}
	\begin{subfigure}{7cm}
		\centering\includegraphics[width=7cm]{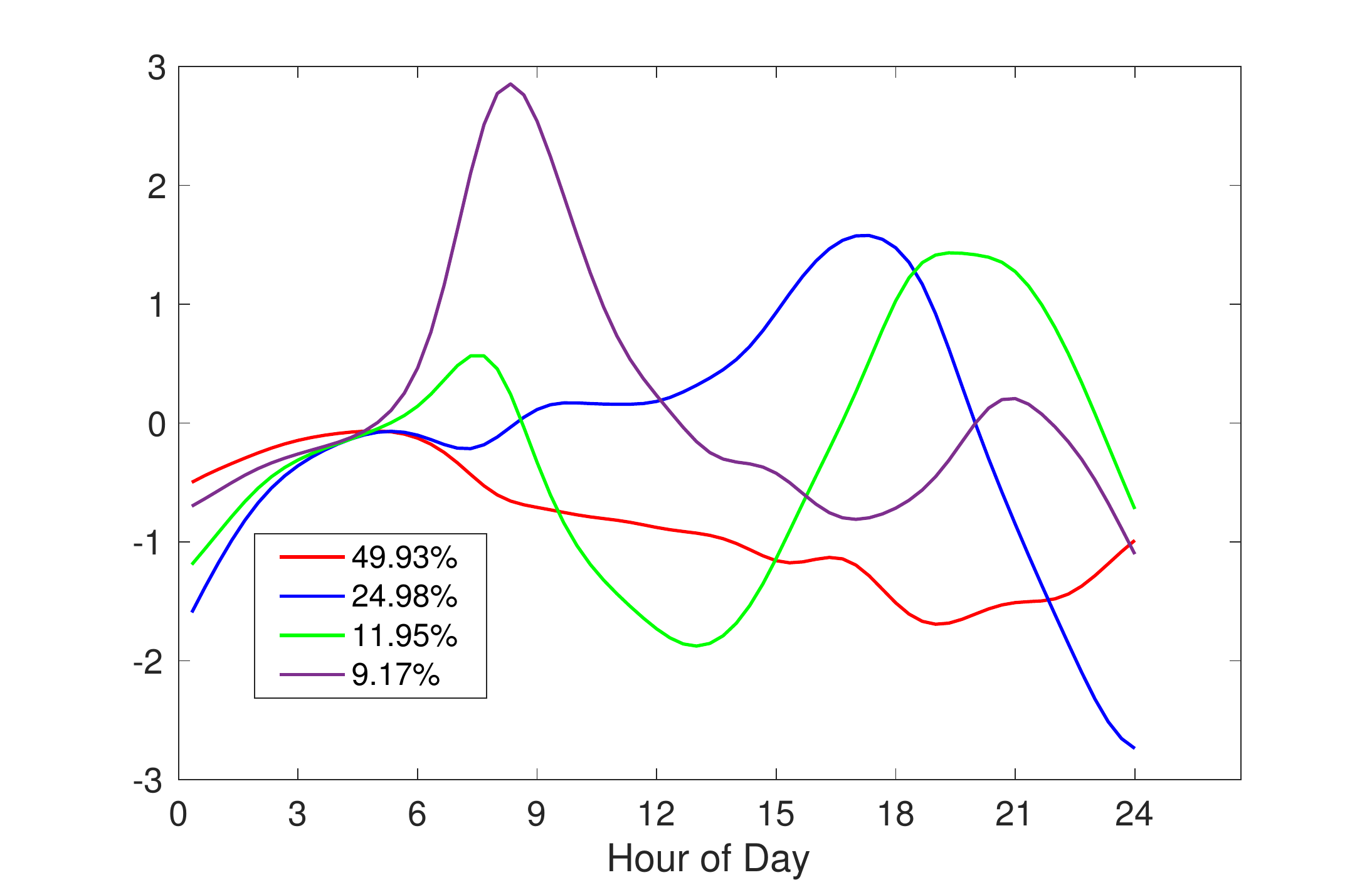}
		\caption{Eigenfunctions}
		\label{fig: f14}
	\end{subfigure}
	\begin{subfigure}{7cm}
		\centering\includegraphics[width=7cm]{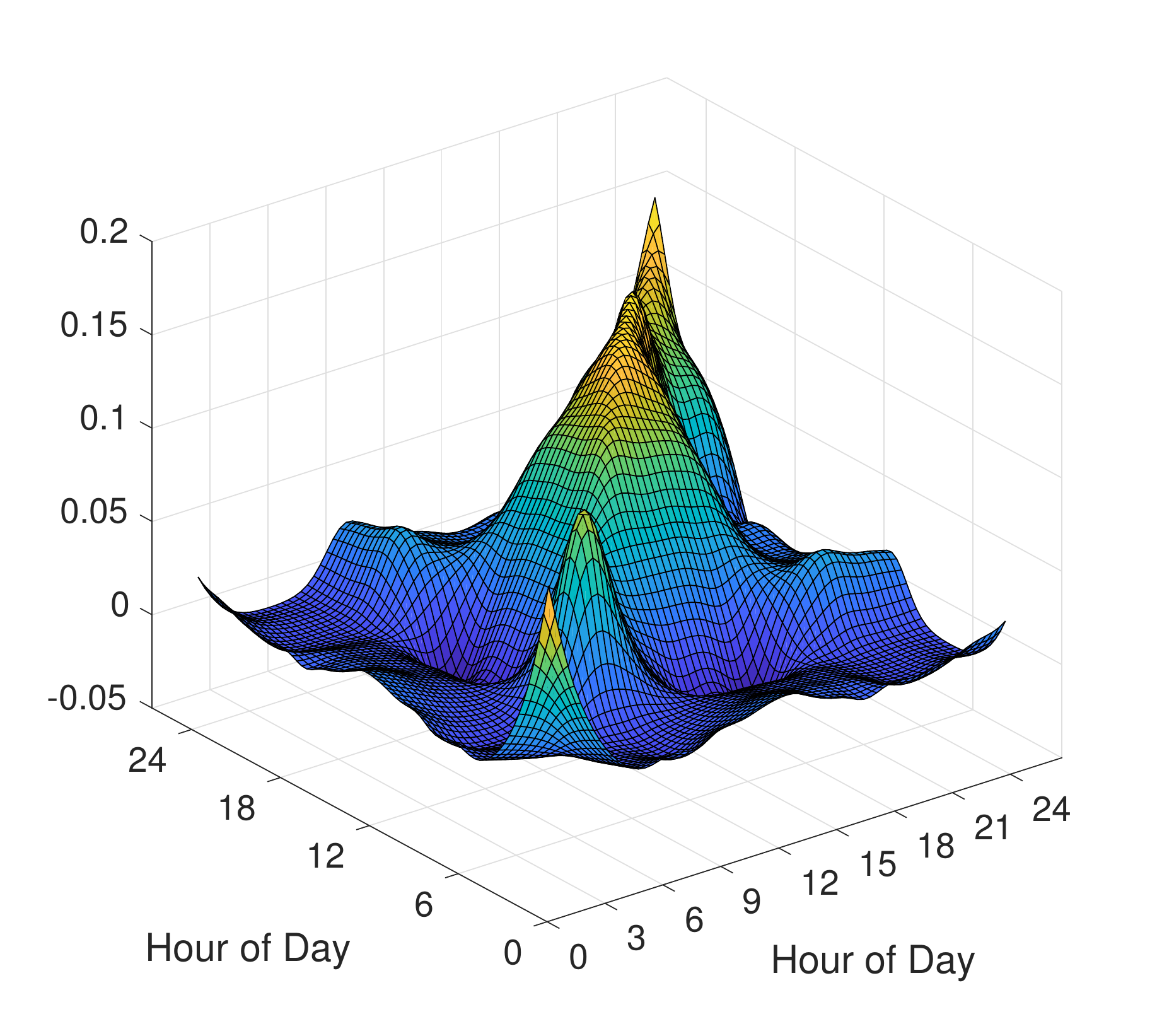}
		\caption{Fridays}
		\label{fig: f15}
	\end{subfigure}
	\hspace{0.2 in}
	\begin{subfigure}{7cm}
		\centering\includegraphics[width=7cm]{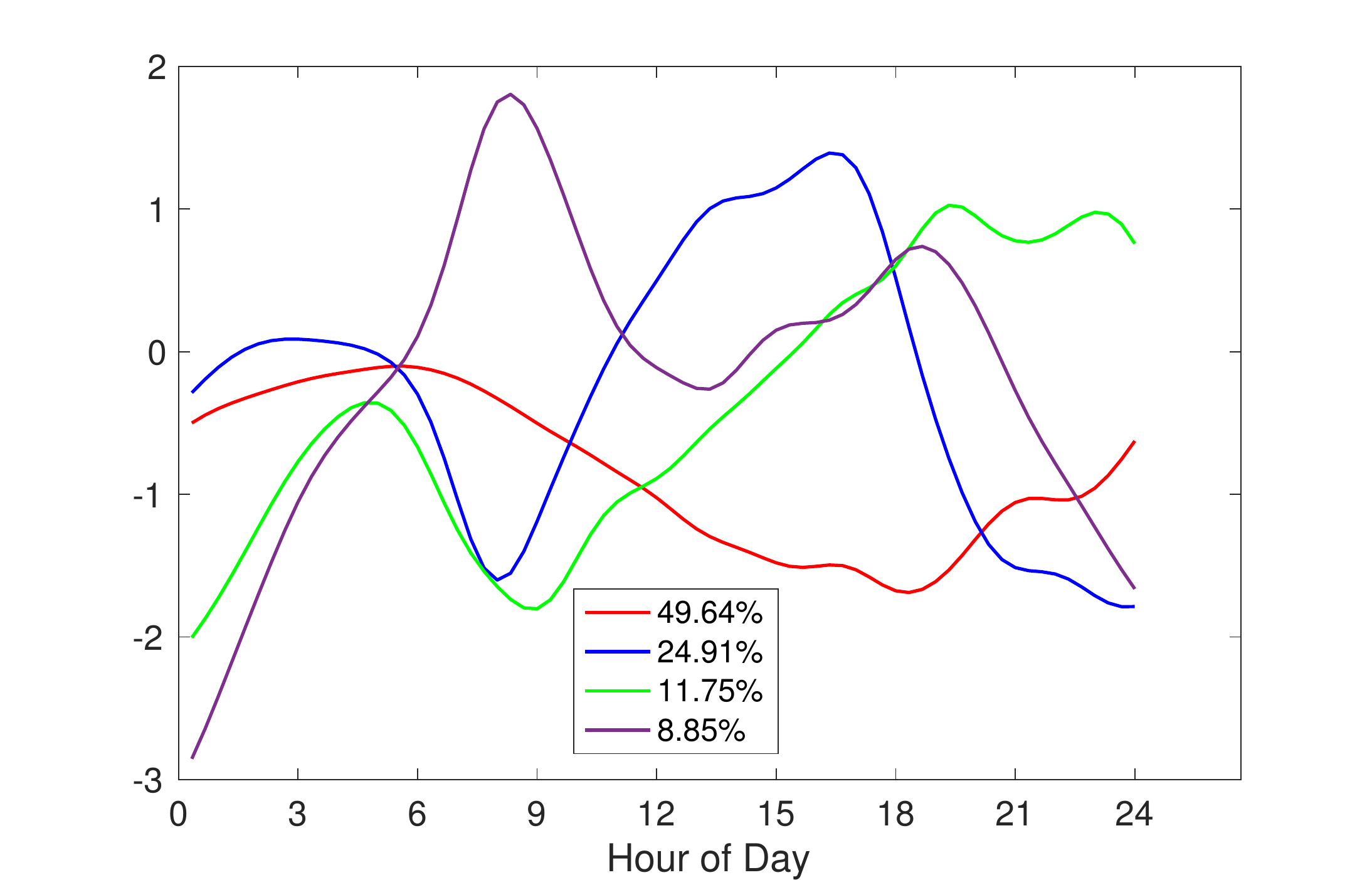}
		\caption{Eigenfunctions}
		\label{fig: f16}
	\end{subfigure}
	\begin{subfigure}{7cm}
		\centering\includegraphics[width=7cm]{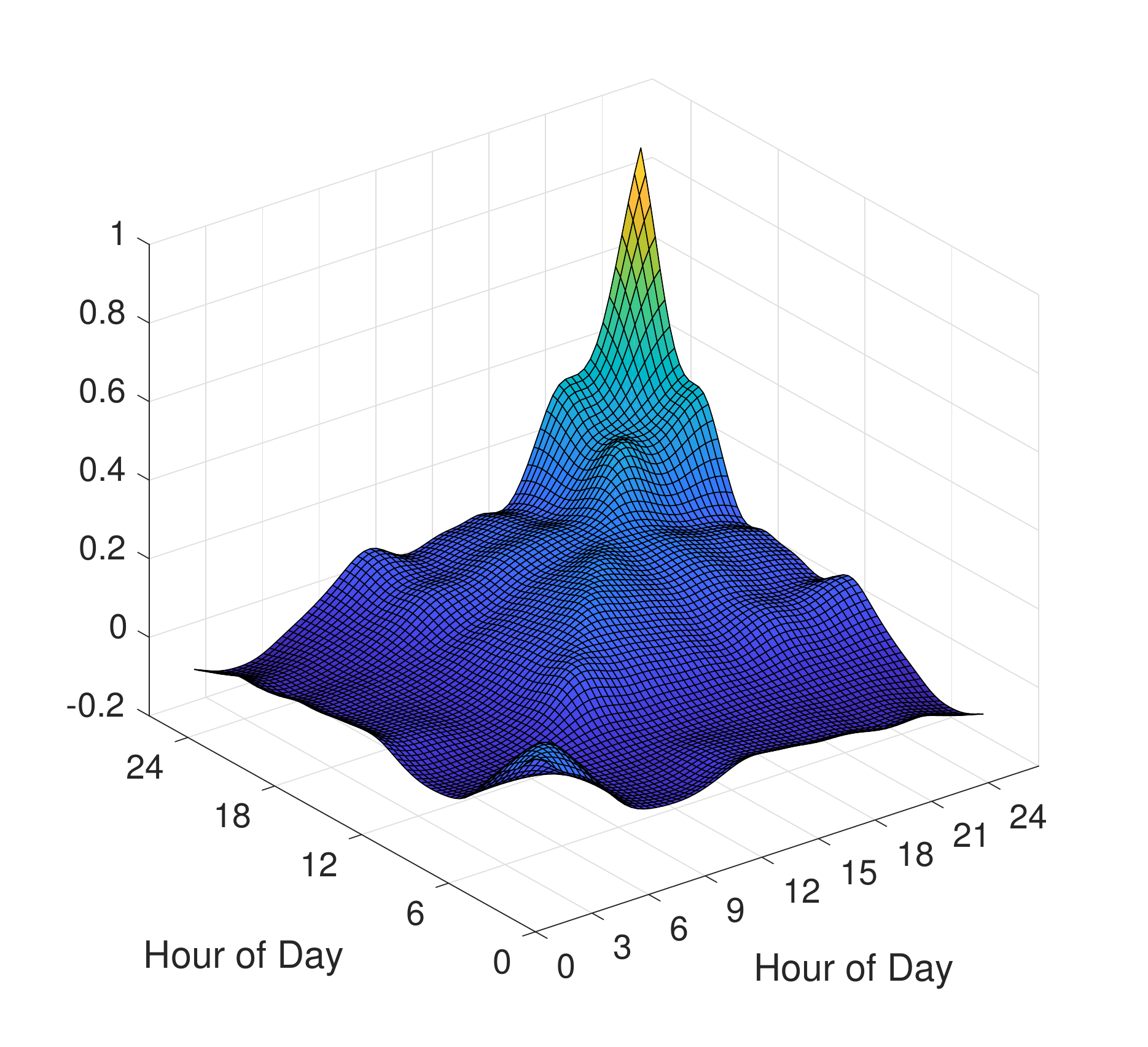}
		\caption{Weekends and Holidays}
		\label{fig: f17}
	\end{subfigure}
	\hspace{0.2 in}
	\begin{subfigure}{7cm}
		\centering\includegraphics[width=7cm]{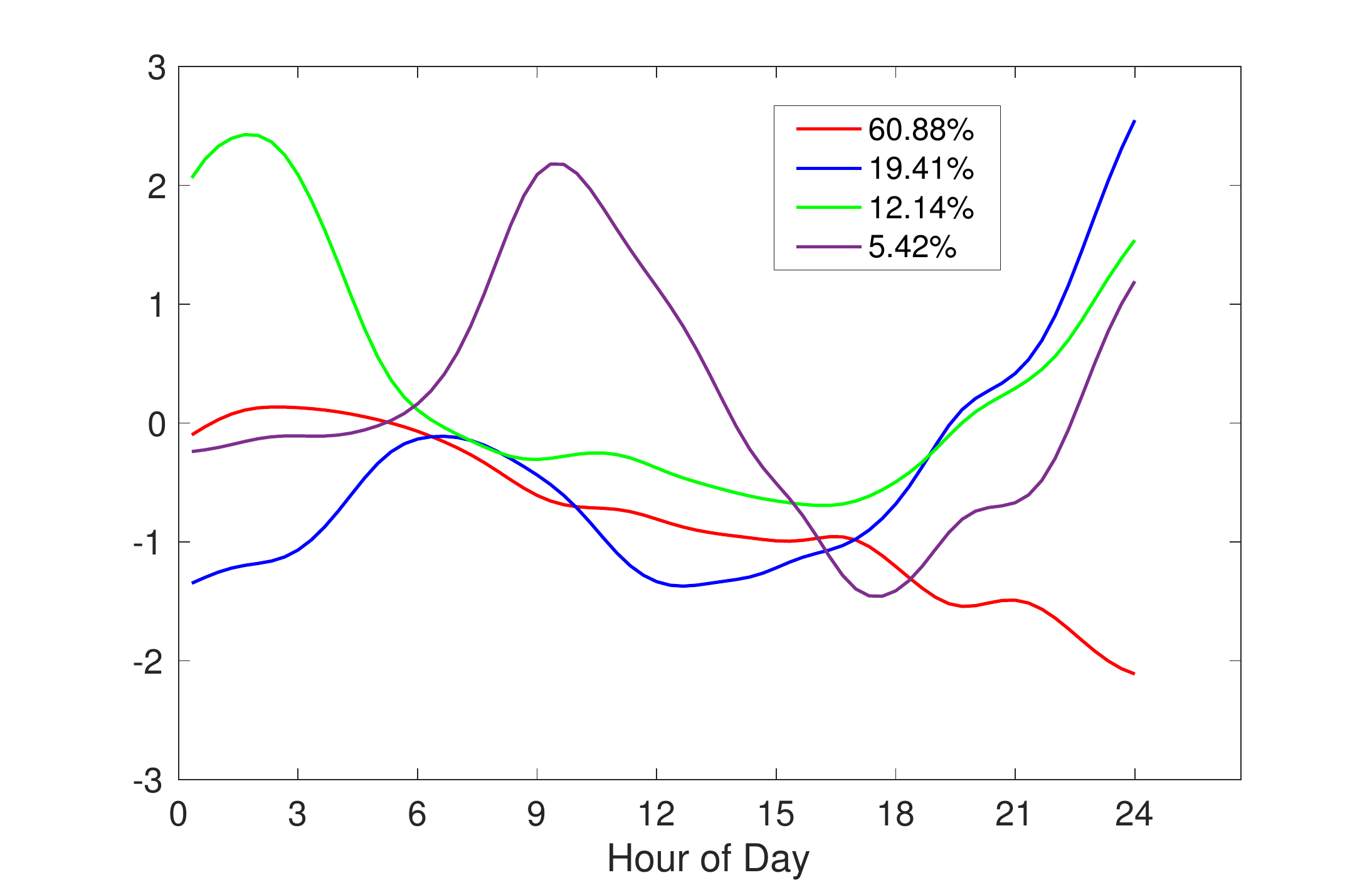}
		\caption{Eigenfunctions}
		\label{fig: f18}
	\end{subfigure}
	\caption{Estimated metric auto-covariance surface \eqref{eq: cov} (left) and associated  eigenfunctions (right), for the New York taxi data, viewed as time-varying networks, separated by groups of days.}
	\label{fig: f12}
\end{figure}

We repeated the analysis of the time-varying networks generated by the New York taxi data separately for three groups of days, namely the weekdays Monday-Thursday (group 1),  Fridays and weekends (group 2) and holidays (group 3).   The results are visualized in  Figure \ref{fig: f12}. This figure clearly  indicates  that the  metric covariance structure and the eigenfunctions differ across the groups. The \F \ integrals for the dominant eigenfunctions reveal different aspects of variation, both within and between daily networks in three groups and are presented in the movies ``week.mov", ``friday.mov" and ``Hol.mov" which are included in the supplementary materials. In the movie frames, the top left panels correspond to the FPCs for the first eigenfunction, the top right panels to those of the second, the bottom left panels to those of the  third and the bottom right panels to those of the fourth eigenfunction. The edge weights in the graphs are proportional to their line widths.

\newpage

\begin{figure}[H]
	\centering
	\includegraphics[scale = .45]{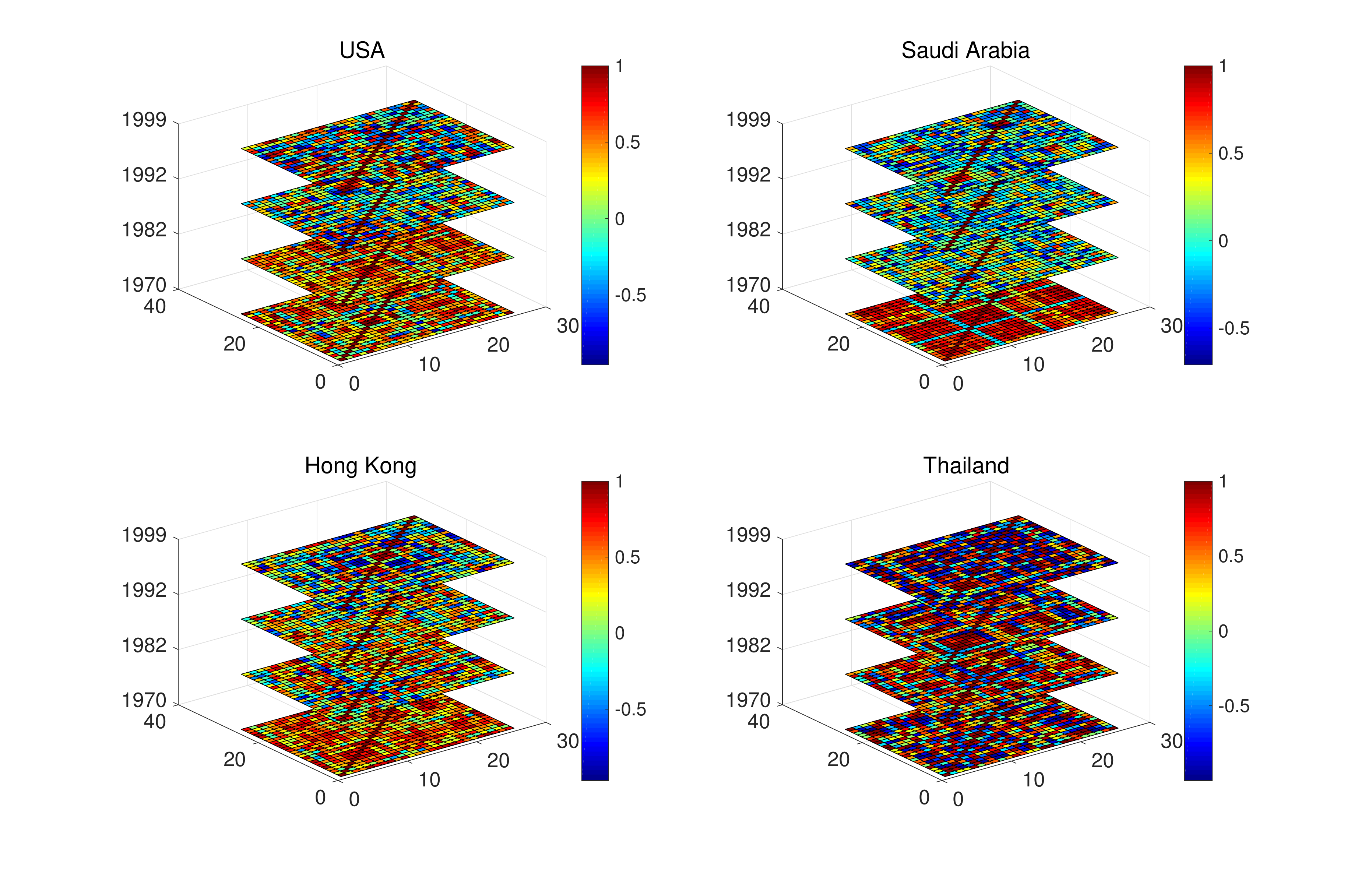}
	\caption{International commodity trade correlation matrices for 1970 (first slice from bottom), 1982 (second slice from bottom), 1992 (third slice from bottom) and 1999 (top slice) for four contries.}
	\label{fig: f23}
\end{figure}

{
	\subsection*{S4. \hs  Additional Visualization for the World Trade Data analysis}
	\label{trade_vis}
	We selected USA, Saudi Arabia, Hong Kong and Thailand and display their time evolving inter-commodity trade correlations as obtained from the fitted model  for the years 1970, 1982, 1992, 1999 (bottom to top) in Figure \ref{fig: f23}.}

\id {We also  computed the \F \ integral covariance matrices for the first four eigenfunctions. For visualization of commodities trade similarities these \F \ integral covariance matrices were converted to correlation matrices that can be viewed in the movie ``trade.mov", available in the online supplement. In the movie frames, the FPCs for the first, second, third and fourth eigenfunctions 
	are at the top left, top right, bottom left and bottom right, respectively.} 

\newpage

{\subsection*{S5. \hs  Description of Movies}}
\label{mov}
The following is a  list of the movies that have been included as supplementary materials and a brief discussion of  their content.

\vspace{2cm}

\begin{center}
	\begin{tabular}{ |p{3 cm}|p{11 cm}|} 
		\hline
		Filename &  Content Description \\ \hline
		{networks.mov} & Object FPCs obtained as \F \ integrals \eqref{eq: int}  for one randomly chosen simulation run for $n=50$,  using the model in \eqref{net_gen} for time-varying networks in section \ref{sec: sim_net}.  \\ \hline 
		mean\_{males}.mov &  Estimated \F \ mean function for males represented as density functions indexed by year, derived from the yearly sample average of the quantile functions of the countries included in the mortality data in section \ref{mort} for each calendar year.   \\ \hline
		mean\_{females}.mov & Estimated \F \ mean function for females. The description is the same as for males.  \\ \hline
		mean\_NY.mov &  Estimated \F \ mean function represented as time varying network adjacency matrices, obtained for each 20 minute time interval by averaging the network adjacency matrices over 363 daily networks for the New York taxi data as described in section \ref{taxi}.  \\ \hline
		week.mov &  \F \ integrals represented as network adjacency matrices for the dominant eigenfunctions,  obtained from the analysis of the New York taxi data as described in section \hyperref[NY_groups]{S3} of the supplement,  for  weekdays. \\ \hline
		friday.mov &  Same as previous, for Fridays. \\ \hline 
		Hol.mov &  Same as previous,  for weekends and special holidays. \\ \hline 
		trade.mov &  \F \ integrals represented as covariance matrices for the dominant eigenfunctions for the trade dataset as described in section \ref{trade}.\\ \hline
	\end{tabular}
\end{center}

\newpage

\subsection*{S6. \hs  Data Descriptions}
\label{data_des}
\subsubsection*{S6.1 Zones in Manhattan, New York}
New York City Taxi and Limousine Commission (NYC TLC) provides records on pick-up and drop-off dates/times, pick-up and drop-off latitudes and longitudes, trip distances, itemized fares, rate types, payment types, and driver-reported passenger counts for yellow and green taxis. The data are  available at \url{http://www.nyc.gov/html/tlc/html/about/trip_record_data.shtml}. The polygon shape  files available at this website represent the boundaries zones for taxi pickups as delimited by the New York City Taxi and Limousine Commission (TLC). The latitudes and longitudes in New York are split into 6 boroughs: Bronx, Brooklyn, Newark Liberty International Airport, Manhattan, Queens and Staten Island. Since yellow taxis operate predominantly in Manhattan, we consider so-called  towns in Manhattan which are further grouped into 10 zones as described in the following Table. 
We excluded the islands from our study. For a description of towns, we  refer to Figure \ref{fig: f2}.

\vspace{-2cm}

\begin{center}
	\begin{tabular}{ |l|p {14 cm}|} 
		\hline
		Zone & Towns \\  \hline \hline 
		1 & Inwood, Fort George, Washington Heights, Hamilton Heights,
		Harlem, 
		East Harlem \\  
		2 & Upper West Side, Morningside Heights,
		Central Park
		\\  
		3 &  Yorkville, Lenox Hill, Upper East Side
		\\ 
		4 &  Lincoln Square, Clinton, Chelsea, Hell's Kitchen \\ 
		5 &  Garment District, Theater District\\ 
		6 & Midtown \\ 
		7&  Midtown South\\ 
		8 & Turtle Bay, Murray Hill, Kips Bay, Gramecy Park,
		Sutton, Tudor, Medical City, Stuy Town
		\\ 
		9 & Meat packing district, Greenwich Village, West Village, Soho,
		 Little Italy,  ChinaTown, Civic center, Noho
		\\
		10 & Lower East Side, East Village, ABC Park, Bowery, Two Bridges,
		Southern tip, White Hall, Tribecca, Wall Street
		\\ 
		\hline
	\end{tabular}
	\label{tab:B-1}
\end{center}
\begin{figure}[H]
	\centering
	\includegraphics[scale = .8]{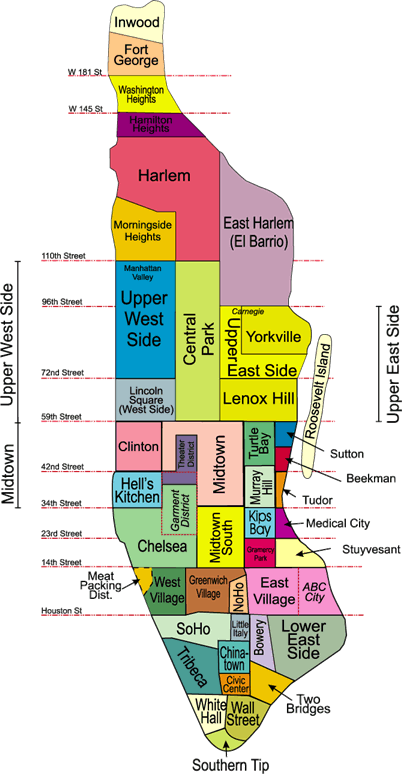}
	\caption{Towns in Manhattan, New York.}
	\label{fig: f2}
\end{figure}
\subsubsection*{S6.2 Trade Data}

The countries chosen for the analysis were Morocco, Tunisia, Egypt, Canada, USA, Argentina, Brazil, Chile, Mexico, Venezuela, Dominican Republic, Israel, Japan, Cyprus, Lebanon, Saudi Arabia, United Arab Emirates, Turkey, Hong Kong, Indonesia, Korea Republic, Malaysia, Philippines, Singapore, Thailand, Taiwan, China, Belgium-Luxemburg, Denmark, France, Greece, Ireland, Italy, Netherlands, Portugal, Spain, UK, Austria, Finland, Norway, Sweden, Switzerland, Malta, Bulgaria, Australia and New Zealand.  A list of the traded commodities can be found in the following table, 

\newpage
\begin{center}
	\begin{tabular}{ |l|p {10 cm}|} 
		\hline
		Number &  Products\\  \hline \hline
		1 & Sugar, Honey\\ 
		2 & Road Vehicles \\
		3 & Fruits and Vegetables\\
		4 & Non metallic Minerals Manufactures \\
		5 & Coffee, Tea, Cocoa, Spices \\
		6 & Tobacco and Tobacco Manufactures\\
		7 & Textiles, Yarns, Fabrics \\
		8 & Printed Books, Maps, Charts, Paper, Stationery \\
		9 & Beverages \\
		10 & Chemical Materials, Products \\
		11 & Machineries \\
		12 & Transport Equipments \\
		13 & Rubber\\
		14 & General Industrial Machinery \\
		15 & Dairy Products, Eggs \\
		16 & Fish and Seafood \\
		17 & Cereals \\
		18 & Petroleum \\
		19 & Dye \\
		20 & Medicines \\
		21 & Oil, Perfumes, Toilet, Cleansing \\
		22 & Paper, Paper Board, Articles \\
		23 & Iron and Steel \\
		24 & Manufacture of Metals \\
		25 & Power Generating Machinery \\
		26 & Telecommunications, Sound Recording, Reproducing Equipments \\
		\hline
	\end{tabular}
\end{center}
\end{document}